\documentclass[]{aa} 

\usepackage{natbib,twoopt}
\usepackage[breaklinks=true]{hyperref} 
\bibpunct{(}{)}{;}{a}{}{,} 
\makeatletter
\newcommandtwoopt{\citeads}[3][][]{\href{http://adsabs.harvard.edu/abs/#3}%
{\def\hyper@linkstart##1##2{}%
\let\hyper@linkend\@empty\citealp[#1][#2]{#3}}}
\newcommandtwoopt{\citepads}[3][][]{\href{http://adsabs.harvard.edu/abs/#3}%
{\def\hyper@linkstart##1##2{}%
\let\hyper@linkend\@empty\citep[#1][#2]{#3}}}
\newcommandtwoopt{\citetads}[3][][]{\href{http://adsabs.harvard.edu/abs/#3}%
{\def\hyper@linkstart##1##2{}%
\let\hyper@linkend\@empty\citet[#1][#2]{#3}}}
\newcommandtwoopt{\citeyearads}[3][][]%
{\href{http://adsabs.harvard.edu/abs/#3}
{\def\hyper@linkstart##1##2{}%
\let\hyper@linkend\@empty\citeyear[#1][#2]{#3}}}
\renewcommand*\aa@pageof{, page \thepage{} of \pageref*{LastPage}}
\usepackage{graphicx}
\usepackage{txfonts}
\usepackage[dvipsnames]{xcolor}

\begin{document}

   \title{Understanding of the properties of neural network approaches for transient light curve approximations \thanks{Table \ref{tab:peak_ztf_sk} is available in electronic form at the CDS via anonymous ftp to cdsarc.cds.unistra.fr (130.79.128.5) or via \url{https://cdsarc.cds.unistra.fr/cgi-bin/qcat?J/A+A/}.}}
   

   \author{Mariia Demianenko\inst{1},
          Konstantin Malanchev\inst{2,3},
          Ekaterina Samorodova\inst{4},
          Mikhail~Sysak\inst{5},
          Aleksandr~Shiriaev\inst{6},
          Denis~Derkach\inst{7},
          Mikhail~Hushchyn\inst{7} \fnmsep\thanks{mhushchyn@hse.ru}}
   \institute{Max-Planck-Institut für Astronomie, Königstuhl 17, 69117 Heidelberg, Germany,
              \email{sekretariat@mpia.de}
         \and
             Department of Astronomy, University of Illinois at Urbana-Champaign, 1002 West Green Street, Urbana, IL 61801, USA
         \and
             Lomonosov Moscow State University, Sternberg Astronomical Institute, Universitetsky pr. 13, Moscow 119234, Russia
         \and
             Lomonosov Moscow State University, Department of Mechanics and Mathematics, Leninskie gory 1, Moscow 119234, Russia
         \and
             Moscow Institute of Physics and Technology, Institutskii Pereulok 9, Dolgoprudny, Moscow Region 141700, Russia
          \and 
             Moscow Polytechnic University, Tverskaya street, 11, Moscow 125993, Russia
          \and
             HSE University, 11 Pokrovsky Bulvar, Moscow 101000, Russia
             }

   \date{Received October , 2022; accepted }

 
\abstract{Modern-day time-domain photometric surveys collect a lot of observations of various astronomical objects and the coming era of large-scale surveys will provide even more information on their properties. Spectroscopic follow-ups are especially crucial for transients such as supernovae and most of these objects have not been subject to such studies. }{Flux time series are actively used as an affordable alternative for photometric classification and characterization, for instance, peak identifications and luminosity decline estimations. However, the collected time series are multidimensional and irregularly sampled, while also containing outliers and without any well-defined systematic uncertainties. This paper presents a search for the best-performing methods to approximate  the observed light curves over time and wavelength for the purpose of generating time series with regular time steps in each passband.}{We examined several light curve approximation methods based on neural networks such as multilayer perceptrons, Bayesian neural networks, and normalizing flows to approximate observations of a single light curve. Test datasets include simulated PLAsTiCC and real Zwicky Transient Facility Bright Transient Survey light curves of transients.}{The tests demonstrate that even just a few observations are enough to fit the networks and improve the quality of approximation, compared to state-of-the-art models. The methods described in this work have a low computational complexity and are significantly faster than Gaussian processes. Additionally, we analyzed the performance of the approximation techniques from the perspective of further peak identification and transients classification. The study results have been released in an open and user-friendly {\tt\string Fulu} Python library available on GitHub for the scientific community.}{}

   \titlerunning{Fulu: light curves neural network approximation}
   \authorrunning{Demianenko et al.}
   \maketitle
%

\section{Introduction}

\label{sec:intro}

Time-domain photometric surveys such as the All-Sky Automated Survey (ASAS)~\citep{2002397}, MASTER Robotic Network~\citep{2010AdAst2010E..30L}, the Zwicky Transient Facility (ZTF)~\citep{Bellm_2019}, Young Supernova Experiment (YSE)~\citep{2021ApJ...908..143J}, and others~\citep{Drake_2009,2018PASP..130f4505T} collect numerous observations of various astronomical objects. These photometric data are analyzed both in near-real-time and in archives by survey teams, alert brokers, such as ANTARES~\citep{2021AJ....161..107M}, Alerce~\citep{2021AJ....161..242F}, Fink~\citep{2021MNRAS.501.3272M}, and others, and the scientific community in order to detect, classify, and characterize variable astronomical sources. Most of the objects have never received spectroscopic, multiwavelength, or any other follow-up observations, which are especially crucial for transients such as supernovae, making the photometric classification and characterization methods significant. Besides that, the upcoming Vera C. Rubin Observatory Legacy Survey of Space and Time (LSST)~\citep{2019ApJ...873..111I} will have a  transient detection rate that is 10-100 times higher than all existing surveys combined, which makes this issue even more acute.

Photometric time series collected by ground-based optical surveys usually have the following properties: 1) irregular sampling due to seasonal factors, weather conditions, and survey strategies; 2) multidimensionality if the objects are observed in multiple passbands; 3) numerous outliers caused by overlapping sources and photometric pipeline artifacts; and 4) poorly defined systematic uncertainties and/or underestimated stochastic uncertainties. The main challenge revealed in the era of big data research is how to extract useful information most efficiently from imperfect available data. In this respect, the application of machine learning and artificial intelligence methods is a promising solution.

Supernovae Type Ia (SNe~Ia) light curves are commonly used to train machine learning models applied to SNe classification and characterization problems. SNe~Ia were found to be standard candles via the relation between their peak luminosity and the luminosity decline rate~\citep{Rust1974, Pskovskii1977,1993ApJ...413L.105P, Riess_etal1996}. This property of SNe~Ia is widely used for luminosity distance measurements and consequent cosmological analysis~\citep{1998AJ....116.1009R,1999ApJ...517..565P,2019ApJ...881...19J}. In the upcoming era of large photometric surveys, it becomes crucial to have a robust pipeline which finds SNe~Ia among other transient and variable sources, and extracts its peak flux, decline rate, and color.

Recently, a variety of machine learning approaches for transients and variable stars classification have been presented. Participants of the Supernova Photometric Classification Challenge~\citep{Kessler_2010} used many different machine-learning and classification approaches including random forests (RF) and neural networks. \citet{Richards_2011} applied the RF algorithm for a variable star classification, where the light curves were represented by a set of periodic and non-periodic features. A list of different statistical parameters such as the mean, skewness, standard deviation, and kurtosis are presented in~\citet{refId0_Ferreira} and~\citet{10.1111/j.1365-2966.2011.18575.x} for the classification with RF.~\cite{10.1093/mnras/stx3222} solved the problem using support vector machines (SVM), neural networks and RF classifiers with 18 input features. This paper describes the scatter and correlation between observations of light curves. Other similar approaches are presented in~\citet{2010IJMPD..19.1049B} and \citet{baron2019machine}.

The further evolution of this approach is closely related to the development of neural network models. The method proposed in~\citet{8280984} includes a transformation of the light curves into 2D histograms, where each dimension corresponds to a difference between magnitudes and timestamps for each pair of observations in the light curve. The histograms are treated as images and used as inputs of a convolutional neural network (CNN) for the model training. Similar differences are used in~\citet{10.3389/fspas.2021.718139} as inputs for a 1D CNN and long short-term memory (LSTM) model for the variable stars classification. \citet{10.1093/mnras/sty2836} take the difference between consecutive time and magnitude values in a light curve. These vectors of differences form a 1D image with two channels, which is passed to a 1D CNN model for solving the classification problem. A similar approach is described in~\citet{10.1093/mnras/staa350} and \citet{2020MNRAS.491.4277M}, where the matrix of the differences in time and magnitude are considered as sequences and used as inputs for a recurrent neural network (RNN) classifier. In~\citet{Dobryakov2021} a linear interpolation of light curve values is used as well as the differences in magnitude for each consecutive pair of data points as inputs for various classification algorithms. An RNN autoencoder in~\citep{Naul_2017} learns a latent representation of the light curves. It then processes the curve observations as a sequence. The obtained representation is used as an input of a RF algorithm for object classification.

One more method is based on different approximation approaches of the light curves and uses the best-fit parameters of phenomenological functions as features for classification models. In~\citet{refId0},~\citet{2019ApJ...884...83V}, and~\citet{2021AJ....161..141S} light curves are approximated using specific parameterized models. The fitted parameters of these models as well as their uncertainties are used as inputs for different classification algorithms. One of the first applications of neural networks for supernovae classification based on this approach is described in~\citet{10.1093/mnras/sts412}. A similar model is used in~\citet{10.1111/j.1365-2966.2011.18514.x} for the classification of simulated Dark Energy Survey (DES) light curves using boosting decision trees (BDTs) and kernel density estimation (KDE) methods. Other functions for light curve approximations are demonstrated in~\citet{refId0_Guy},~\citet{10.1111/j.1365-2966.2011.18514.x}, and~\citet{refId0_Taddia}. Different parameterization methods and wavelet decomposition approaches for input feature generation are presented in~\citet{Lochner_2016} and tested in photometric supernova classification using a list of machine learning algorithms.

A poor approximator choice can negatively affect the classification quality and the results of the analysis. Parametric functions usually are not connected with physical processes in transients, and also have limited generalizability. Therefore, it is inefficient to use this approach for datasets of different object types since each class requires its own custom parametric function fit. However, a Gaussian process (GP) model is used in~\citet{10.1093/mnras/stx2109} as an alternative for the light curve approximation in each passband. A range of parameterizations and classification approaches are considered in the Photometric LSST Astronomical Time-series Classification Challenge (PLAsTiCC,~\citealp{hlozek2020results}). The winner of the challenge demonstrates that approximation of the light curves using GP may lead to better results than other methods~\citep{2019AJ....158..257B}. Currently, this approach is the state-of-the-art method for the SN light-curve pre-processing (see, e.g., \citealt{2021AJ....162...67Q,2022ApJS..258...23A,10.1093/mnras/stac3672} for classification and \citealt{2019MNRAS.489.3591P,2021ApJS..255...24V,2021A&A...650A.195I,10.1093/mnras/stac2582} for anomaly detection) and we use it as a baseline result.

GP has also become a popular method of SN light curve characterization since it was first used for standardization of the peak SN~Ia absolute magnitude \citep{2013ApJ...766...84K}. \citet{2022MNRAS.512.3266M} proposed a GP instead of commonly used light-curve templates for the data-driven SN~Ia fitting. \citet{10.1093/mnras/stac3523} use a GP-approximated light curve to get the rising time and the decline rate of SN Type~II and IIb and study their morphology and genealogy. \citet{10.1007/978-3-030-96600-3_5} proposed using 2D GP for light-curve forecasting to optimize the follow-up strategy.

In this work, we consider different neural network models such as multilayer perceptron (MLP), Bayesian neural networks (BNN), and normalizing flows (NF) as alternatives to the light curve approximation. We demonstrate that these models trained on observations of a single light curve provide a high-quality approximation. To prove it, we compare the approximation metrics of the neural network methods with the GP. In addition, we explore how the models influence the metrics of indirect physical-motivated tasks such as binary classification for various transients including SNe~Ia as well as the identification of the light-curve peak. As~\citet{Dobryakov2021} showed a deterioration in quality metrics when applying machine learning models to real data compared to simulated data, we test our models on both the simulated PLAsTiCC and real ZTF Bright Transient Survey (BTS) datasets.

While neural network methods could be adopted for catalog-level studies aggregating the data from different objects into a single model, we focus our research on methods for the multipassband light curve approximation of individual objects.
Aggregation models potentially could improve the average quality of catalog-level light curve approximation, but they also introduce a bias for every single object.
This bias may affect consequent studies, such as classification, because the behavior of the light curves of the type dominating the data could be transferred to the objects of other types. 
The single-object approach makes our algorithms work out-of-the-box for different domains and object classes with predictable performance.
We present open-source Python~3 package \texttt{Fulu}\footnote{\url{https://github.com/HSE-LAMBDA/fulu}}, implementing the Neural Network light-curve approximation models described in this paper.

\section{Problem statement}

We investigate the problem of approximation of irregular time series consisting of non-simultaneous observations in different photometric passbands.
The main requirement for all models is considered to be the ability to find correlations among the observations both over time and across different passbands, as well as to calculate an approximation of the light curve within the time and wavelength ranges of the input data.
Within this approach, a single training dataset consists of the light curve of a single object only.
Thus, we prepared an independent approximation model for each object, ignoring any other objects.

We describe each point of a light curve by an observed flux value, $y_i$, its error, $\varepsilon_i$, a timestamp, $t_i$, and a photometric passband characterized by an effective wavelength, $\lambda_i$. For simplicity, we defined an input feature vector, $x_i = (t_i, \lambda_i)^{T}$. Then, the light curve is represented by a conditional probability density distribution, $p(y|x)$. Here, we assume that the flux $y$ is a random variable and its values, $y_i$, are sampled from the distribution below:

\begin{equation}
y_i \sim p(y|x_i).
\end{equation}

\begin{figure*}
\centering
\includegraphics[width=0.47\linewidth]{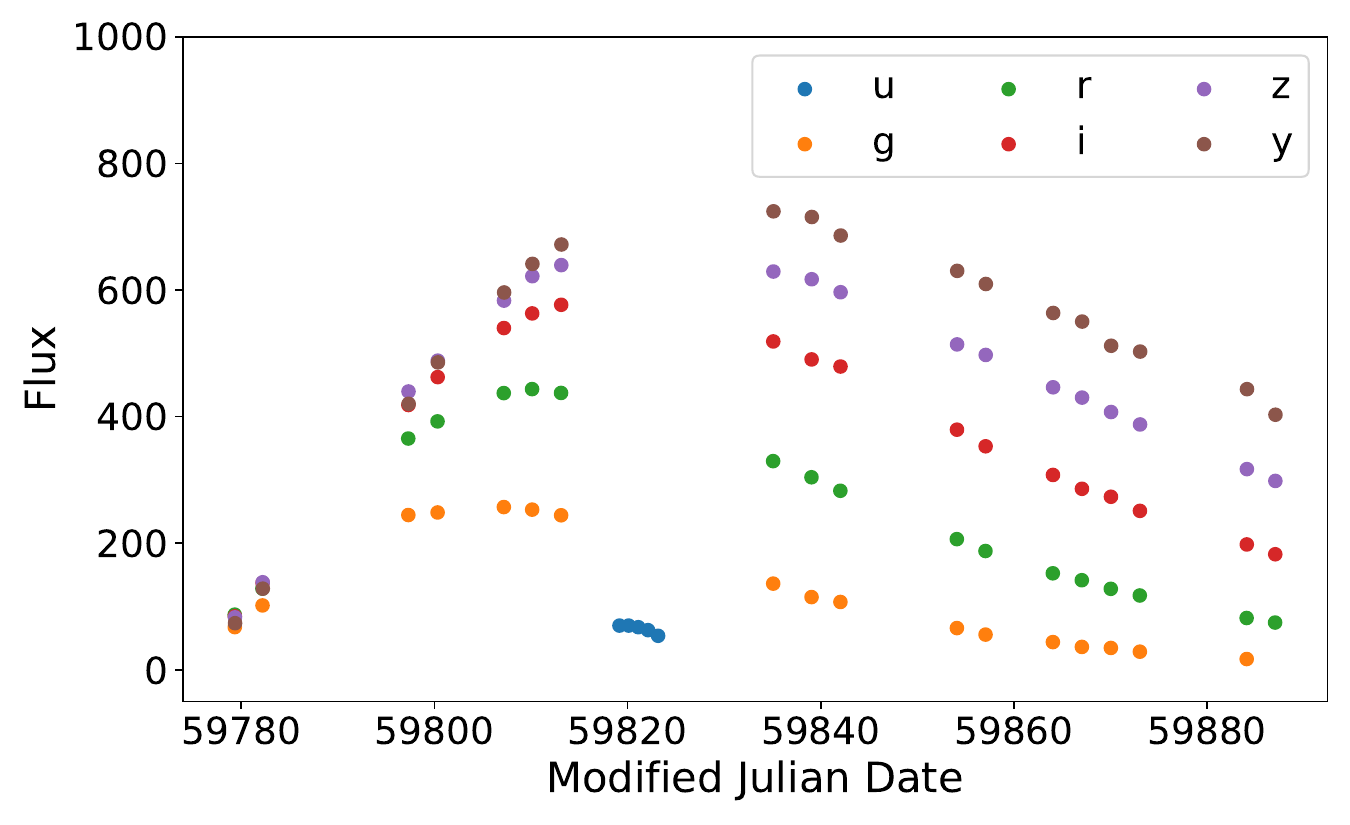}{(a)}
\includegraphics[width=0.47\linewidth]{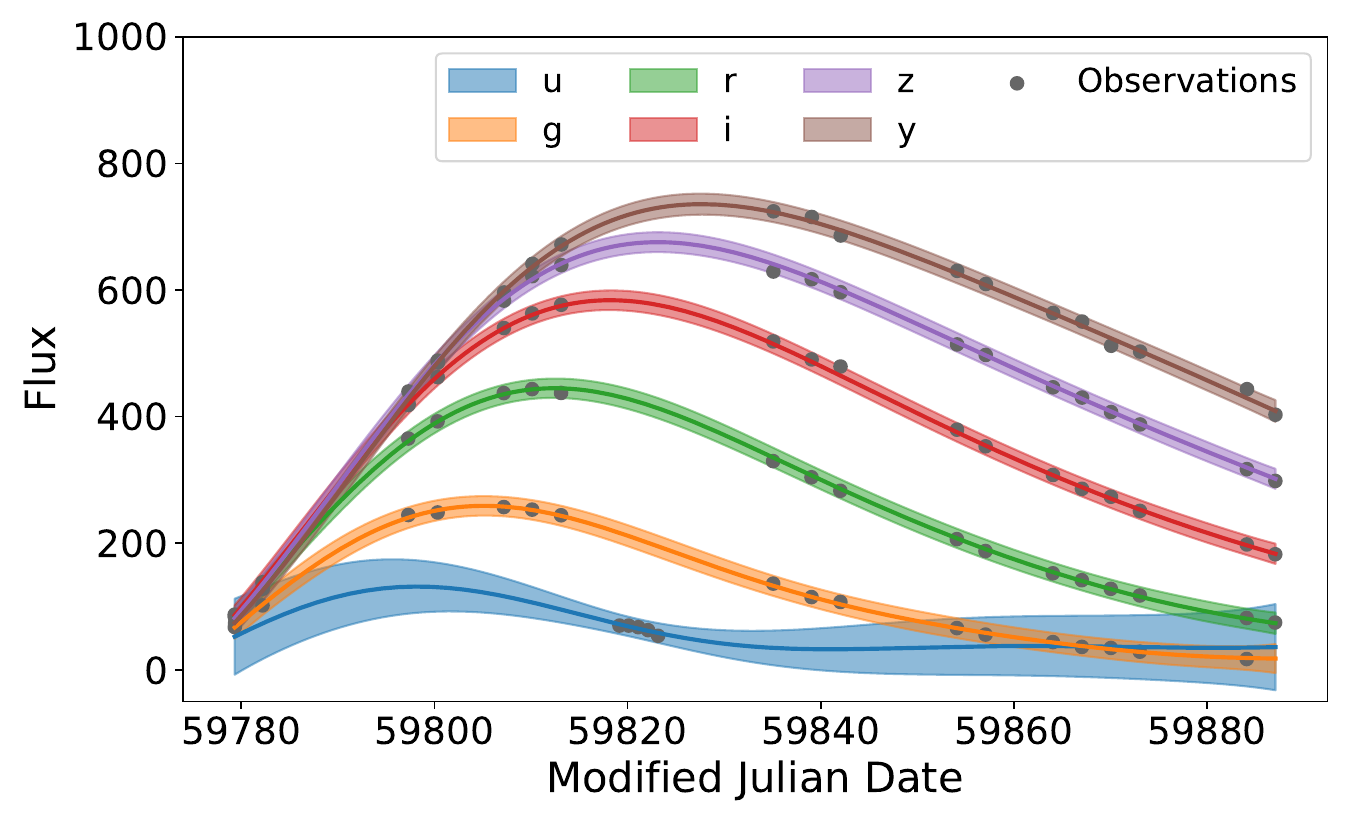}{(b)}
\caption{ Light curve of SN~Ibc (ID 34299) from the PLAsTiCC dataset (a) before approximation. The points represent measurements in the corresponding passbands. (b) The light curve after approximation using the GP method. The solid lines are the estimated mean $\mu(x)$ values. The shaded areas represent the $\pm 3\sigma(x)$ uncertainty band for the light curve approximation.
\label{fig:problem_example}}
\end{figure*}

Generally, it is hard to find the distribution for a given sample, which, however, is not necessary for many practical cases. In this respect, the goal of our work is to estimate the mean flux $\mu(x)$ and the standard deviation $\sigma(x)$ for the given features $x$ of the light curve. At the next stage, we use these calculated functions for the light curve interpolation. Figure~\ref{fig:problem_example} shows an example of a light curve for a SN~Ibc from the PLAsTiCC dataset, measured in the $\{u,g,r,i,z,y\}$ passbands, before and after the approximation. The light curve approximation problem can be solved using several machine learning algorithms. We describe a set of such models in the following section. 

The radiation flux in \texttt{mJy} units for the ZTF BTS catalog observations can be calculated from the $\{g,r\}$ magnitudes as follows:
\begin{equation}
    \text{Flux}~
    [\text{mJy}] =  10^{-0.4(m - 16.4)},
\end{equation}
where $m$ is ZTF BTS magnitude.
The PLAsTiCC uses calibrated flux in units of \texttt{FLUXCAL} of SNANA \citep{2009PASP..121.1028K} having 27.5 AB-magnitude as zero point. Therefore, we adapted it to have flux values in the same \texttt{mJy} units as we used for the ZTF BTS catalog:
\begin{equation} 
\text{Flux} [\text{mJy}] = \texttt{FLUXCAL} \times 10^{-4.44},
\end{equation}
The timestamps in the light curve observations are measured in Modified Julian Date (MJD) units.

\section{Approximation methods}

In our study, we tested various neural networks based on different ideas for the light curve approximation. The most obvious solution is to take regression models based on MLP. They learn to predict the mean flux value for the given timestamp and passband. To be precise, we use two different implementations: {\tt\string PyTorch} \citep{NEURIPS2019_9015} and {\tt\string scikit-learn} \citep{scikit-learn}. The {\tt\string PyTorch} library uses computation graphs with an automatic differentiation which significantly simplifies building models and architectures. {\tt\string Scikit-learn} is based on {\tt\string NumPy} arrays with manual definitions of all derivatives. It makes this library less flexible, but faster than {\tt\string PyTorch}, which is also demonstrated in this work.

We also explored BNN, which learn the optimal distributions of each weight of the model. Injecting weight uncertainty helps to estimate the variation in the flux predictions. In theory, BNNs are powerful predictors with peculiar properties, however,  in practice it is too complicated to fit them properly. In addition, we tested normalizing flow (NF) as a representative of generative models. The NF estimates a distribution of the light curve observations. The knowledge of this distribution allows us to calculate the mean and variance of the flux. The following subsections describe the details of all models used in this work.

\subsection{Regression model}

A simple model that can be used for light curve approximation is a MLP for a regression task. It assumes that the flux $y$ is normally distributed with mean $\mu(x|w)$ and constant standard deviation, $\sigma(x)=\sigma$, values:

\begin{equation}
p(y_i|x_i, w) = \frac{1}{\sigma \sqrt{2\pi}}e^{-\frac{(y_i - \mu(x_i|w))^2}{2\sigma^2}},
\label{eq:mlp_prob}
\end{equation}

where $\mu(x_i|w)$ is estimated by the neural network with weights $w$. The model takes $\{x_i\}$ as inputs and learns its weights by optimizing the log-likelihood function:

\begin{equation}
L(w) = \frac{1}{n} \sum_{i=1}^{n} \log p(y_i|x_i, w) \to \max_{w},
\end{equation}

where $n$ is the number of observations in the light curve. Let us denote the optimal weights $w_{ML}=\arg \max L(w)$. According to Section 3.1.1 in~\citet{10.5555/1162264}, this optimization problem is equivalent to the MSE loss minimization:

\begin{equation}
w_{ML} = \arg \max_{w} L(w) = \arg \min_{w} \frac{1}{n} \sum_{i=1}^{n} (y_i - \mu(x_i|w))^2.
\end{equation}

The optimal solution provides the following functions for the light curve approximation:

\begin{equation}
\mu(x) = \mu(x|w_{ML}),
\end{equation}

\begin{equation}
\sigma(x) = \sigma = \sqrt{ \frac{1}{n} \sum_{i=1}^{n} (y_i - \mu(x_i))^2 }.
\end{equation}

In this work, we use two MLP models implemented in two frameworks: {\tt\string PyTorch} and {\tt\string scikit-learn}. The {\tt\string PyTorch} model is a one-layer neural network with 20 neurons, having {\tt\string tanh} as the activation function, and is trained with an {\tt\string Adam} optimizer. The {\tt\string scikit-learn} model has two layers with 20 and ten neurons, the activation function {\tt\string tanh}, which is trained with the optimizer {\tt\string LBFGS}. {\tt\string LBFGS} is a second-order optimizer, which converges to the optimum faster than first-order algorithms such as {\tt\string Adam}. Furthermore, in the text and tables, these two models are designated based on the framework on which they are implemented: MLP (sklearn) and MLP (pytorch). The results provided in the following sections show that the {\tt\string scikit-learn} implementation is dramatically faster than other considered approximation algorithms. An example of an SN~II light curve approximation using MLP (sklearn) and MLP (pytorch) is shown in Fig.~\ref{fig:approx_example}~(c) and (d).

\subsection{Bayesian neural network}

The BNN uses the same observation model described in Equation~\ref{eq:mlp_prob} as the MLP models. In addition, in the BNN model, it is assumed that its weights $w$ is a random variable with a distribution defined by the light curve observations, $D = \{x_1, x_2, ..., x_i, ..., x_n \}$. Let us define a prior distribution $p(w)$ of the weights:

\begin{equation}
p(w) = \mathcal{N}(w|0, \alpha^2 I) =\prod_{j=1}^{m} \frac{1}{\alpha \sqrt{2\pi}}e^{-\frac{w_{j}^2}{2\alpha^2}},
\label{eq:bnn_prior}
\end{equation}

where $m$ is the number of weights of the network and $\alpha$ is an adjustable parameter. Applying Bayes' rule to Equation~\ref{eq:mlp_prob} and Equation~\ref{eq:bnn_prior}, we derive the posterior probability density distribution $p(w|D)$ of the BNN model's weights given the light curve observations:

\begin{equation}
p(w|D) \propto p(w)\prod_{i=1}^{n} p(y_i|x_i, w) \to \max_{w},
\label{eq:bnn_posterior}
\end{equation}

where the objective of the model is to maximize this distribution with respect to the weights. However, it is difficult to estimate the exact expression of the prior distribution. \citet{blundell2015weight} suggest approximating it with a normal distribution $q(w)$, which has a diagonal covariance matrix $\Sigma$:

\begin{equation}
p(w|D) \approx q(w) = \mathcal{N}(w|\mu_w, \Sigma_w) = \prod_{j=1}^{m} \mathcal{N}(w_j|\mu_{wj}, \sigma_{wj}),
\label{eq:bnn_vi}
\end{equation}

where $\mu_{wj}$ and $\sigma_{wj}$ are the mean and the standard deviation for the weight $w_j$ of the BNN model. Following~\citet{blundell2015weight}, the optimization problem in Equation~\ref{eq:bnn_posterior} is approximated by maximizing the lower variational bound:

\begin{equation}
L(\mu_{wj}, \sigma_{wj}) = \mathop{\mathbb{E}}_{q(w)} \left[ \sum_{i=1}^{n} \log p(y_i|x_i, w) \right] - KL(q(w)||p(w)) \to \max_{\mu_{wj}, \sigma_{wj}},
\label{eq:bnn_elbo}
\end{equation}

where $KL(q(w)||p(w))$ is the Kullback–Leibler divergence between the distributions $q(w)$ and $p(w)$. The mean $\mu(x)$ and the standard deviation $\sigma(x)$ for the light curve approximation are defined as

\begin{equation}
\mu(x) = \mathop{\mathbb{E}}_{w \sim q(w)} \left[ \mu(x|w) \right],
\end{equation}

\begin{equation}
\sigma(x) = \sqrt{ \mathop{\mathbb{V}}_{w \sim q(w)} \left[ \mu(x|w) \right] }.
\end{equation}

As shown in Section 5.7.1 of~\citet{10.5555/1162264}, the variance of the approximation can be represented as the sum of two terms:

\begin{equation}
\sigma^2(x) = \sigma^2 + \sigma_{model}^2(x).
\end{equation}

where $\sigma^2$ is the variance that arises from the intrinsic noise in the flux $y$ as shown in Equation~\ref{eq:mlp_prob}. The second term stands for the uncertainty in the BNN model due to the uncertainty of the weights.

\begin{figure*}
\includegraphics[width=0.47\linewidth]{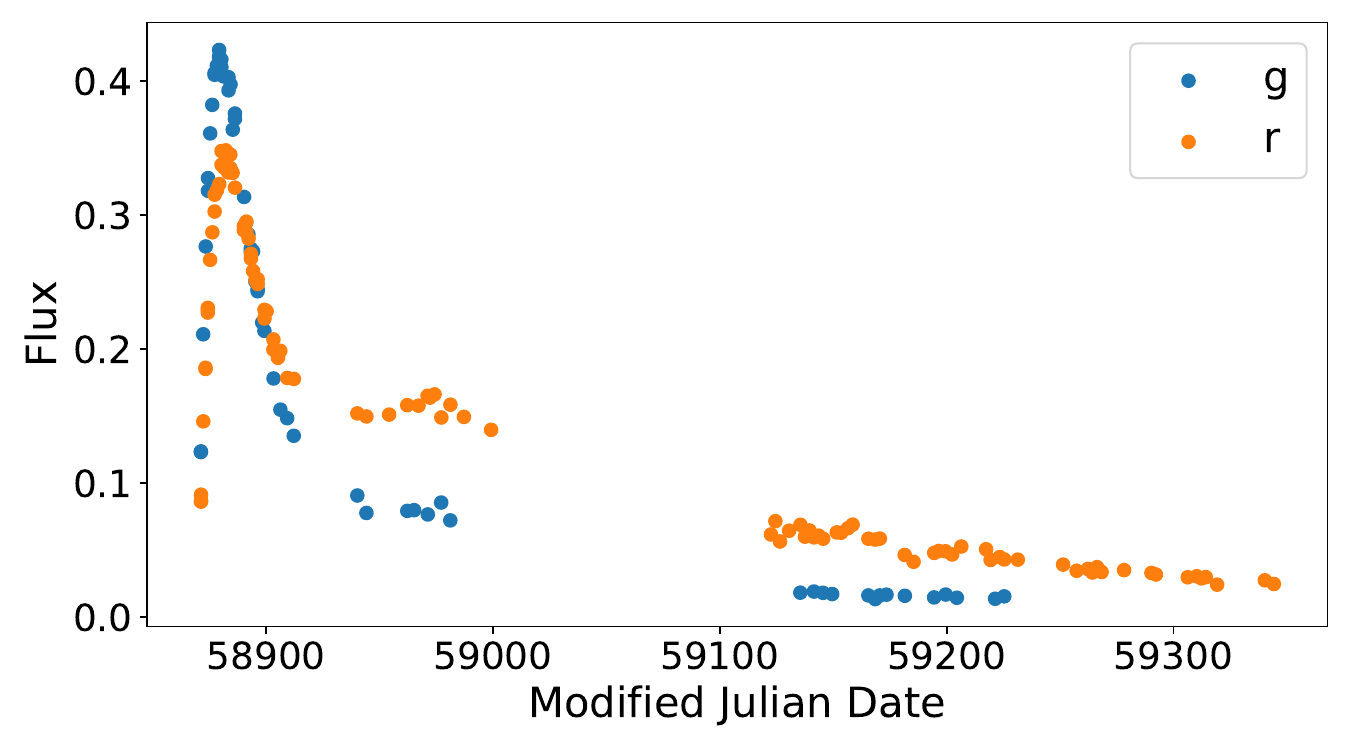}{(a)}
\includegraphics[width=0.47\linewidth]{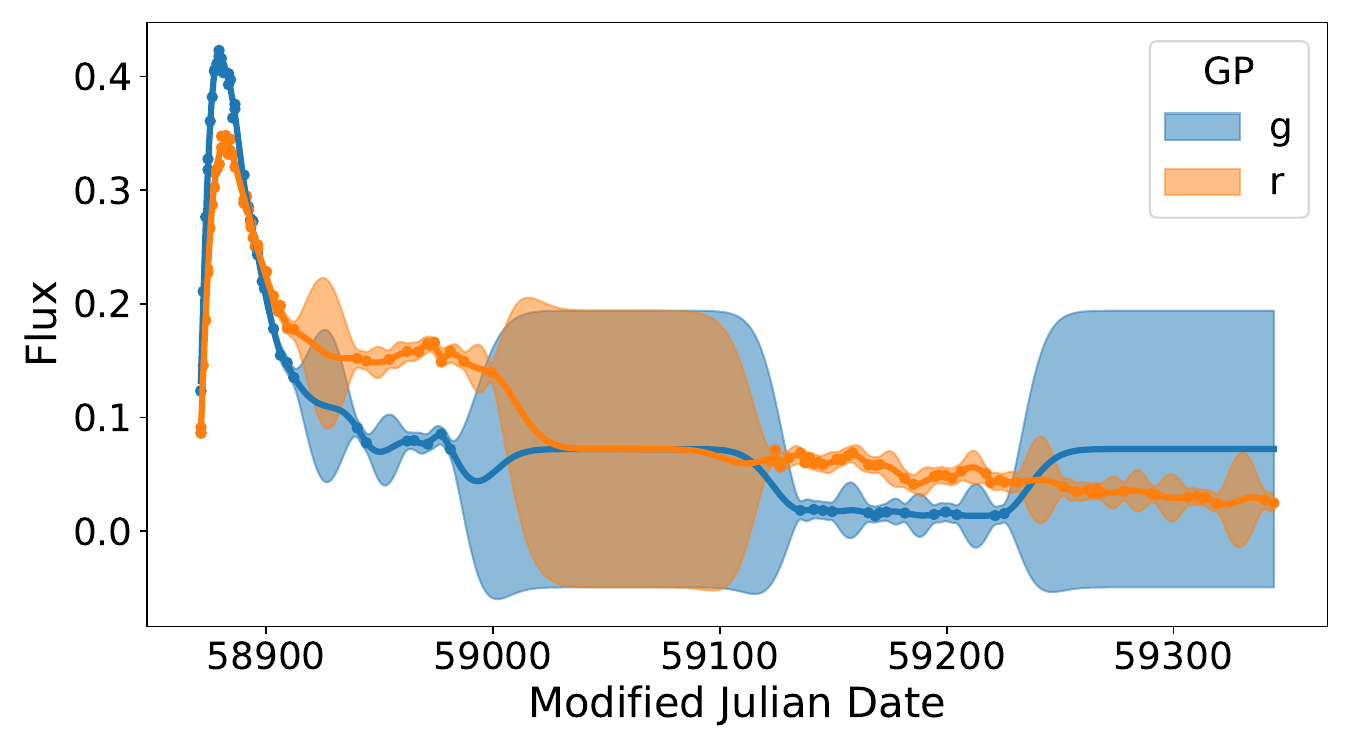}{(b)}
\includegraphics[width=0.47\linewidth]{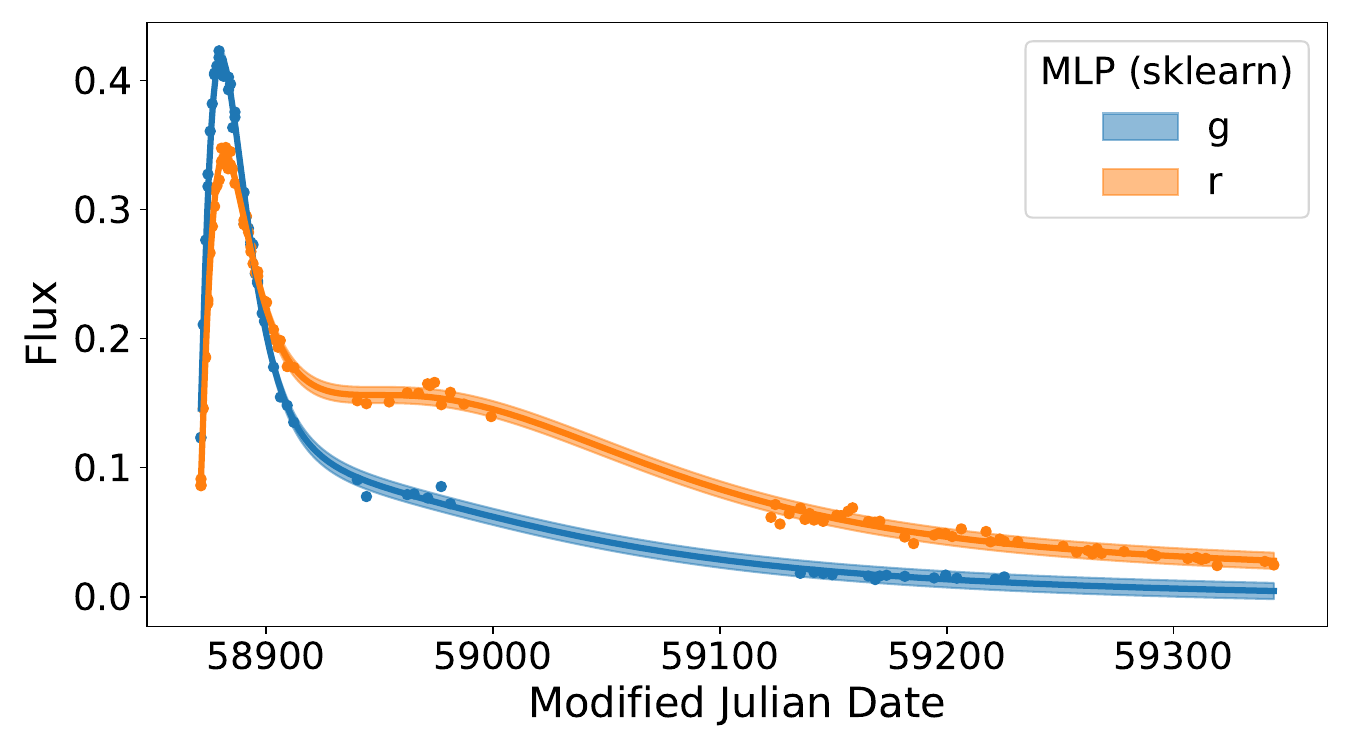}{(c)}
\includegraphics[width=0.47\linewidth]{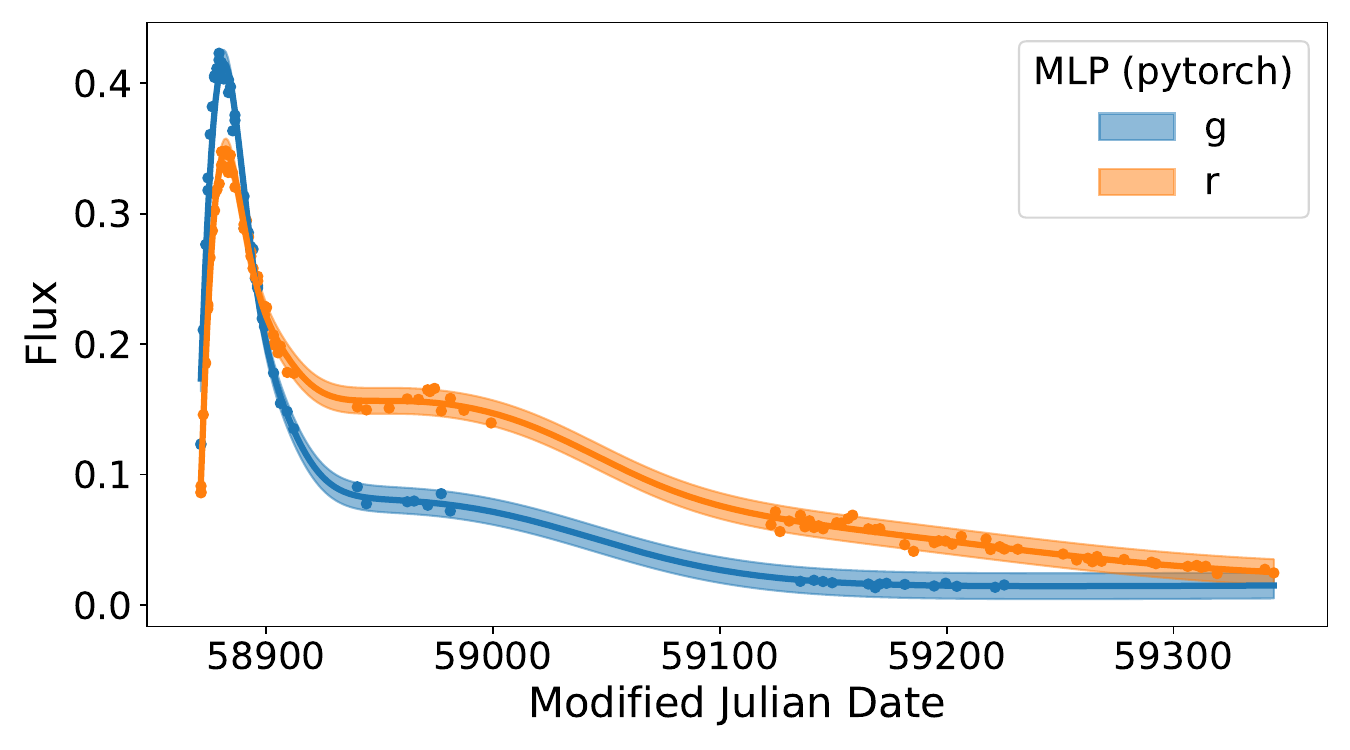}{(d)}
\includegraphics[width=0.47\linewidth]{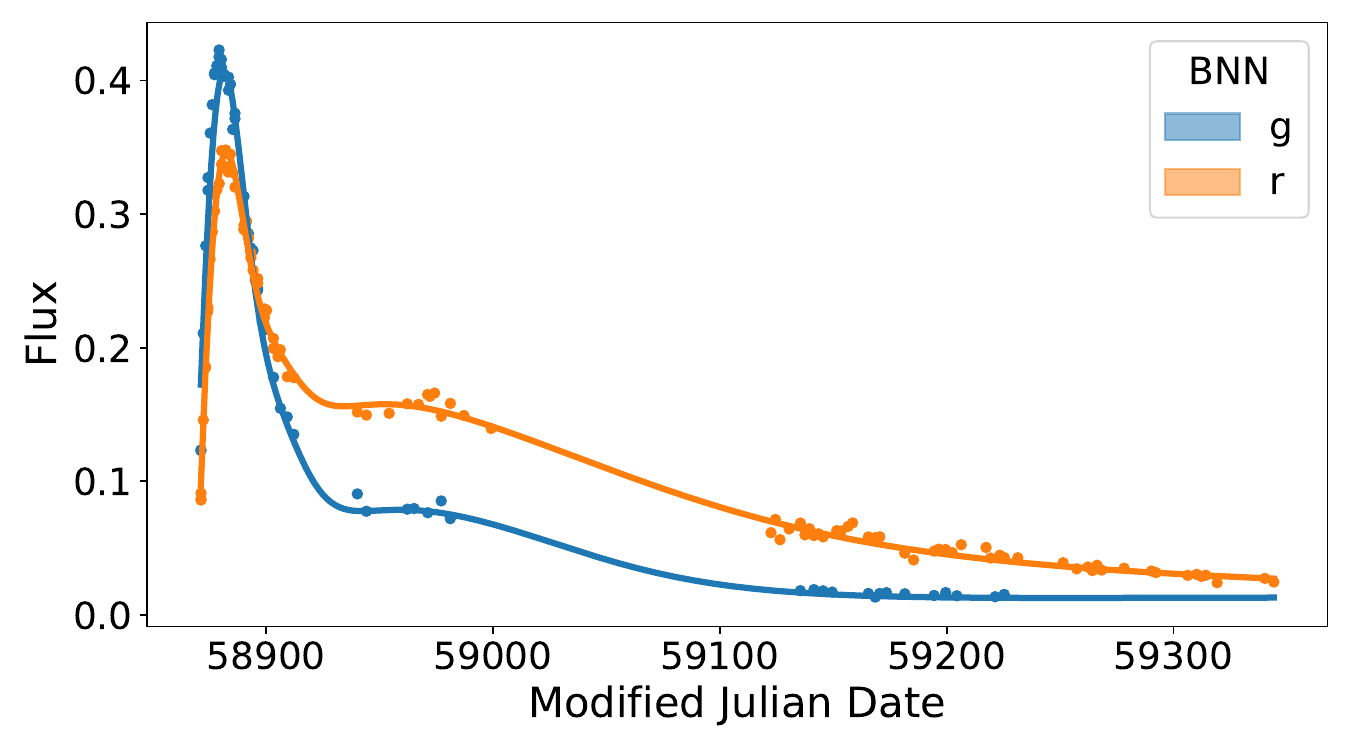}{(e)}
\includegraphics[width=0.47\linewidth]{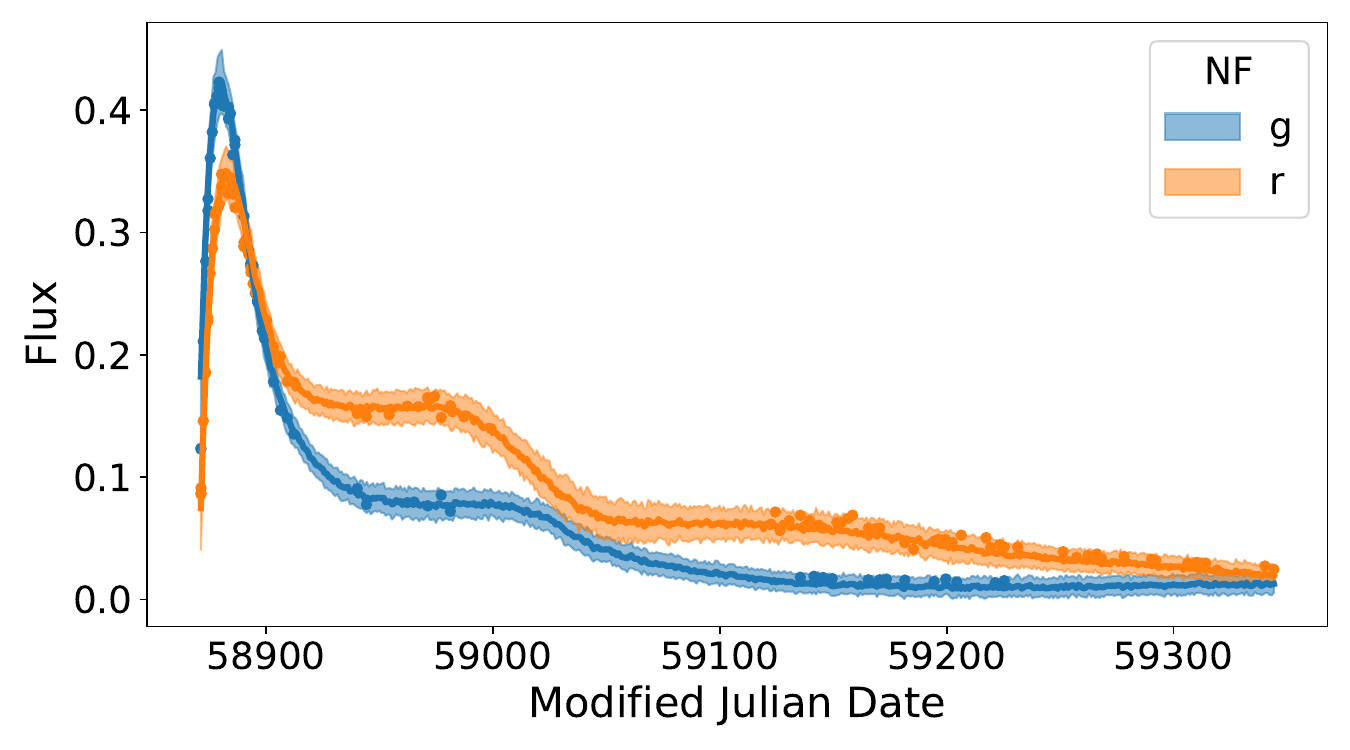}{(f)}
\caption{Examples of SN~II ZTF20aahbamv light curve approximations using different methods: (a) the light curve before approximation, (b) GP, (c) MLP (sklearn), (d) MLP (pytorch), (e) BNN, and (f) NF. The points represent measurements in the corresponding passbands. The solid lines are the estimated mean $\mu(x)$ values. The shaded areas represent the $\pm 1\sigma(x)$ uncertainty band.
\label{fig:approx_example}}
\end{figure*}

In this work, we use the BNN model with one linear layer of 20 neurons, and weight priors in the form of standard normal distributions $\mathcal{N}(0, 0.1)$. In such a model, the optimizer {\tt\string Adam} and the activation function {\tt\string tanh} are used. The example of a SN~II light curve approximation with the use of BNN is shown in Fig.~\ref{fig:approx_example}~(e).

\subsection{Normalizing flows}

We start by defining a latent random variable, $z,$ with the standard normal distribution, $q(z) = \mathcal{N}(z|0, I)$. The goal of a normalizing flow (NF) model~\citep{NF1, NF2} is to find an invertible function which transforms the flux $y_i$ into the latent variable $z_i$ with the given $x_i$:

\begin{equation}
z_i = f(y_i, x_i, w),
\end{equation}

where $w$ is the vector of learnable parameters of the function. Then, the change of variables theorem defines the following relation between the flux $p(y_i|x_i, w)$ and the latent variable $q(z)$ distributions:

\begin{equation}
p(y_i|x_i, w) = q(f(y_i, x_i, w)) \left| \det \frac{\partial f(y_i, x_i, w)}{\partial y_i} \right|.
\end{equation}

In this work, we use real-valued non-volume preserving (real NVP) NF model~\citep{dinh2017density}, where the function $f(y_i, x_i, w)$ is designed using neural networks with weights, $w$. The optimal weights, $w_{ML}$, are estimated by maximizing the log-likelihood function:

\begin{equation}
L(w) = \frac{1}{n} \sum_{i=1}^{n} \log p(y_i|x_i, w) \to \max_{w}.
\end{equation}

The main advantage of this model is that it can estimate any probability density function of the flux $p(y_i|x_i, w)$. Then, the mean $\mu(x)$ and the standard deviation $\sigma(x)$ for the light curve approximation are defined as:

\begin{equation}
\mu(x) = \mathop{\mathbb{E}}_{z \sim q(z)} \left[ f^{-1}(z, x, w_{ML}) \right],
\end{equation}

\begin{equation}
\sigma(x) = \sqrt{ \mathop{\mathbb{V}}_{z \sim q(z)} \left[ f^{-1}(z, x, w_{ML}) \right] }.
\end{equation}

We used eight real NVP transformations, where two simple fully connected neural networks are exploited in each transformation. The model uses the {\tt\string Adam} optimizer and the {\tt\string  tanh} activation function. The example of a SN~II light curve approximation using NF is shown in Fig.~\ref{fig:approx_example} (f).

\subsection{Gaussian processes}

Let us consider a light curve with observations $\{(x_1, y_1), (x_2, y_2), ..., (x_n, y_n)\}$. The GP~\citep{NIPS1995_7cce53cf} model assumes that the joint probability density function of the observations is multivariate normal with a zero mean and the covariance matrix, $K$:

\begin{equation}
p(Y) = \mathcal{N}(0, K),
\label{eq:gp_model}
\end{equation}

with

\begin{equation}
K = 
\left(\begin{array}{ccc}
k(x_1, x_1) & \cdots & k(x_1, x_n)\\
\vdots & \ddots & \vdots \\
k(x_n, x_1) & \cdots & k(x_n, x_n)
\end{array}\right),
\end{equation}

where $Y = [y_1, y_2, \cdots, y_n]^T$ and $k(x_i, x_j)$ is a covariance function and is considered as the model hyperparameter. In this work, we use combinations of {\tt\string  RBF}, {\tt\string  Matern}, and {\tt\string  White} kernels as the covariance function. The best combination of these kernels for each dataset is estimated by the optimization procedure. In its turn, parameters of the kernels are estimated by maximizing the log-likelihood function:

\begin{equation}
L(w) = \log p(Y) = -\frac{1}{2}Y^T K^{-1} Y - \frac{1}{2} \log |K| - \frac{n}{2} \log 2\pi \to \max_{w}.
\end{equation}

With Equation~\ref{eq:gp_model}, we can find the joint distribution with a new pair $(x, y)$ as the following form:

\begin{equation}
p(y, Y) = \mathcal{N} \left (0, 
\left(\begin{array}{cc}
K & K_{*}^T\\
K_{*} & K_{**}
\end{array}\right)
\right ),
\label{eq:gp_model_new}
\end{equation}

where $K_{*} = [k(x, x_1), \cdots, k(x, x_n)]$, and $K_{**} = k(x, x)$. Following~\citet{NIPS1995_7cce53cf}, the posterior probability density function for the prediction of $y$ is also assumed normal:

\begin{equation}
p(y|Y) = \mathcal{N}(y|\mu(x), \sigma(x)),
\end{equation}

where the mean $\mu(x)$ and the standard deviation $\sigma(x)$ functions are used for the light curve approximation and are defined as follows:

\begin{equation}
\mu(x) = K_{*}K^{-1}Y,
\end{equation}

\begin{equation}
\sigma(x) = K_{**} - K_{*}K^{-1}K_{*}^{T}.
\end{equation}

~\citet{2019AJ....158..257B} demonstrates that the approximation of light curves using GP provides better results than other methods. Currently, this approach is the state-of-the-art method and we consider it to be as a baseline in this work. The example of a SN~II light curve approximation using GP is shown in Fig.~\ref{fig:approx_example}~(b).

\subsection{Other models}

In addition to the models described above, we also analyzed other regression models in machine learning such as decision trees, random forest, gradient boosting decision trees, support vector machines, and linear regression. However, the tests show poor approximation quality for all of these models. Therefore, in this paper, we do not consider them further. Moreover, we tested other generative models such as Variation Autoencoders (VAE) and Generative Adversarial Networks (GAN). However, they also show unsatisfactory approximation metrics and we do not consider them further.

\section{Quality metrics}

\subsection{Approximation quality}
\label{sec:qm_approx}

We measured the quality of the proposed approximation methods using direct and indirect approaches. In the direct approach, we measured the difference between the models' predictions and observations of the light curves. In order to achieve this, the observations were divided into train and test samples using a time-binned split. Consider a light curve as an object with $n$ points $\{(x_1, y_1), (x_2, y_2), ..., (x_n, y_n) \}$ at timestamps of $\{t_1, t_2, ..., t_n \}$, where $t_1 \le t_2 \le ... \le t_n$. We then define the width, $\Delta t,$ of a time bin and split the observations into $k$ bins, $\{b_1, b_2, ..., b_k \}$:

\begin{equation}
b_j = \{i : j \Delta t \le t_i < (j+1) \Delta t\}.
\end{equation}

We randomly selected two nonempty bins $b_{test}$, with the exclusion of the first $b_1$ and the last $b_k$, to be the test set and to measure the approximation quality. All observations in the other bins were used for training. We note that we did not limit the bins selection by passbands, so the test split could contain observations in a passband that was not present in the training set. To estimate the quality, we calculated the following metrics: root mean squared error (RMSE), mean absolute error (MAE), mean absolute percentage error (MAPE), relative squared error (RSE), and relative absolute error (RAE), defined as follows:

\begin{equation}
\text{RMSE} = \sqrt{ \frac{1}{m} \sum_{i \in b_{test}} (y_i - \mu(x_i))^2 },
\end{equation}

\begin{equation}
\text{MAE} = \frac{1}{m} \sum_{i \in b_{test}} |y_i - \mu(x_i)|,
\end{equation}

\begin{equation}
\text{MAPE} = \frac{100}{m} \sum_{i \in b_{test}} \left |\frac{y_i - \mu(x_i)}{y_i} \right |,
\end{equation}

\begin{equation}
\text{RSE} = \sqrt{ \frac{\sum_{i \in b_{test}} (y_i - \mu(x_i))^2}{\sum_{i \in b_{test}} (y_i - \bar{y})^2} },
\end{equation}

\begin{equation}
\text{RAE} = \frac{\sum_{i \in b_{test}} |y_i - \mu(x_i)|}{\sum_{i \in b_{test}} |y_i - \bar{y}|},
\end{equation}

where $\mu(x_i)$ is a prediction of an approximation model, $m$ is the number of observations inside the test bins, $b_{test}$, and $\bar{y}$ is the average of flux measurements $y_i$ inside the bins.

The deviation of the true curve from the predicted one can be important in the case of particular physically motivated measurements. In order to characterize the set of estimated deviations we used the following metrics to measure the quality of the flux standard deviation $\sigma(x)$ estimated by the models. These metrics are  negative log predictive density (NLPD), normalized root mean squared error based on observed error (nRMSEo), normalized root mean squared error based on predicted error (nRMSEp)~\citep{10.1007/11736790_1}, and prediction interval coverage probability (PICP)~\citep{SHRESTHA2006225}, which are defined as follows:

\begin{equation}
\text{NLPD} = \frac{1}{2} \log(2\pi) + \frac{1}{m} \sum_{i \in b_{best}} \left[\log(\sigma(x_i)) + \frac{(y_i - \mu(x_i))^2}{2\sigma(x_i)^2} \right],
\end{equation}

\begin{equation}
\text{nRMSEo} = \frac{1}{m} \sum_{i \in b_{best}} \left[\frac{(y_i - \mu(x_i))^2}{2\varepsilon_i^2} \right],
\end{equation}

\begin{equation}
\text{nRMSEp} = \frac{1}{m} \sum_{i \in b_{best}} \left[\frac{(y_i - \mu(x_i))^2}{2\sigma(x_i)^2} \right],
\end{equation}

\begin{equation}
\text{PICP}_{\alpha} = \frac{1}{m} \sum_{i \in b_{best}} I \left[  PL_{i}^{L} \le y_{i} \le PL_{i}^{U} \right] \times 100\%,
\end{equation}

where $\varepsilon_i$ is the observed flux error, $PL_{i}^{L}$ and $PL_{i}^{U}$ are the lower and upper prediction limits for the $i$-th observation, respectively. The limits are symmetric and estimated based on the normal distribution, $\mathcal{N}(\mu(x_i), \sigma(x_i)),$ for a given coverage probability, $\alpha$. We calculated these metrics for each light curve and then averaged them over all curves in the dataset. The values obtained are used as the final quality metrics of the approximation. 

While the metrics mentioned above give us an accurate estimation of the quality of the obtained approximation, this cannot be used as an ultimate check. In order to check that these approximations do not change the physical features of the light curve, we solve two practical astrophysical problems: supernova classification and peak identification.

\subsection{Supernova classification}
\label{sec:sup_class}

\begin{figure*}
\centering
\includegraphics[width=0.47\linewidth]{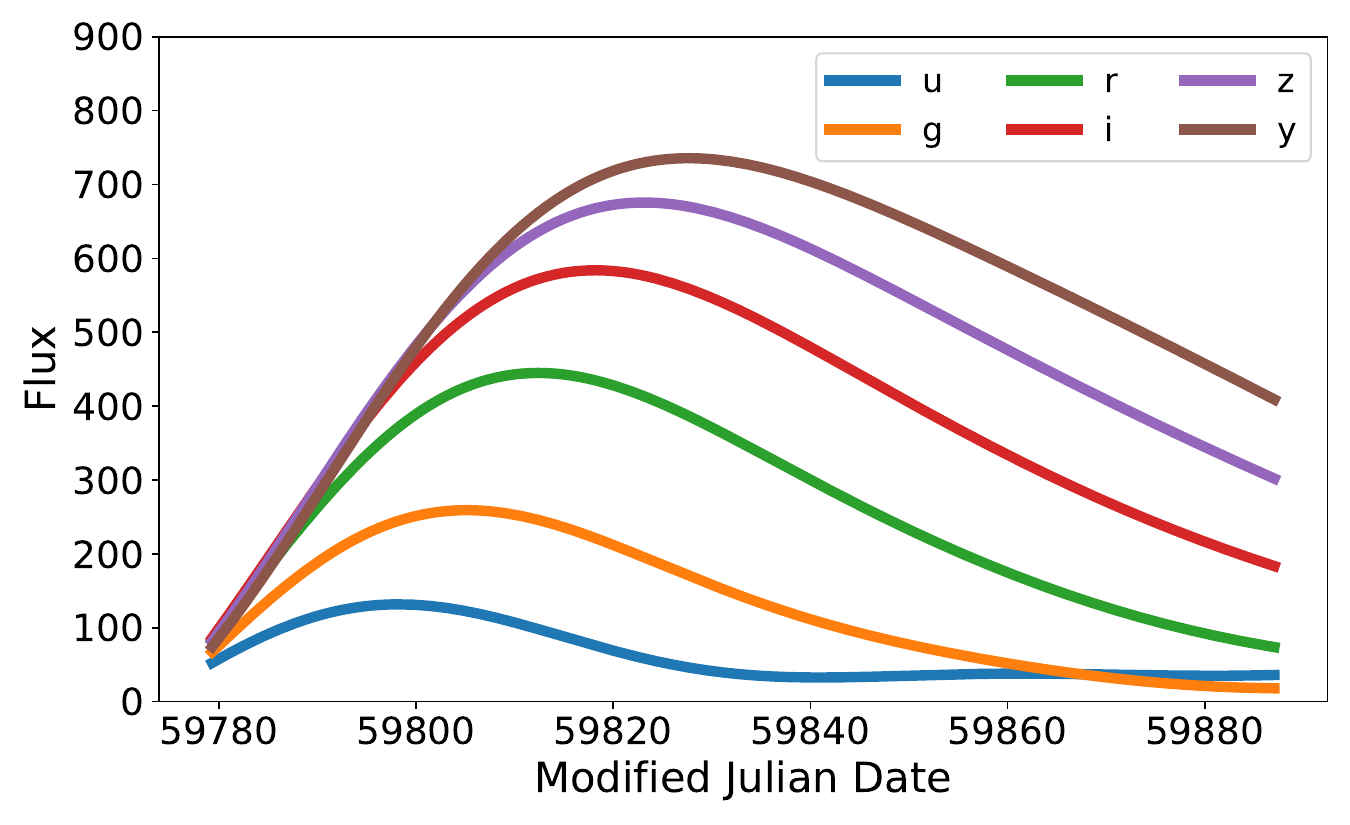}{(a)}
\includegraphics[width=0.47\linewidth]{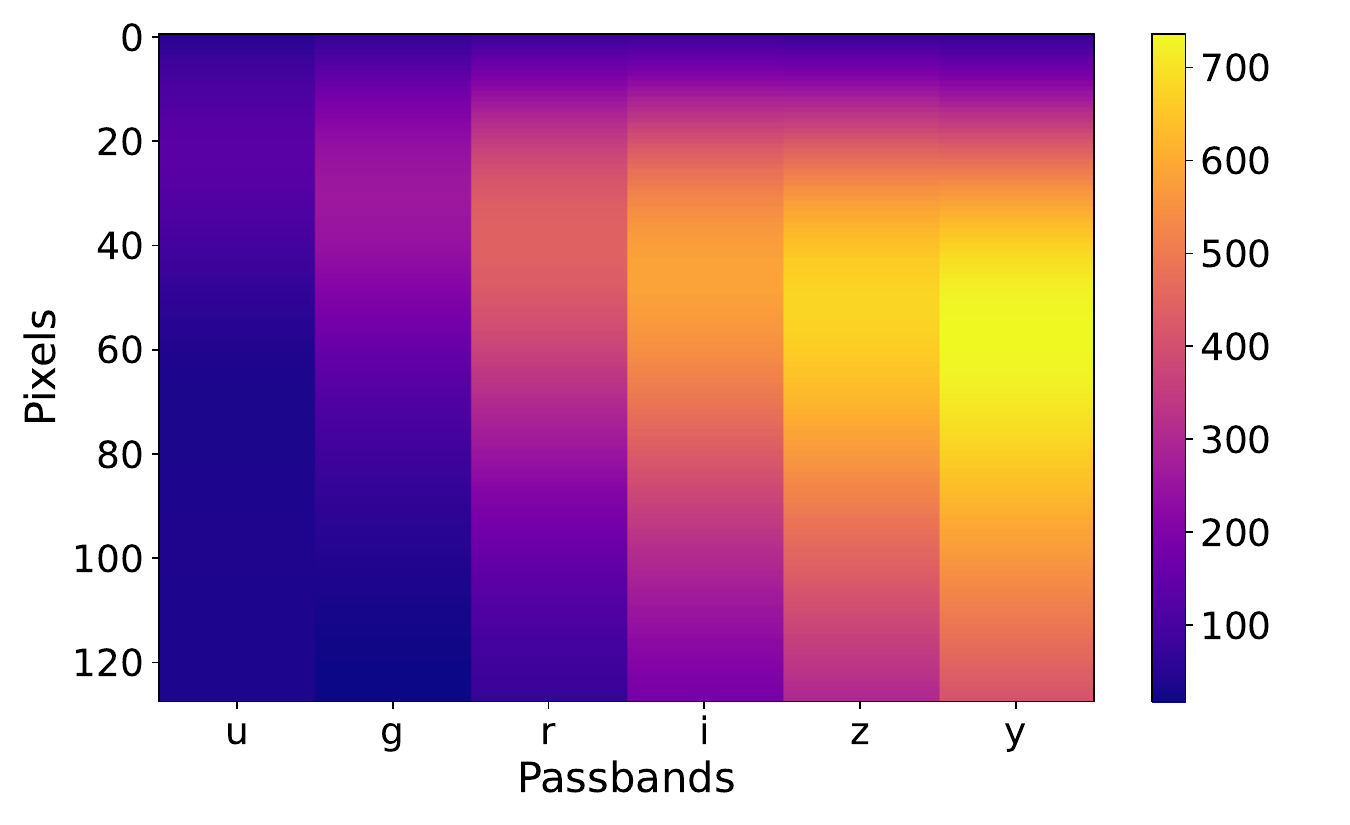}{(b)}
\caption{Example of 1D image $I$ with $128 \times 1$ pixels and 6 channels, and corresponding light curve: (a) Light curve after approximation. The solid lines are the estimated mean $\mu(x)$ values; (b) Example of the light curve transformation into a 1D image with 128 pixels and 6 channels, which is used as input of the CNN.
\label{fig:clf_preproc}}
\end{figure*}

In this work, we evaluated metrics of a supernova classification task to measure the light curve approximation quality indirectly. 
Similarly to the direct approach, we consider a light curve as an object with $n$ points at timestamps $\{t_1, t_2, ..., t_n \}$, where $t_1 \le t_2 \le ... \le t_n$, and with $c$ passbands $\{\lambda_1, \lambda_2, ..., \lambda_c \}$. All these observations were used to fit an approximation model to estimate the $\mu(x)$ and $\sigma(x)$ functions. Then, we generated $k=128$ new regular timestamps $\{\tau_1, \tau_2, ..., \tau_k \}$ between $t_1$ and $t_2$ with the step $\Delta \tau = (t_n - t_1) / k$. Using the approximation function $\mu(x)$, we created the following matrix of size $(128, c)$:

\begin{equation}
I = 
\left(\begin{array}{cccc}
\mu(\lambda_1, \tau_1) & \mu(\lambda_2, \tau_1) & \cdots & \mu(\lambda_c, \tau_1)\\
\mu(\lambda_1, \tau_2) & \mu(\lambda_2, \tau_2) & \cdots & \mu(\lambda_c, \tau_2)\\
\vdots & \vdots & \ddots & \vdots \\
\mu(\lambda_1, \tau_k) & \mu(\lambda_2, \tau_k) & \cdots & \mu(\lambda_c, \tau_k)\\
\end{array}\right),
\label{eq:lc_image}
\end{equation}

where the notation $x=(\lambda, t)$ is used. We consider this matrix $I$ as a one-dimensional image with $k \times 1$ pixels and $c$ channels. An example of such an image is demonstrated in Figure~\ref{fig:clf_preproc}.

\begin{figure*}
\centering
\includegraphics[width=1.\linewidth]{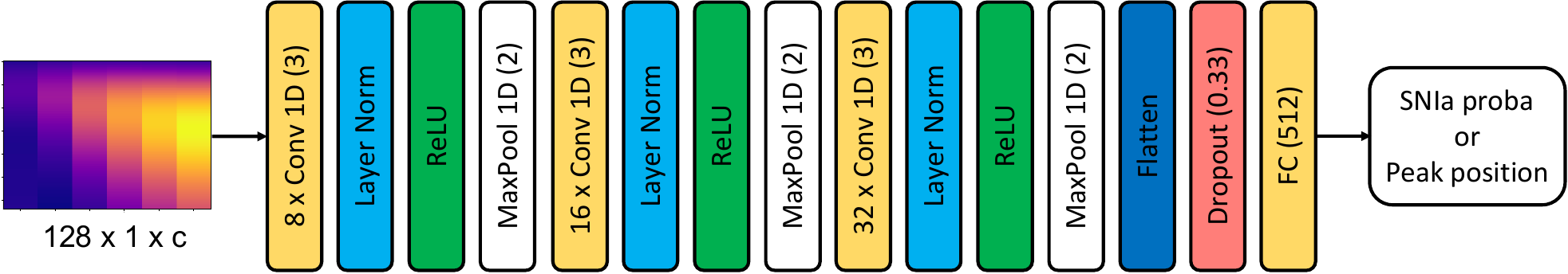}
\caption{Architecture of the CNN model for supernovae type classification and peak identification.
\label{fig:cnn_type}}
\end{figure*}

We used these images as inputs to a CNN classifier, which is trained to separate Type~Ia~SNe from all other types of objects. The network contains three 1D convolutional layers with a {\tt\string  ReLU} activation function, a {\tt\string  max pooling} layer, and a {\tt\string  dropout} layer followed by a fully connected layer, as shown in Figure~\ref{fig:cnn_type}. The classifier returns the probability that the input image corresponds to a Type~Ia~SN. This model is fitted using a binary cross-entropy loss function.

Light curves in the dataset were divided into train and test samples. The first one was used to train the CNN model, while the second one was used to calculate the following classification quality metrics: area under the ROC curve (ROC AUC), area under the precision-recall curve (PR AUC), Accuracy, Recall, and Precision~\citep{James2013}. We used these metrics  of physically motivated classification problem as an indirect approach for estimation of the light curve approximation quality.

\subsection{Peak identification}
\label{sec:peak_finding}

Peak identification is an additional indirect approach to the approximation quality of measurements. We used two methods to estimate the peak time stamp. In the first method, we took the light curve image defined in Equation~\ref{eq:lc_image}, Fig.\ref{fig:clf_preproc} and calculated the sum of fluxes for all passbands. Then, the peak position, $\tau_{peak}$, was estimated as the timestamp with maximum total flux value with equation:

\begin{equation}
\tau_{peak} = \arg \max_{\tau} \sum_{i} \mu(\lambda_i, \tau).
\end{equation}

In the second method, we used a CNN regression model. It takes the light curve matrices, $I,$ as inputs and predicts the peak position. Following the terminology used for convolution neural networks, these matrices are considered as 1D images with $k \times 1$ pixels, representing flux for each time stamp and $c$ channels, representing passbands. The network has the same architecture as for the classification. The model is fitted using the MSE loss function.

The light curves in a dataset were divided into train and test samples. The first one is used to train the CNN model, while the second one was used to calculate the regression quality metrics such as RMSE, MAE, RSE, RAE, and MAPE for both of the estimation methods. We use them for indirect measurements of the light curve approximation quality.

\section{PLAsTiCC Dataset}

\subsection{Approximation quality}

First, we applied the approximation methods to the PLAsTiCC dataset~\citep{https://doi.org/10.48550/arxiv.1810.00001}. This dataset is produced as a simulation of the upcoming observations from the Vera C. Rubin Observatory. It contains 199434 simulated light curves of transient astronomical events of different classes. The light curves represent measurements of an object's flux in six different passbands $\{u,g,r,i,z,y\}$, covering ultra-violet, optical, and infrared wavelength ranges.

For our tests, we selected 10764 light curves based on the following criteria. Firstly, we kept only objects with positive fluxes. Also, we checked the number of observations in each filter and kept only the light curves having at least seven observations per passband in at least three different passbands. Finally, we took into consideration points from all passbands and selected only the light curves with time gaps between neighboring observations in any filter smaller than 50 days. Examples of light curves, which are the closest to the limit of our selection criteria, before and after approximation are shown in Fig.~\ref{fig:plastic_4551_appendix}, Fig.~\ref{fig:plastic_31087708_appendix}, and Fig.~\ref{fig:plastic_68731_appendix}. After that, points of each selected light curve were split into train and test sets to fit analyzed models and measure the quality of all the approximation methods, as described in Sec.~\ref{sec:qm_approx}. The optimal hyperparameter values of the methods were estimated using a grid search with the MAPE objective function. The obtained quality metrics values are presented in Table~\ref{tab:regression_metrics_plastic1} for approximated fluxes and Table~\ref{tab:regression_metrics_plastic2} for predicted errors for the fluxes.

The results in Table~\ref{tab:regression_metrics_plastic1} demonstrate that MLP (sklearn) model outperforms GP by MAE and MAPE metrics and has the same quality evaluated by RMSE. Overall, NF is better by RSE and RAE metrics. This fact proves that neural networks are effective even if trained on a small dataset from one light curve and can be used for single light curve approximation in astronomy. In particular, neural network approximation models are helpful when a researcher does not have prior information or a full physical model of a studied object. Neural networks have a stronger generalizing ability compared to GP, wherein we would need to choose a covariance matrix manually. 

The results presented in Tab.~\ref{tab:regression_metrics_plastic2} illustrate the ability of the methods to estimate the uncertainty of the flux measurements. GP shows better quality metrics than other methods for all metrics except nRMSEo. Neural networks tend to underestimate uncertainties. NF demonstrates the best results in uncertainty estimation among neural network models.

The PLAsTiCC dataset includes light curves with no observations in one or several passbands. All approximation methods we use in this study are able to predict the flux in these missing filters. The models learn the flux dependency on both time and passband central wavelength. Similarly to approximation over time, the models interpolate and extrapolate the learned dependencies for missing passbands. To demonstrate this, we took a PLAsTiCC SN Ibc (ID 34299) light curve and removed all observations in the $r$ and $y$ passbands. Examples of light curve approximations using different methods are shown in Fig.~\ref{fig:empty_passpands_plastic}. All presented methods interpolate the flux values for $\{r\}$ and extrapolate them for  the $\{y\}$ band. The quality depends on the number of observations in each filter. A minimal theoretical requirement for the interpolation and extrapolation through passbands is  the presence of observations in at least two passbands.

\begin{table*}
\centering
\caption{Regression quality metrics for approximation models applied to the PLAsTiCC. Estimation quality of mean fluxes. \label{tab:regression_metrics_plastic1}}
\begin{tabular}{cccccc}
\hline \hline
Model & RMSE, $\mu$Jy & MAE, $\mu$Jy & RSE & RAE & MAPE \\
\hline
GP             & \boldmath{$3.5 \pm 0.3$} & \boldmath{$2.8 \pm 0.3$} & $1.2 \pm 0.1$            & $1.1 \pm 0.1$              & $19.9 \pm 0.3$ \\
MLP (sklearn)  & \boldmath{$3.5 \pm 0.3$} & \boldmath{$2.8 \pm 0.3$} & $1.2 \pm 0.1$            & $1.1 \pm 0.1$              & \boldmath{$17.9 \pm 0.2$} \\
MLP (pytorch)  & $3.8 \pm 0.4$            & $3.1 \pm 0.3$            & $1.3 \pm 0.2$            & $1.2 \pm 0.1$              & $21.6 \pm 0.3$ \\
BNN            & $4.2 \pm 0.3$            & $3.4 \pm 0.3$            & $1.6 \pm 0.3$            & $1.5 \pm 0.2$              & $26.4 \pm 0.3$ \\
NF             & $3.6 \pm 0.4$            & \boldmath{$2.8 \pm 0.3$} & \boldmath{$1.1 \pm 0.1$} & \boldmath{$1.03 \pm 0.09$} & $18.5 \pm 0.3$ \\
\hline
\end{tabular}
\end{table*}

\begin{table*}
\centering
\caption{Regression quality metrics for approximation models applied to the PLAsTiCC. Estimation quality of the flux uncertainties. \label{tab:regression_metrics_plastic2}}
\begin{tabular}{cccccc}
\hline \hline
Model & NLPD & nRMSEo & nRMSEp & $\text{PICP}_{68}$, \% & $\text{PICP}_{95}$, \% \\
\hline
GP             & \boldmath{$5.25 \pm 0.09$} & $6.6 \pm 0.1$            & \boldmath{$1.26 \pm 0.01$} & \boldmath{$67.5 \pm 0.3$} & \boldmath{$86.5 \pm 0.2$} \\
MLP (sklearn)  & $20 \pm 2$                 & \boldmath{$6.4 \pm 0.1$} & $3.33\pm 0.05$             & $45.6 \pm 0.3$            & $65.4 \pm 0.3$ \\
MLP (pytorch)  & $18 \pm 1$                 & $7.6 \pm 0.1$            & $3.25 \pm 0.04$            & $45.1 \pm 0.3$            & $64.8 \pm 0.3$ \\
BNN            & $26.4 \pm 0.3$             & $9.1 \pm 0.1$            & $2.04 \pm 0.02$            & $51.9 \pm 0.3$            & $73.7 \pm 0.3$ \\
NF             & $10.9 \pm 0.7$             & $6.5 \pm 0.1$            & $2.05 \pm 0.03$            & $56.3 \pm 0.3$            & $80.8 \pm 0.3$ \\
\hline
\end{tabular}
\end{table*}

\subsection{Supernova classification}

We also tested the approximation methods in application to the supernova classification problem. To accomplish that, we took the same light curves as described in the previous section and divided them in two classes. To be more precise, objects with \{90, 67, 52\} class labels in the PLAsTiCC which are SN~Ia, SN~Ia-91bg, and SN~Iax supernova types respectively, were considered as the positive class. All other objects were labeled as a negative class.  

As a next step, we used the approximation methods for light curve augmentation and transformed the light curves into 1D images with 6 channels as it is described in Section~\ref{sec:sup_class}. Here we used the same optimal hyperparameter values estimated in the previous section. The images were used as an input for the CNN classifier model. About 60\% of the light curves were used to train the CNN and the other 40\% were used to estimate the classification quality. The metric values are demonstrated in Tab.~\ref{tab:classification_metrics_plastic}.

The results show that GP model is the best approximation method if we compare all the quality metrics. However, the NF model can be ranked second with differences from the GP model within the margin of the error. In general, all methods demonstrate similar results. In conclusion, it should be stated that the choice of the approximation method does not have a significant impact on the supernova classification results.

\begin{table*}
\centering
\caption{Supernova classification quality metrics for approximation models applied to the PLAsTiCC. \label{tab:classification_metrics_plastic}}
\begin{tabular}{ccccccc}
\hline \hline
Model & ROC AUC & PR AUC & Log Loss & Accuracy & Recall & Precision \\
\hline
GP             & \boldmath{$0.996 \pm 0.001$} & \boldmath{$0.993 \pm 0.001$} & \boldmath{$0.078 \pm 0.007$} & \boldmath{$0.975 \pm 0.002$ } & \boldmath{$0.961 \pm 0.005$} & \boldmath{$0.964 \pm 0.005$} \\
MLP (sklearn)  & $0.993 \pm 0.001$            & $0.986 \pm 0.002$            & $0.10 \pm 0.01$              & $0.970 \pm 0.003$             & $0.952 \pm 0.006$            & $0.959 \pm 0.005$ \\
MLP (pytorch)  & $0.993 \pm 0.001$            & $0.988 \pm 0.002$            & $0.100 \pm 0.008$            & $0.968 \pm 0.003$             & $0.960 \pm 0.005$            & $0.948 \pm 0.006$ \\
BNN            & $0.994 \pm 0.001$            & $0.988 \pm 0.002$            & $0.092 \pm 0.007$            & $0.970 \pm 0.003$             & $0.956 \pm 0.005$            & $0.954 \pm 0.005$ \\
NF             & $0.995 \pm 0.001$            & $0.992 \pm 0.001$            & $0.083 \pm 0.008$            & $0.973 \pm 0.002$             & $0.957 \pm 0.005$            & $0.963 \pm 0.005$ \\
\hline
\end{tabular}
\end{table*}

\subsection{Peak identification}

For the peak identification test, we considered only Type~Ia SNe. Similar to the classification task, we used approximation methods to convert light curves into 1D images. After that, we use them to estimate the time of peak applying two approaches described in Section~\ref{sec:peak_finding}: direct and CNN-based. About 60\% of the light curves were used to train the CNN, while the other 40\% are used to estimate the peak identification quality. The quality metric values are demonstrated in Table~\ref{tab:peak_metrics_plastic_direct} for direct method and Table~\ref{tab:peak_metrics_plastic_cnn} for CNN-based method.

Table~\ref{tab:peak_metrics_plastic_direct} shows that the direct approach to the peak identification
demonstrates better results when it is applied to the light curves approximated with GP model, compared to the neural network-based methods. Table~\ref{tab:peak_metrics_plastic_cnn} illustrates that CNN helps reduce the error of the peak prediction by more than half, depending on the approximation methods. In this case, the best results were achieved with the MLP method (sklearn) and the GP model ranked second. 

\begin{table*}
\centering
\caption{Peak identification quality metrics for approximation models applied to the PLAsTiCC. Direct approach refers to when the peak is determined using the sum of all passbands. \label{tab:peak_metrics_plastic_direct}}
\begin{tabular}{cccccc}
\hline \hline
Model & RMSE, days & MAE, days & RSE & RAE & MAPE, \% \\
\hline 
GP             & \boldmath{$3.14 \pm 0.09$} & \boldmath{$2.04 \pm 0.04$} & \boldmath{$0.0102 \pm 0.0003$} & \boldmath{$0.0075 \pm 0.0002$} & \boldmath{$0.0034 \pm 0.0001$} \\
MLP (sklearn)  & $4.2 \pm 0.1$              & $2.67 \pm 0.06$            & $0.0136 \pm 0.0003$            & $0.0098 \pm 0.0003$            & $0.0044 \pm 0.0001$ \\
MLP (pytorch)  & $5.4 \pm 0.1$              & $3.61 \pm 0.07$            & $0.0174 \pm 0.0004$            & $0.0133 \pm 0.0003$            & $0.0060 \pm 0.0001$ \\
BNN            & $6.1 \pm 0.1$              & $4.39 \pm 0.08$            & $0.0199 \pm 0.0004$            & $0.0162 \pm 0.0004$            & $0.0073 \pm 0.0001$ \\
NF             & $5.8 \pm 0.1$             & $3.95 \pm 0.08$             & $0.0188 \pm 0.0004$            & $0.0146 \pm 0.0004$            & $0.0066 \pm 0.0001$ \\
\hline
\end{tabular}
\end{table*}

\begin{table*}
\centering
\caption{Peak identification quality metrics for approximation models applied to the PLAsTiCC. CNN-based approach is when CNN is trained to predict the true peak position. \label{tab:peak_metrics_plastic_cnn}}
\begin{tabular}{cccccc}
\hline \hline
Model & RMSE, days & MAE, days & RSE & RAE & MAPE, \% \\
\hline
GP             & $1.76 \pm 0.07$            & $1.22 \pm 0.04$            & $0.0057 \pm 0.0002$            & $0.0045 \pm 0.0002$              & $0.00202 \pm 0.00006$ \\
MLP (sklearn)  & \boldmath{$1.67 \pm 0.06$} & \boldmath{$1.15 \pm 0.04$} & \boldmath{$0.0054 \pm 0.0002$} & \boldmath{$0.0042 \pm 0.0002$}   & \boldmath{$0.00192 \pm 0.00006$} \\
MLP (pytorch)  & $1.9 \pm 0.1$              & $1.26 \pm 0.04$            & $0.0061 \pm 0.0004$            & $0.0047 \pm 0.0002$              & $0.00210 \pm 0.00007$ \\
BNN            & $1.99 \pm 0.08$            & $1.34 \pm 0.04$            & $0.0064 \pm 0.0003$            & $0.0049 \pm 0.0002$              & $0.00223 \pm 0.00009$ \\
NF             & $2.17 \pm 0.09$            & $1.51 \pm 0.05$            & $0.0070 \pm 0.0003$            & $0.0056 \pm 0.0002$              & $0.00251 \pm 0.00008$ \\
\hline
\end{tabular}
\end{table*}

\section{ZTF BTS Observations}

\subsection{Approximation quality}

The Zwicky Transient Facility (ZTF) scans the entire visible northern sky with a cadence of three days. The ZTF Bright Transient Survey (BTS)~\citep{https://doi.org/10.48550/arxiv.1910.12973} is a major spectroscopic project that complements the ZTF photometric survey. The objective of the BTS is to spectroscopically classify extra-galactic transients with a $m_{peak} \le 18.5~mag$ in ZTF $\{g, r, i\}$ passbands, and to publicly announce these classifications. Transients discovered by the BTS are predominantly supernovae. It is the largest flux-limited SN survey to date.
 
The ANTARES data broker~\citep{2021AJ....161..107M} was used to download the light curves of the ZTF Bright Transient Survey catalog. With the help of this broker, light curves were downloaded from the ZTF release date by the names of the objects from the Bright Transient Survey. To test the models, all 3374 light curves of the objects presented in the ZTF BTS at the time of 23:29 09/22/2021 were downloaded.

For the approximation method tests, we kept only objects with at least ten observations per each $\{g\}$ and $\{r\}$ passbands. We did not consider measurements in the $\{i\}$ passband, because it contains very few observations and not for all objects. An example of the light curve, which is the closest to our limitation (ten observations in each passband), is demonstrated in Fig.~\ref{fig:ZTF21abvrfsr_22_points}. The total number of selected light curves is 1870. Then, points of each light curve were split into training and testing datasets to fit and measure the quality of all approximation methods, as described in Sec.~\ref{sec:qm_approx}. The optimal hyperparameters of the methods were estimated using a grid search with the MAPE as the objective function. The gained quality metric values are demonstrated in Table~\ref{tab:regression_metrics_ztf1} for approximated fluxes and Table~\ref{tab:regression_metrics_ztf2} for predicted errors of fluxes.

\begin{figure*}
\centering
\includegraphics[width=0.305\linewidth]{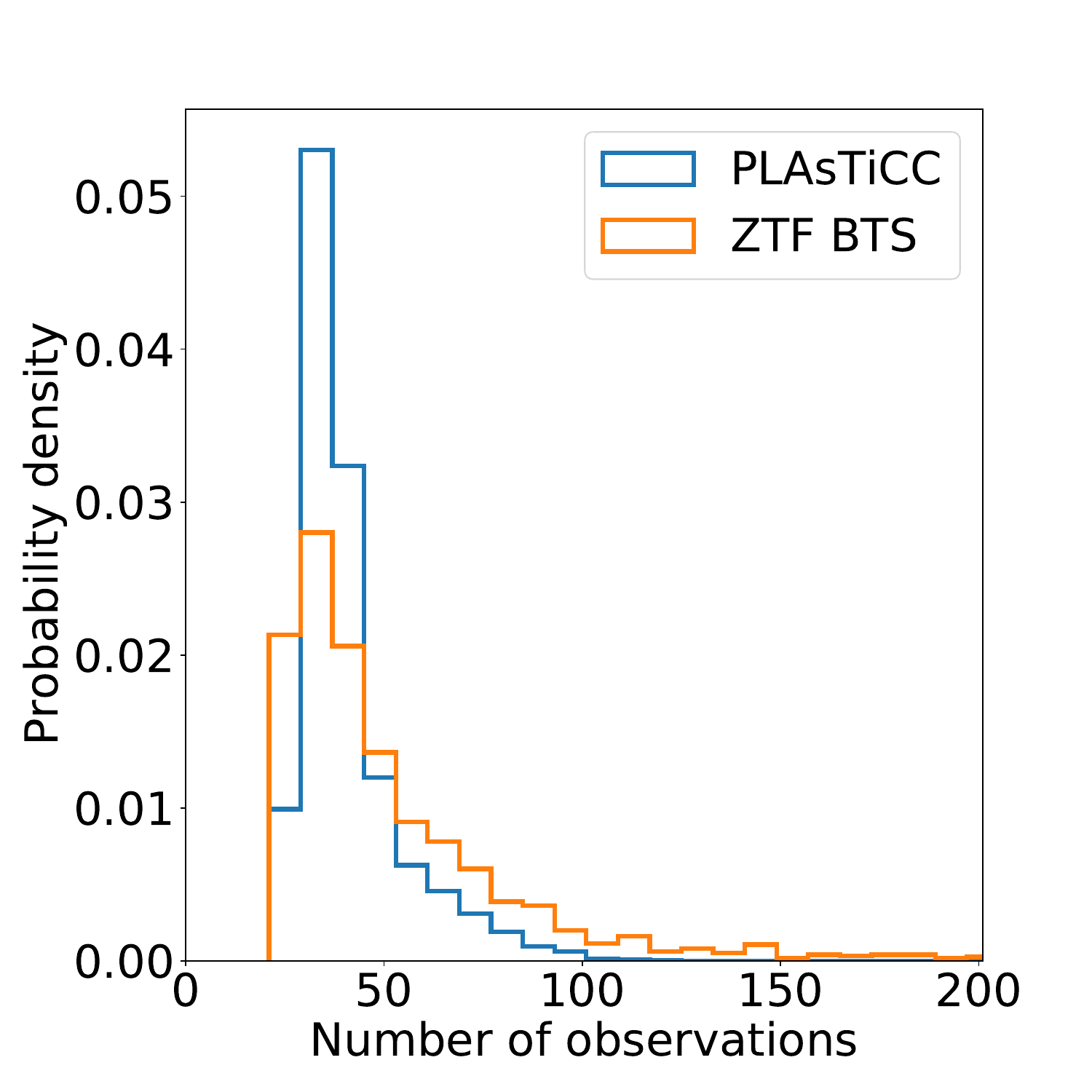}{(a)}
\includegraphics[width=0.305\linewidth]{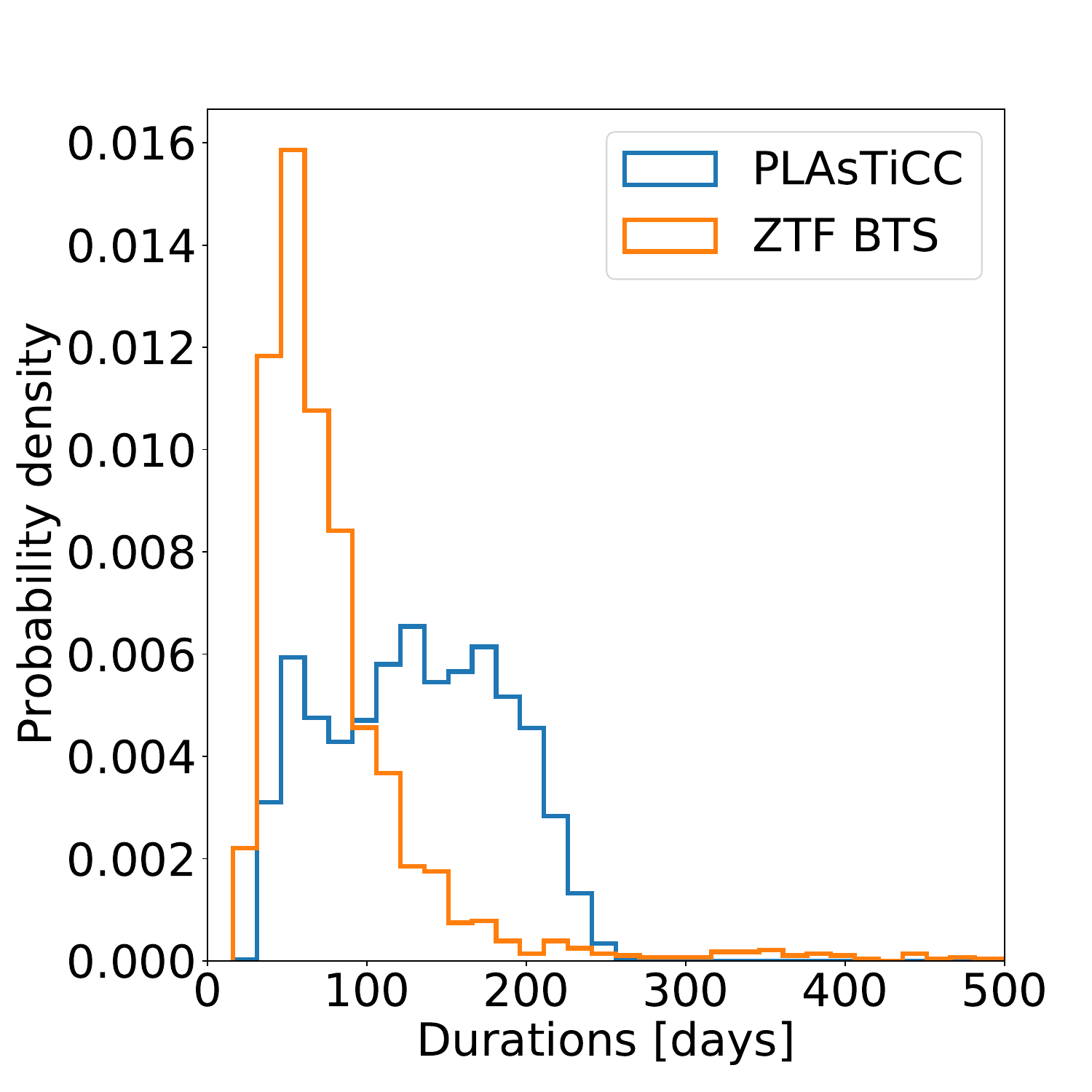}{(b)}
\includegraphics[width=0.305\linewidth]{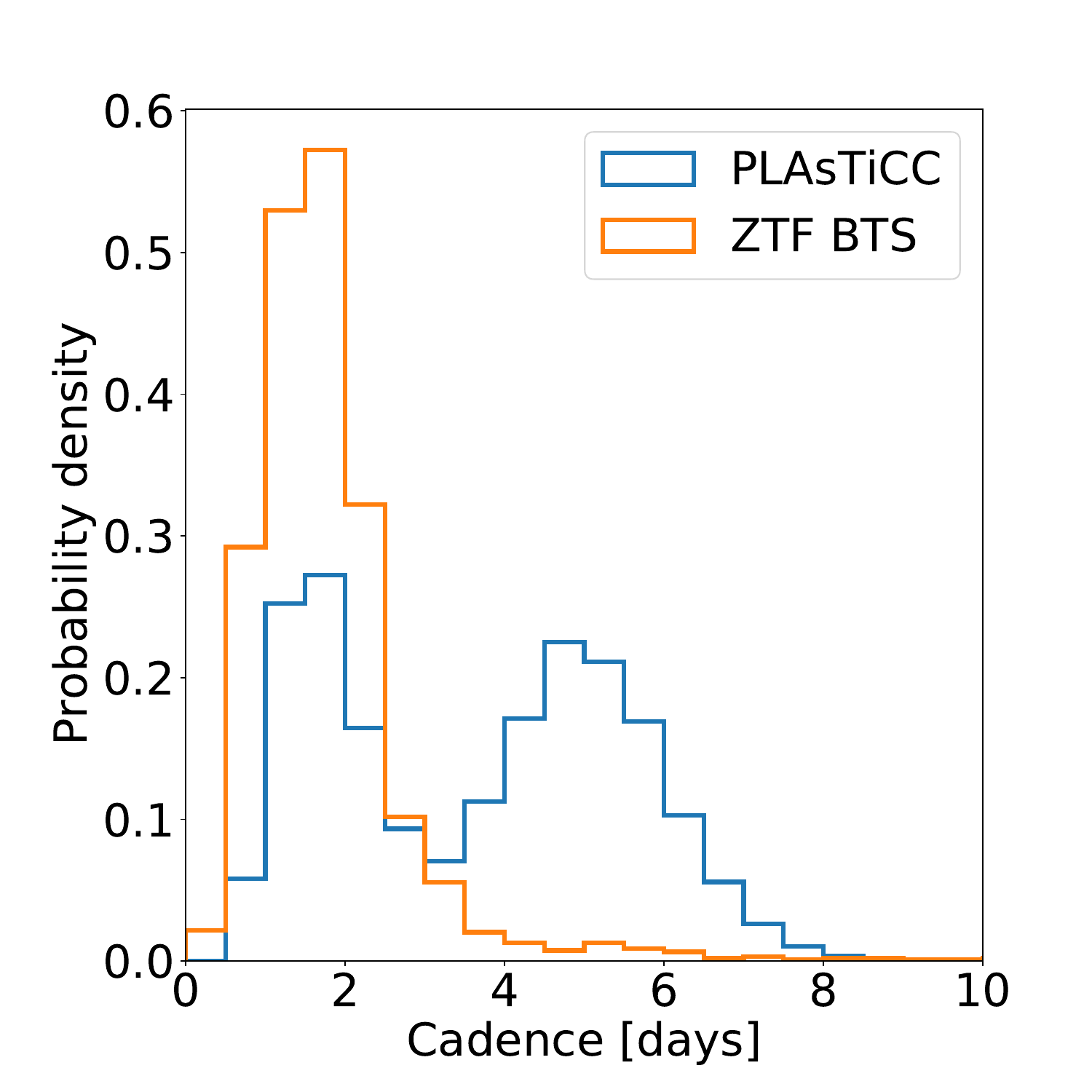}{(c)}
\caption{Probability densities of (a) observations number (b) difference between maximum and minimum time points (c) cadence of the light curves. Selected subsamples of the PLAsTiCC and the ZTF BTS used for our experiments correspond to the blue and orange colors.}
\label{fig:histograms}
\end{figure*}

The results in Table~\ref{tab:regression_metrics_ztf1} show that the MLP (sklearn) model outperforms the GP model by the RAE and MAPE metrics, and has the same values for the RMSE, MAE, and RSE. It demonstrates that neural networks demonstrate excellent performance on real time series and definitely can be used for approximation of the light curves from the ZTF BTS. The reasons for a higher performance of neural network models on real data in comparison with the PLAsTiCC can be the following: (i) the ZTF BTS light curves have a smaller mean cadence. This circumstance is visualized on the cadence histogram Fig. \ref{fig:histograms} (c); (ii) the ZTF BTS sample includes only bright object light curves with a more prominent amplitude of possible features; (iii) the ZTF BTS light curves have a larger number of observations compared to the PLAsTiCC sample in keeping with Fig. \ref{fig:histograms} (a) illustrating probability density distributions; (iv) We did not take poor ZTF\_i passband into account. The PLAsTiCC has six passbands, including passbands with a small number of observations, in contrast to using two passbands (ZTF\_r, ZTF\_g) out of three released in the ZTF BTS sample.

As in the case of the simulations, we quantified the ability of the methods to estimate the uncertainties of the flux measurements as presented in Table~\ref{tab:regression_metrics_ztf2}. It shows that NF model shows better quality metrics than other methods for all metrics except for nRMSEo and PICP$_{68}$. The GP demonstrates similar results. In general, neural networks tend to underestimate the uncertainties on the ZTF BTS as well as on the PLAsTiCC.

\begin{table*}
\centering
\caption{Regression quality metrics for approximation models applied to the ZTF BTS. Estimation quality of the mean fluxes. \label{tab:regression_metrics_ztf1}}
\begin{tabular}{cccccc}
\hline \hline
Model & RMSE, mJy & MAE, mJy & RSE & RAE & MAPE, \% \\
\hline
GP             & \boldmath{$0.018 \pm 0.002$} & \boldmath{$0.015 \pm 0.002$} & \boldmath{$0.66 \pm 0.04$} & $0.66 \pm 0.04$            & $11.0 \pm 0.3$ \\
MLP (sklearn)  & \boldmath{$0.018 \pm 0.001$} & \boldmath{$0.015 \pm 0.001$} & \boldmath{$0.66 \pm 0.05$} & \boldmath{$0.65 \pm 0.05$} & \boldmath{$9.9 \pm 0.2$} \\
MLP (pytorch)  & $0.019 \pm 0.001$            & \boldmath{$0.015 \pm 0.001$} & $0.68 \pm 0.05$            & $0.66 \pm 0.05$            & $10.4 \pm 0.3$ \\
BNN            & $0.021 \pm 0.001$            & $0.018 \pm 0.001$            & $0.77 \pm 0.06$            & $0.75 \pm 0.06$            & $11.4 \pm 0.3$ \\
NF             & $0.021 \pm 0.002$            & $0.017 \pm 0.001$            & $0.72 \pm 0.05$            & $0.72 \pm 0.05$            & $10.6 \pm 0.2$ \\
\hline
\end{tabular}
\end{table*}

\begin{table*}
\centering
\caption{Regression quality metrics for approximation models applied to the ZTF BTS. Estimation quality of the flux uncertainties. \label{tab:regression_metrics_ztf2}}
\begin{tabular}{cccccc}
\hline \hline
Model & NLPD & nRMSEo & nRMSEp & $\text{PICP}_{68}$, \% & $\text{PICP}_{95}$, \% \\
\hline
GP             & $-2.57 \pm 0.05$            & $1.10 \pm 0.03$            & $1.16 \pm 0.02$  & \boldmath{$65.7 \pm 0.6$} & $88.4 \pm 0.5$ \\
MLP (sklearn)  & $3.4 \pm 0.6$               & \boldmath{$1.07 \pm 0.03$} & $2.64 \pm 0.07$             & $45.7 \pm 0.7$            & $68.0 \pm 0.7$ \\
MLP (pytorch)  & $8 \pm 2$                   & $1.13 \pm 0.03$            & $2.99 \pm 0.09$             & $44.7 \pm 0.7$            & $65.8 \pm 0.7$ \\
BNN            & $66 \pm 6$                  & $1.25 \pm 0.03$            & $7.9 \pm 0.2$               & $19.7 \pm 0.5$            & $35.4 \pm 0.7$ \\
NF             & \boldmath{$-2.61 \pm 0.05$} & $1.19 \pm 0.03$            & \boldmath{$0.86 \pm 0.02$}             & $79.0 \pm 0.6$            & \boldmath{$94.4 \pm 0.4$} \\
\hline
\end{tabular}
\end{table*}

\begin{figure*}
\centering
\includegraphics[width=0.47\linewidth]{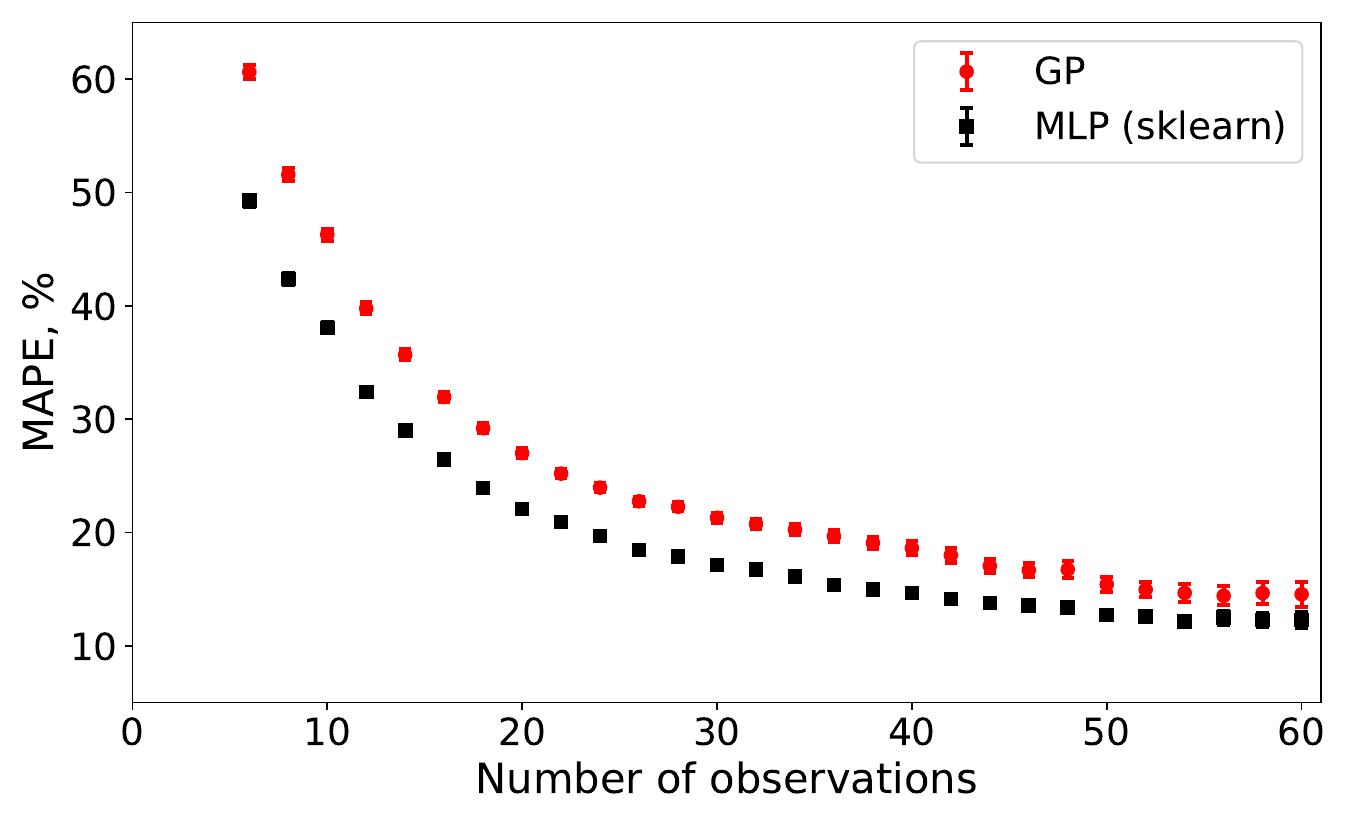}{(a)}
\includegraphics[width=0.47\linewidth]{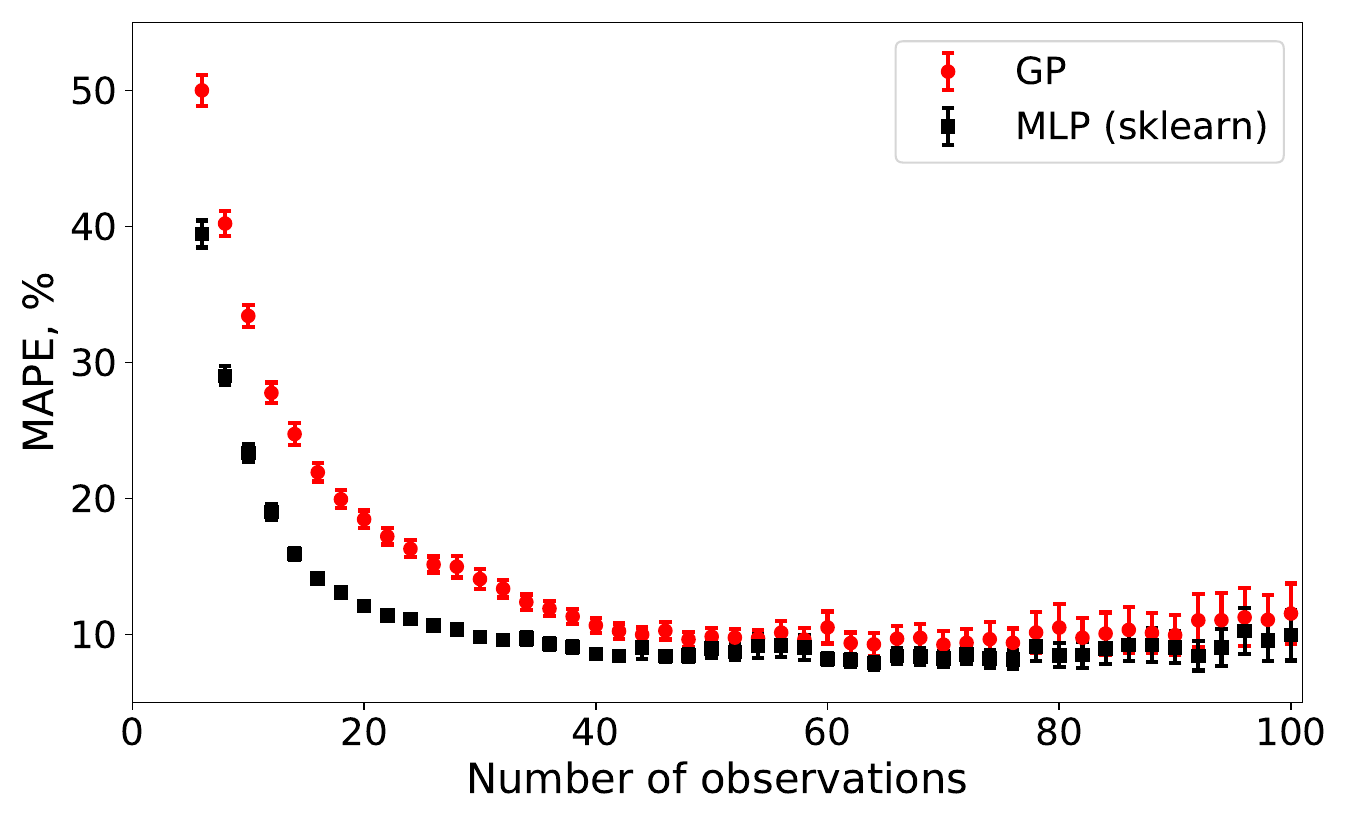}{(b)}
\caption{Dependence of MAPE approximation quality metric on the number of observations in (a) PLAsTiCC and (b) ZTF BTS light curves for GP and MLP (sklearn) methods. The horizontal axis shows how many observations were used to fit each method. The number of light curves in our samples drops with increasing the number of observations. Error bars show the uncertainty of MAPE measurements.}
\label{fig:MAPE_approx}
\end{figure*}

We conducted an additional experiment to measure the dependency of the approximation quality on the number of observations in a light curve. We split each individual light curve in our PLAsTiCC and ZTF BTS datasets into training and testing samples. The testing sample contains points from two time bins with a width of five days. We used them to calculate the MAPE metric values. From the training sample, we randomly selected $N$ observations from each light curve for the approximation fitting. If there were fewer than N points in the curve, it was discarded. We estimated the MAPE mean value and its uncertainty using the bootstrap technique over the selected light curves. This procedure was repeated for $N$ ranging from six to 60 observations for the PLAsTiCC and from six to 100 observations for the ZTF BTS with a step of two. The results are demonstrated in Fig.~\ref{fig:MAPE_approx}. As expected, the plots show that the approximation error significantly drops with the increase in the number of observations. The metric values flatten after the number of observations in the training sample reaches the point of 50. MLP method shows a better approximation quality than GP in all ranges of $N$. We conclude that neural networks reach appropriate results -- even when they are trained on a light curve with a very small number of observations.

\subsection{Supernova classification}

We additionally performed a test of the various approximation methods on the supernova classification problem. For that, we took the same light curves described in the previous section and divided them into two classes. Objects of SN~Ia, SN~Ia-91T, SN~Ia-pec, SN~Ia-91bg, SN~Iax, and SN~Ia-CSM supernova types were labeled as the positive class. All other types: SN~II, SN~IIn, SN~Ic, SN~Ib, SN~IIb, SN~Ib-pec, SN~Ic-BL, TDE, SN~Ib/c, SLSN-I, SN~IIP, LBV, SN~II-pec, SLSN-II, nova, SN~Ibn, ILRT, and others (unclassified in the ZTF BTS), were considered as a negative class.  

We used the approximation methods for the light curve augmentation and transformed them into 1D images with 2 channels as it is described in Section~\ref{sec:sup_class}. Here we used the same optimal hyperparameters as estimated in the previous section. The images were used as input for the CNN classifier model training. About 60\% of the light curves were used to train the CNN, while the other 40\% were used to estimate the classification quality. The metric values are demonstrated in Table~\ref{tab:classification_metrics_ztf}.
The results show that all methods produce similar results. We conclude that choosing a specific approximation method does not have a significant impact on supernova classification results.

\begin{table*}
\centering
\caption{Supernovae classification quality metrics for approximated ZTF BTS light curves. \label{tab:classification_metrics_ztf}}
\begin{tabular}{ccccccc}
\hline \hline
Model & ROC AUC & PR AUC & Log Loss & Accuracy & Recall & Precision \\
\hline
GP             & $0.94 \pm 0.01$            & $0.976 \pm 0.008$            & \boldmath{$0.26 \pm 0.03$} & \boldmath{$0.90 \pm 0.02$} & \boldmath{$0.95 \pm 0.01$} & $0.92 \pm 0.02$ \\
MLP (sklearn)  & $0.95 \pm 0.01$            & $0.981 \pm 0.005$            & \boldmath{$0.26 \pm 0.04$} & \boldmath{$0.90 \pm 0.02$} & $0.93 \pm 0.02$            & $0.93 \pm 0.01$ \\
MLP (pytorch)  & \boldmath{$0.96 \pm 0.01$} & \boldmath{$0.983 \pm 0.005$} & $0.29 \pm 0.05$            & \boldmath{$0.90 \pm 0.02$} & $0.93 \pm 0.02$            & $0.93 \pm 0.01$ \\
BNN            & $0.95 \pm 0.01$            & $0.980 \pm 0.006$            & $0.31 \pm 0.05$            & $0.89 \pm 0.02$            & $0.91 \pm 0.02$            & \boldmath{$0.94 \pm 0.02$} \\
NF             & $0.94 \pm 0.01$            & $0.97 \pm 0.01$              & $0.39 \pm 0.07$            & $0.89 \pm 0.02$            & $0.91 \pm 0.02$            & \boldmath{$0.94 \pm 0.01$} \\
\hline
\end{tabular}
\end{table*}

\subsection{Peak identification}

\begin{figure*}
\includegraphics[width=0.47\linewidth]{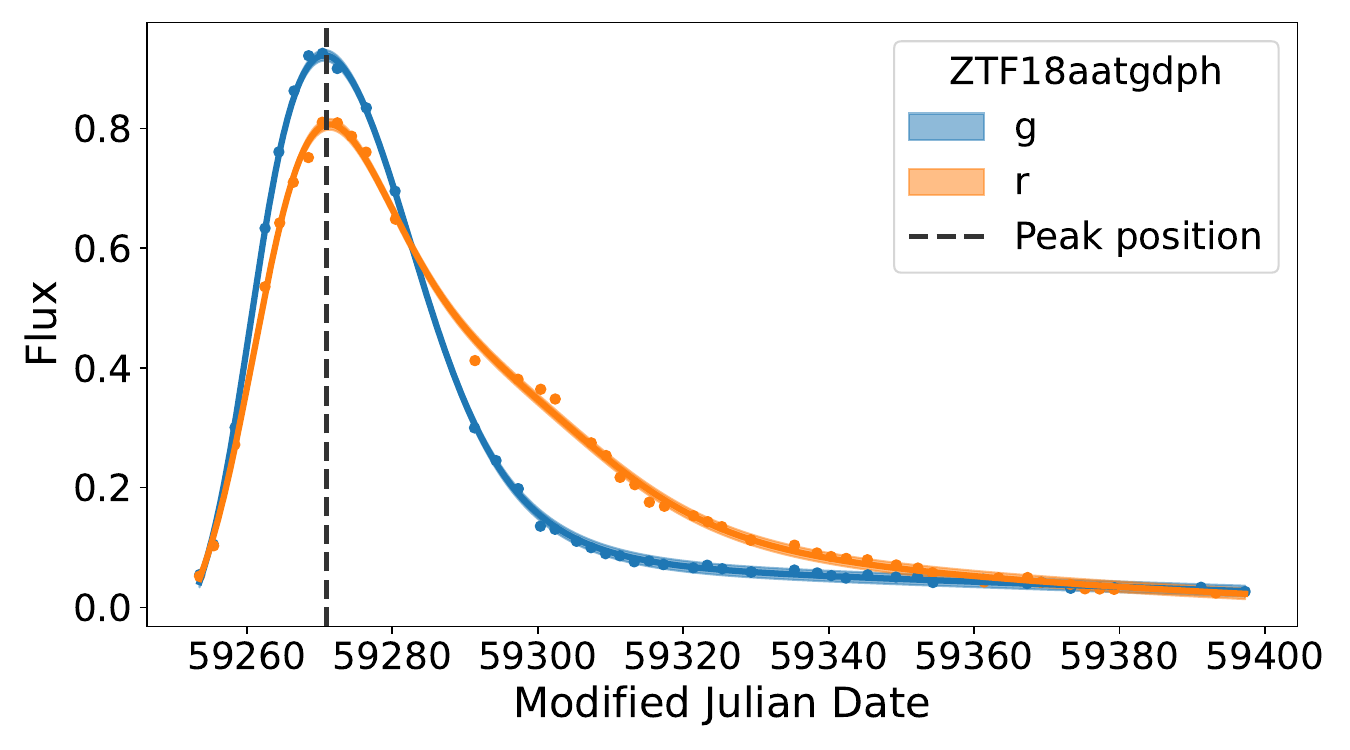}{(a)}
\includegraphics[width=0.47\linewidth]{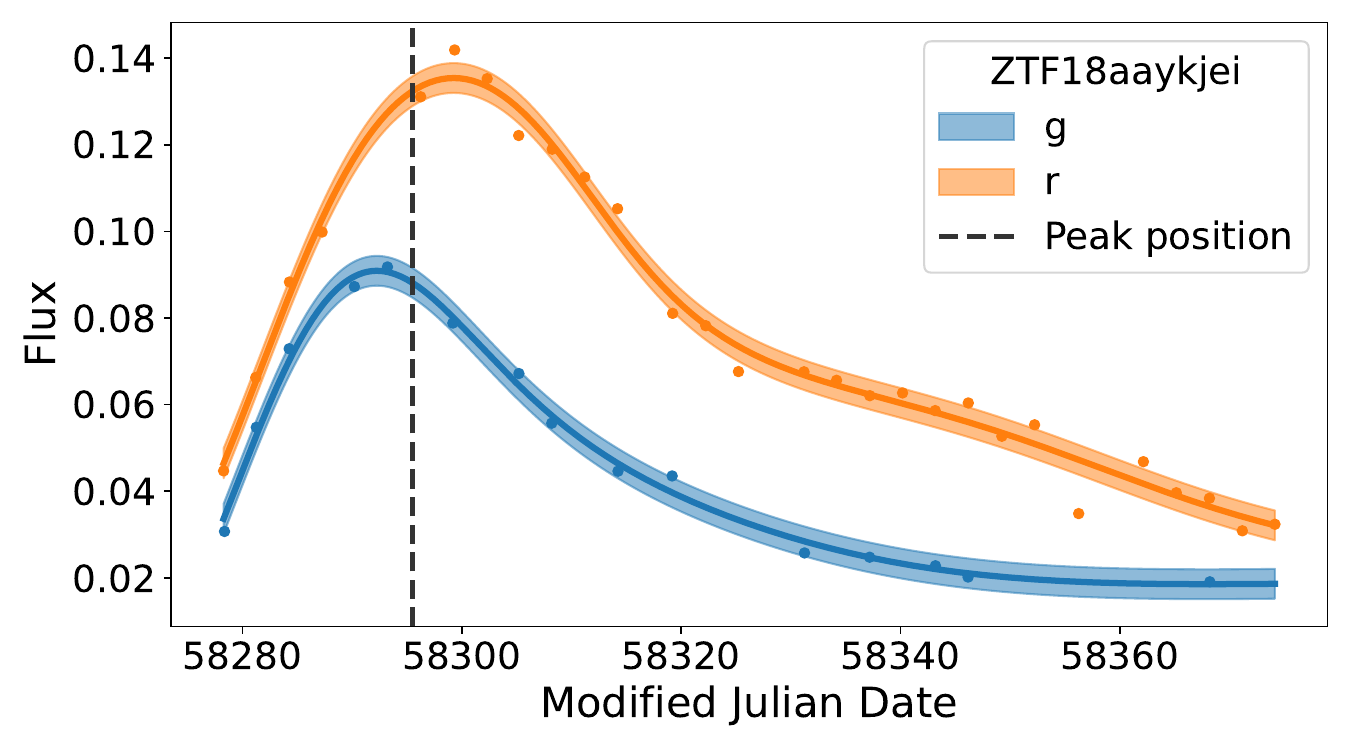}{(b)}
\includegraphics[width=0.47\linewidth]{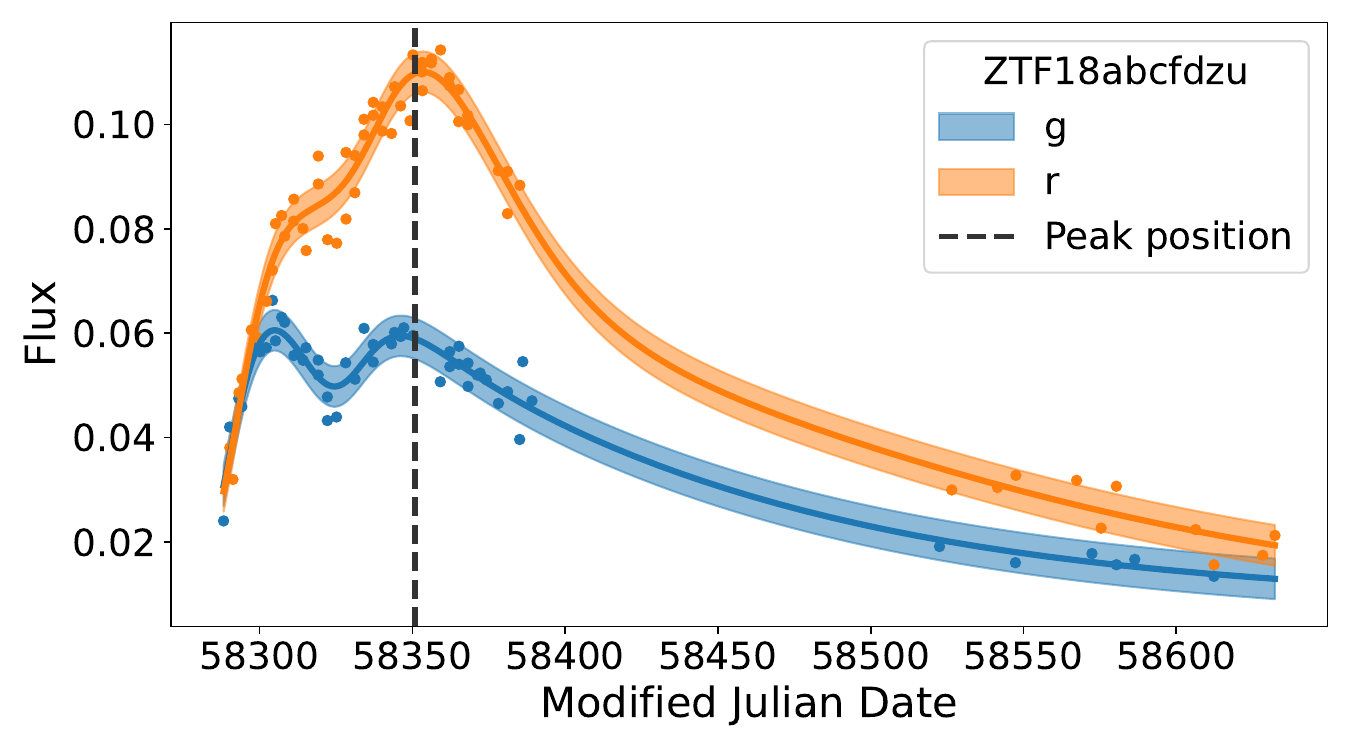}{(c)}
\includegraphics[width=0.47\linewidth]{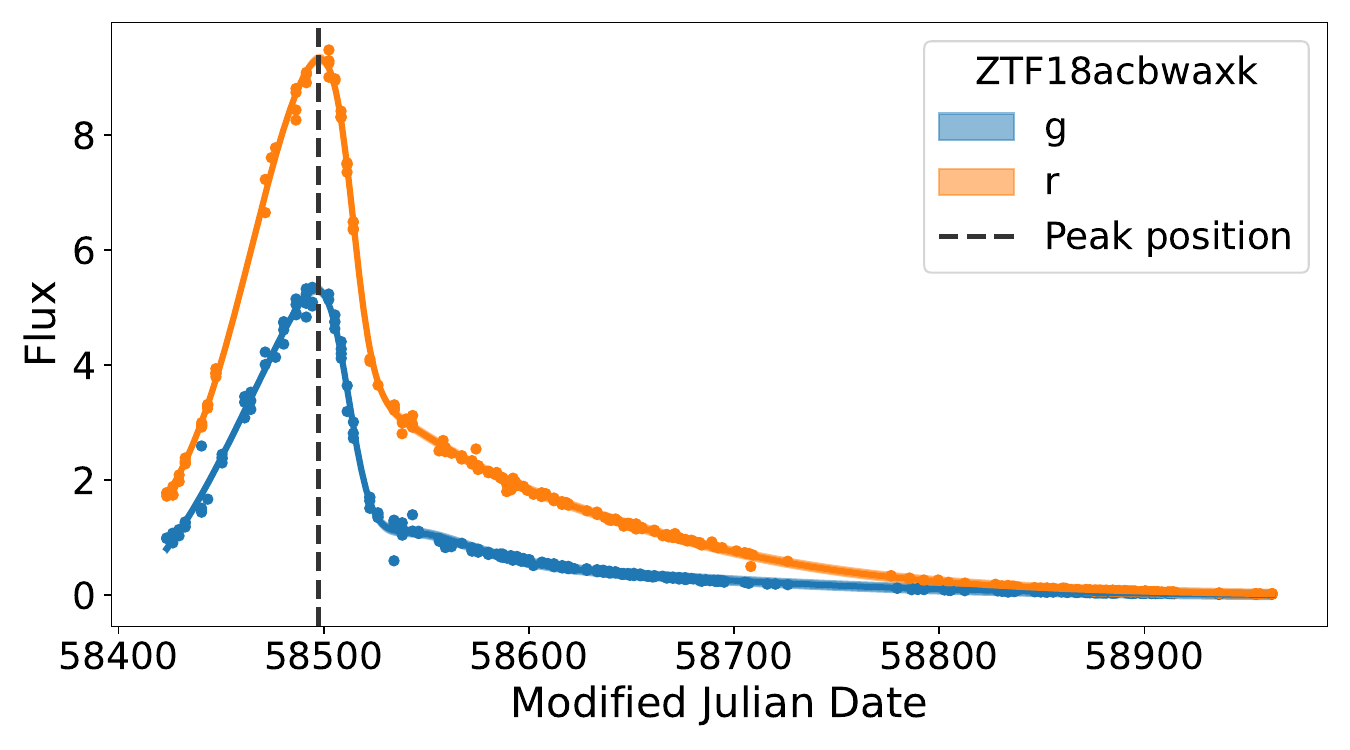}{(d)}
\includegraphics[width=0.47\linewidth]{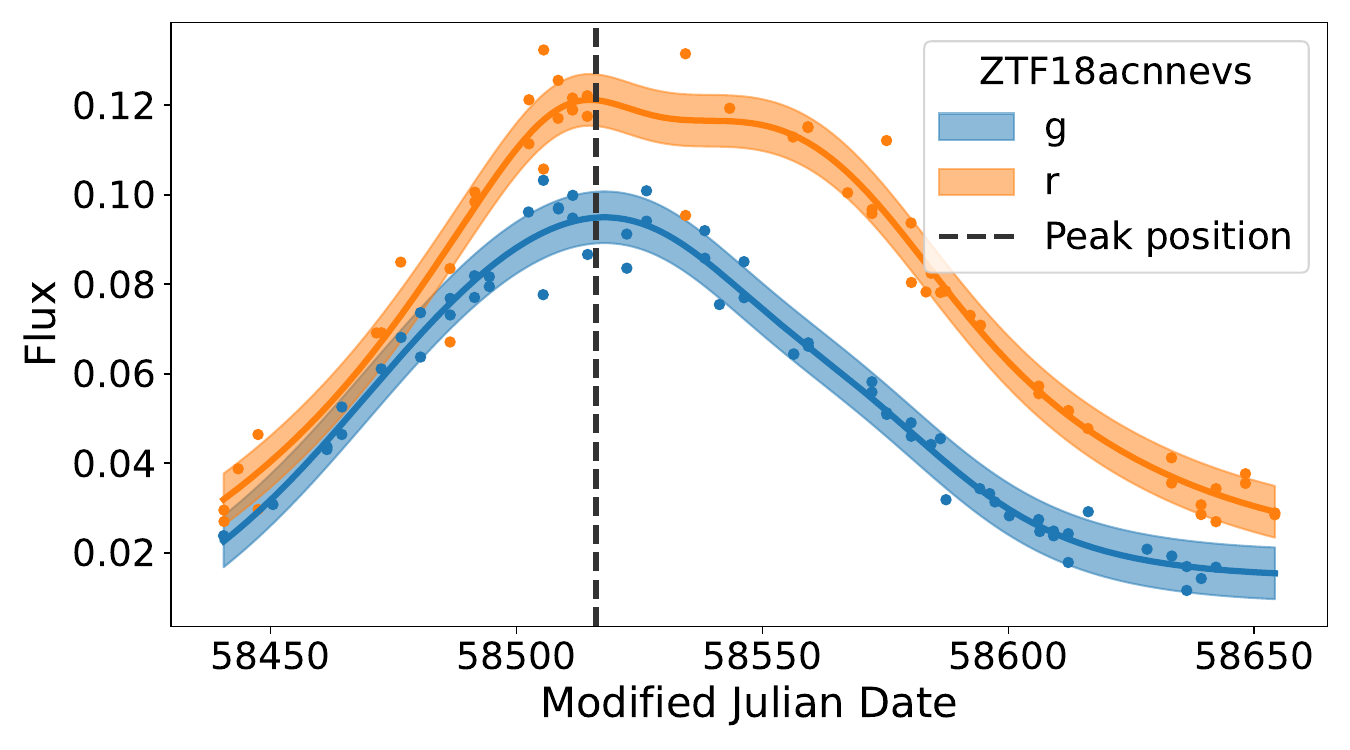}{(e)}
\includegraphics[width=0.47\linewidth]{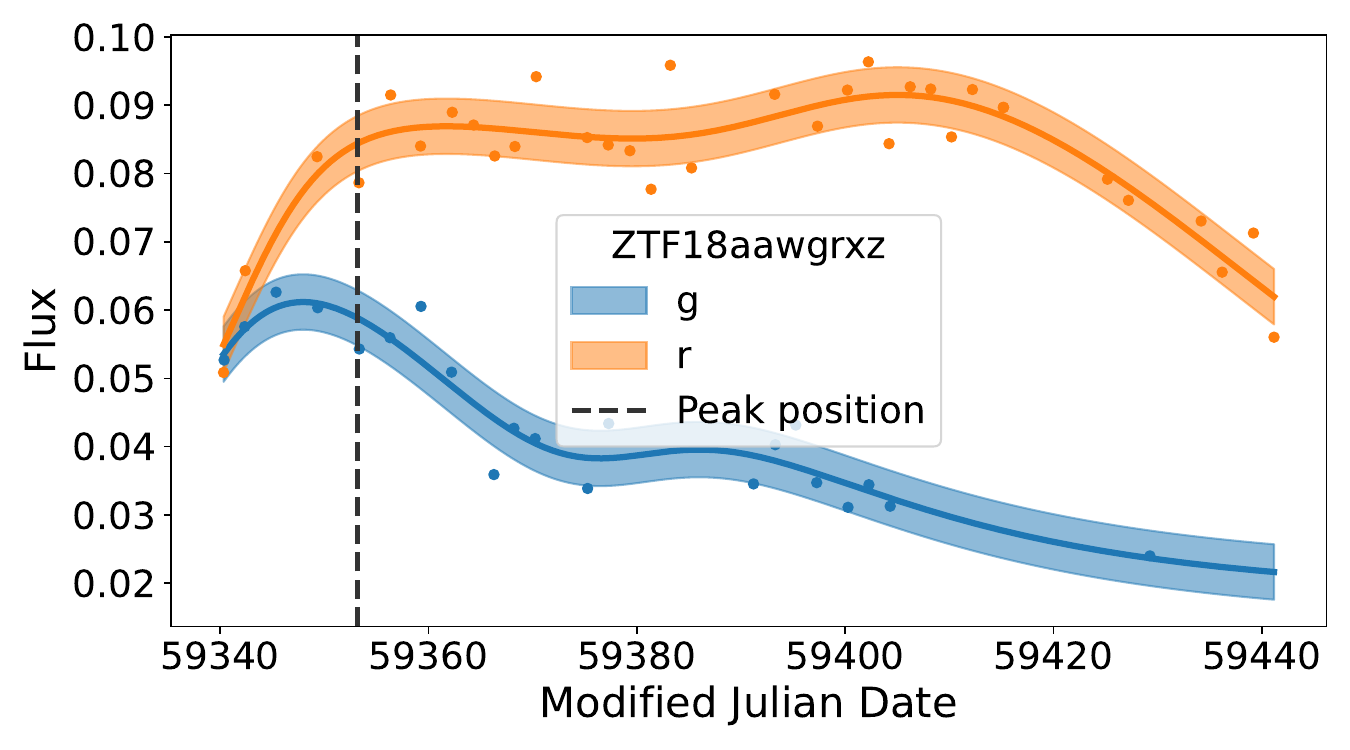}{(f)}

\caption{Examples of peak direct estimation using MLP (sklearn) approximations for a range of the ZTF BTS objects: (a) ZTF18aatgdph; (b) ZTF18aaykjei; (c) ZTF18abcfdzu; (d) ZTF18acbwaxk; (e) ZTF18acnnevs; (f) ZTF18aawgrxz. The points represent measurements in the corresponding passbands. The solid lines are the estimated mean $\mu(x)$ values. The shaded areas represent $\pm 1\sigma(x)$ uncertainty band.
\label{fig:peak_example}}
\end{figure*}

Similarly to the classification task, we approximated light curves into a regular high-cadence time grid. Then, we used them to estimate the time of peaks using the direct approach described in Section~\ref{sec:peak_finding}. Examples of ZTF BTS light curves with estimated peaks are shown in Fig.~\ref{fig:peak_example}. Table~\ref{tab:peak_ztf_sk} shows the peak positions and magnitudes of the ZTF BTS objects.

\begin{table*}
\begin{center}
\caption{Estimated peaks for the ZTF BTS objects. Approximated using MLP (sklearn) model. The full version for each model is available \href{http://lc-dev.voxastro.org/}{online}.} \label{tab:peak_ztf_sk}
\begin{tabular}{cccccccc}
\hline \hline
ZTF name & SN name & $m_{g}$ & $t_{g}$ & $m_{r}$ & $t_{r}$ & $m_{\Sigma}$ & $t_{\Sigma}$ \\
\hline
ZTF21abwxaht &  SN 2021xqc & 18.71 & 59469.60250 & 18.52 & 59471.90505 & 17.87 & 59470.37001 \\
ZTF21abjrwhq &  SN 2021sdv & 18.25 & 59410.74629 & 18.17 & 59411.12950 & 17.46 & 59411.12950 \\
ZTF21abcgaln &  SN 2021mxx & 16.55 & 59363.71318 & 16.16 & 59366.15413 & 15.60 & 59364.93366 \\
ZTF21aaufvpg &  SN 2021jap & 17.01 & 59330.54799 & 17.14 & 59332.12218 & 16.32 & 59331.07272 \\
ZTF21aapkfjx &  SN 2021fqb & 16.91 & 59301.93222 & 16.98 & 59301.93222 & 16.19 & 59301.93222 \\
ZTF21aagqcnl & SN 2021biz & 15.71 & 59258.66084 & 15.83 & 59258.66084 & 15.02 & 59258.66084 \\
ZTF20acwcfvx &  SN 2020acaf & 18.57 & 59198.98497 & 18.62 & 59200.19335 & 17.84 & 59199.79055 \\
ZTF20acobevy &  SN 2020yub & 18.25 & 59158.45014 & 18.32 & 59160.87008 & 17.54 & 59159.41812 \\
ZTF20acedqis &  SN 2020ufx & 18.24 & 59122.09822 & 18.30 & 59126.93300 & 17.59 & 59125.72430 \\
ZTF20abuqali &  SN 2020rht & 19.04 & 59080.45034 & 18.99 & 59080.45034 & 18.26 & 59080.45034 \\
ZTF20abhlncz &  SN 2020nke & 18.96 & 59035.17439 & 18.74 & 59038.62660 & 18.12 & 59036.32513 \\
ZTF20aayjxdv & SN 2020jge & 18.10 & 58986.15285 & 18.16 & 58987.64490 & 17.38 & 58986.89887 \\
ZTF20aakbtyz & SN 2020bjg & 18.64 & 58888.80773 & 18.68 & 58889.20067 & 17.91 & 58888.80773 \\
ZTF19acxyumq &  SN 2019wvc & 16.79 & 58844.91564 & 16.87 & 58844.91564 & 16.08 & 58844.91564 \\
ZTF19abyjjxp &  SN 2019qcj & 16.97 & 58746.44892 & 16.87 & 58746.44892 & 16.17 & 58746.44892 \\
ZTF19abhzelh &  SN 2019lrv & 17.67 & 58701.41276 & 17.73 & 58702.18964 & 16.95 & 58701.41276 \\
ZTF19aaugaam & SN 2019evh & 17.95 & 58623.55538 & 18.01 & 58637.28421 & 17.27 & 58626.78569 \\
ZTF19aadnxog &  SN 2019vb & 16.71 & 58514.63476 & 16.66 & 58521.57320 & 15.96 & 58516.36937 \\
ZTF18abdfydj &  SN 2018dzr & 18.30 & 58310.05330 & 18.69 & 58308.08598 & 17.73 & 58309.39752 \\
ZTF18aatgdph &  SN 2021cgl & 16.49 & 59270.91968 & 16.63 & 59272.37189 & 15.81 & 59270.91968 \\
ZTF18aaykjei &  SN 2018crl & 19.02 & 58292.80533 & 18.57 & 58299.58256 & 18.04 & 58295.70986 \\
ZTF18abcfdzu &  SN 2018dfa &  19.45 & 58305.60431 & 18.79 & 58354.25815 & 18.33 & 58350.78288 \\
ZTF18acbwaxk &  SN 2018hna & 14.56 & 58494.24002 & 13.99 & 58499.68246 & 13.49 & 58494.24002 \\
ZTF18acnnevs &  SN 2018lng & 18.98 & 58516.02974 & 18.67 & 58526.82415 & 18.07 & 58520.3475 \\
ZTF18aawgrxz &  SN 2021lmp & 19.47 & 59349.49727 & 19.0 & 59405.52445 & 18.51 & 59354.59065 \\
\hline
\end{tabular}
\end{center}
\end{table*}

\section{Conclusions}

This work shows the results of the light curve approximation over time and wavelength using neural networks on the PLAsTiCC and the ZTF BTS datasets. The outcomes have led us to draw the following conclusions:
\begin{itemize}
  \item Neural networks trained on a single multipassband light curve are a reasonable alternative to other approximation methods (including Gaussian processes). This approach is suitable for both single-object and catalog-level studies.
  \item Neural networks require $O(N)$ operations for the approximation fitting, compared to $O(N^3)$ for the GP method, where $N$ is the number of observations in a light curve. The real computational time significantly depends on specific implementations of the methods. In this respect, we recommend using {\tt\string  LBFGS} optimizer for any approximation model.
  \item Approximation models can be used to generate additional observations for the light curves. These synthetic observations help to estimate various physical properties of astronomical objects, such as the peak time and the supernova type.
  \item The supernova type identification and peak identification using CNNs do not show any dependence on the approximation methods. Neural networks and GP demonstrate similar qualities.
  \item \texttt{Fulu} Python library for the light-curve approximation which implements neural network methods for light curves having as small as $\sim\,10$ observations. The quality of individual approximations can be judged using built-in metrics or further physics-based analysis.
\end{itemize}

The evolution of light-curve approximations has significantly progressed with advancements in mathematical and software tools: from manually drawing curves through data to interpolating with polynomials and splines, through approximating observations with uncertainties using smoothing splines, and up to fitting the data  with Gaussian processes.
This work presented a suite of innovative methods for light-curve approximation, utilizing the power of neural networks.
These methods have been made into the publicly available Python package, \texttt{Fulu}, which provides a user-friendly interface for precise approximation.
The methodologies described in this work could find applications in data reduction, machine learning, or detailed analysis of specific astronomical objects, particularly within the field of time-domain astronomy.

\section*{Acknowledgement}
KM's work on the data preparation is supported by the RFBR and CNRS according to the research project No. 21-52-15024. DD, MH, and MD are supported by the HSE Basic Research Fund in the design, construction, and testing of the data augmentation techniques. We express our appreciation to Matvey Volkov, who developed a website for visualizing the results of this paper. We appreciate Dr. Jacob Isbell's contribution as a native speaker editor of our paper language, Dr. Anton Afanasyev as language editor, and an anonymous referee for valuable comments and suggestions.

\textit{Facilities:} This research was supported in part through computational resources of HPC facilities at HSE University~\citep{Kostenetskiy_2021}.

\textit{Data availability:}
Python library is available in the GitHub repository {\tt\string  Fulu}\footnote{\url{https://github.com/HSE-LAMBDA/fulu}}. Scripts of all our experiments in this work are provided in the GitHub repository\footnote{\url{https://github.com/HSE-LAMBDA/light_curve_approx}} as well. It also contains five CSV tables with the results of peak identification experiments for the ZTF BTS for different approximation models. The website with interactive plots for approximation experiments on the ZTF BTS \footnote{\url{http://lc-dev.voxastro.org/}} can help to understand which model is more useful for each object in this catalog. Additionally, Table \ref{tab:peak_ztf_sk} is available in electronic form at the CDS via anonymous ftp to cdsarc.cds.unistra.fr (130.79.128.5) or via \url{https://cdsarc.cds.unistra.fr/cgi-bin/qcat?J/A+A/}.

\clearpage
\newpage

\bibliographystyle{aa}
\bibliography{aa_template}
\begin{appendix}
\section{Light curve examples}
\begin{figure*}
\includegraphics[width=0.47\linewidth]{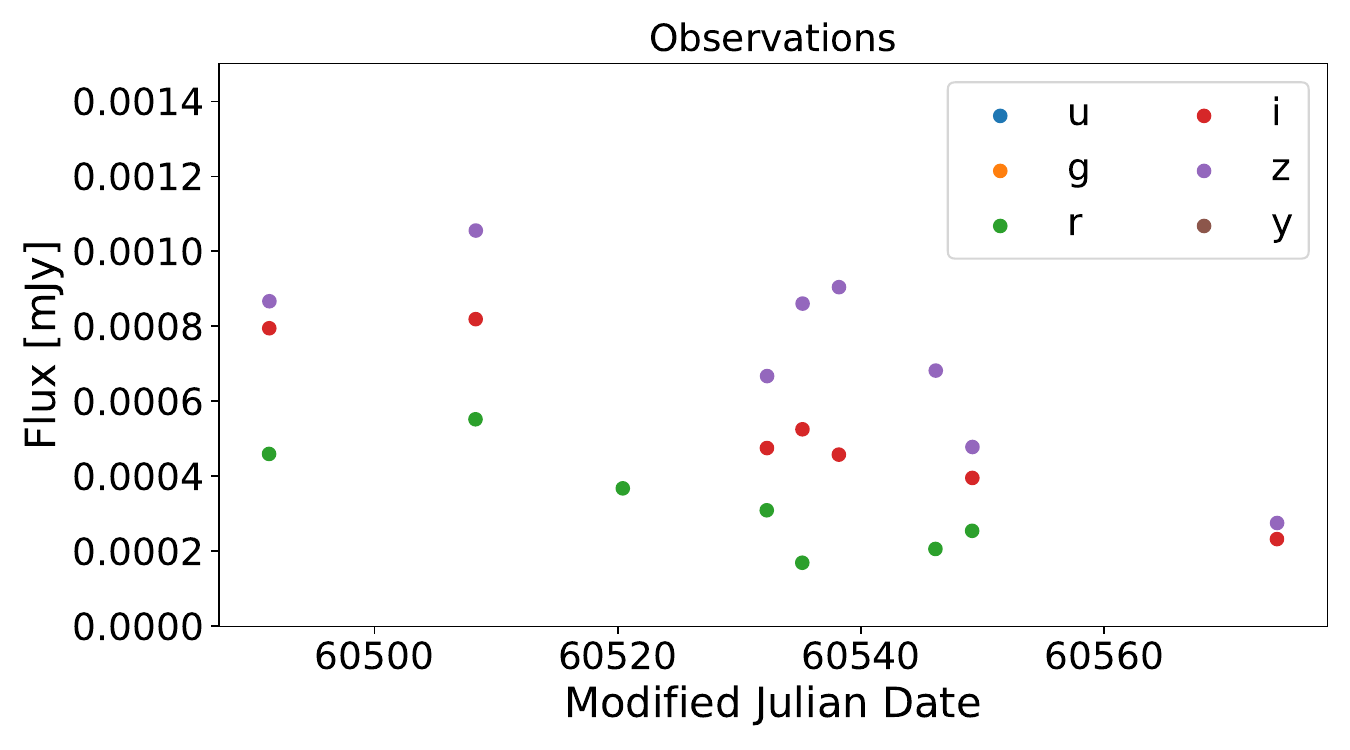}{(a)}
\includegraphics[width=0.47\linewidth]{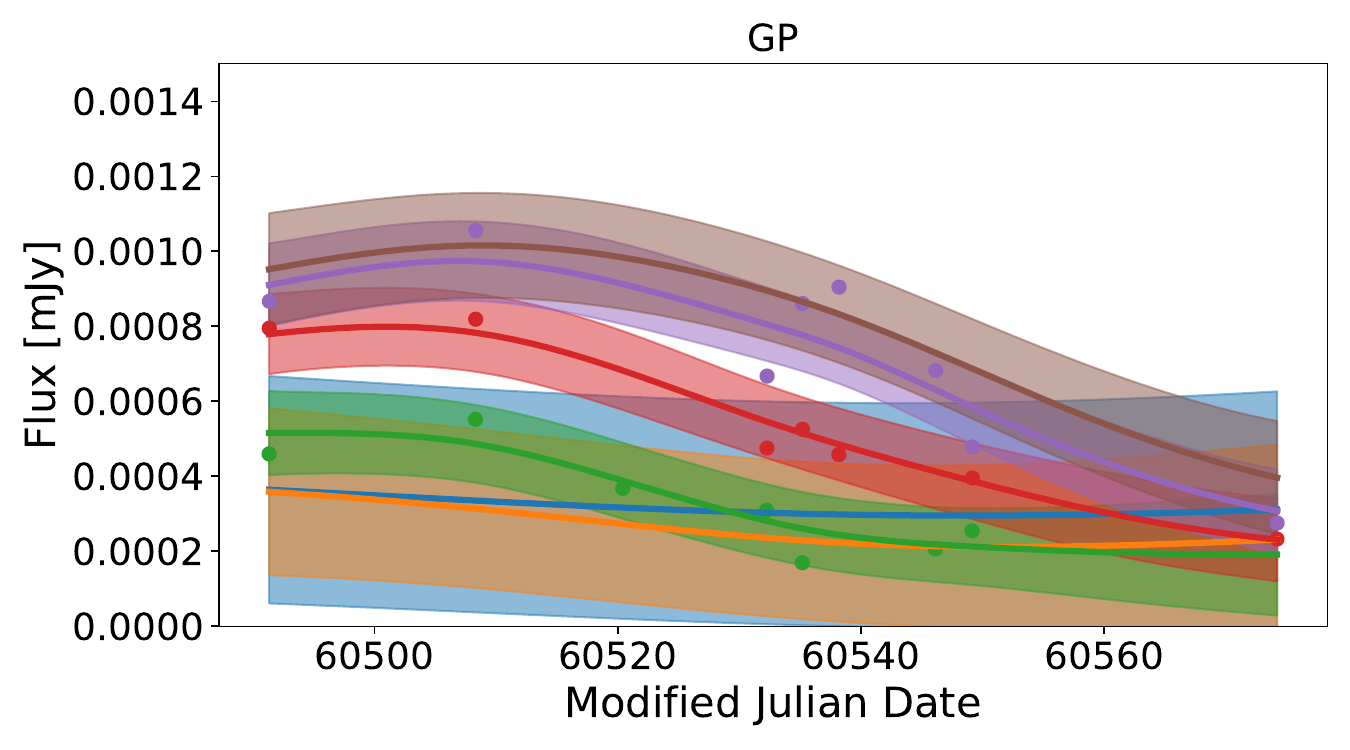}{(b)}
\includegraphics[width=0.47\linewidth]{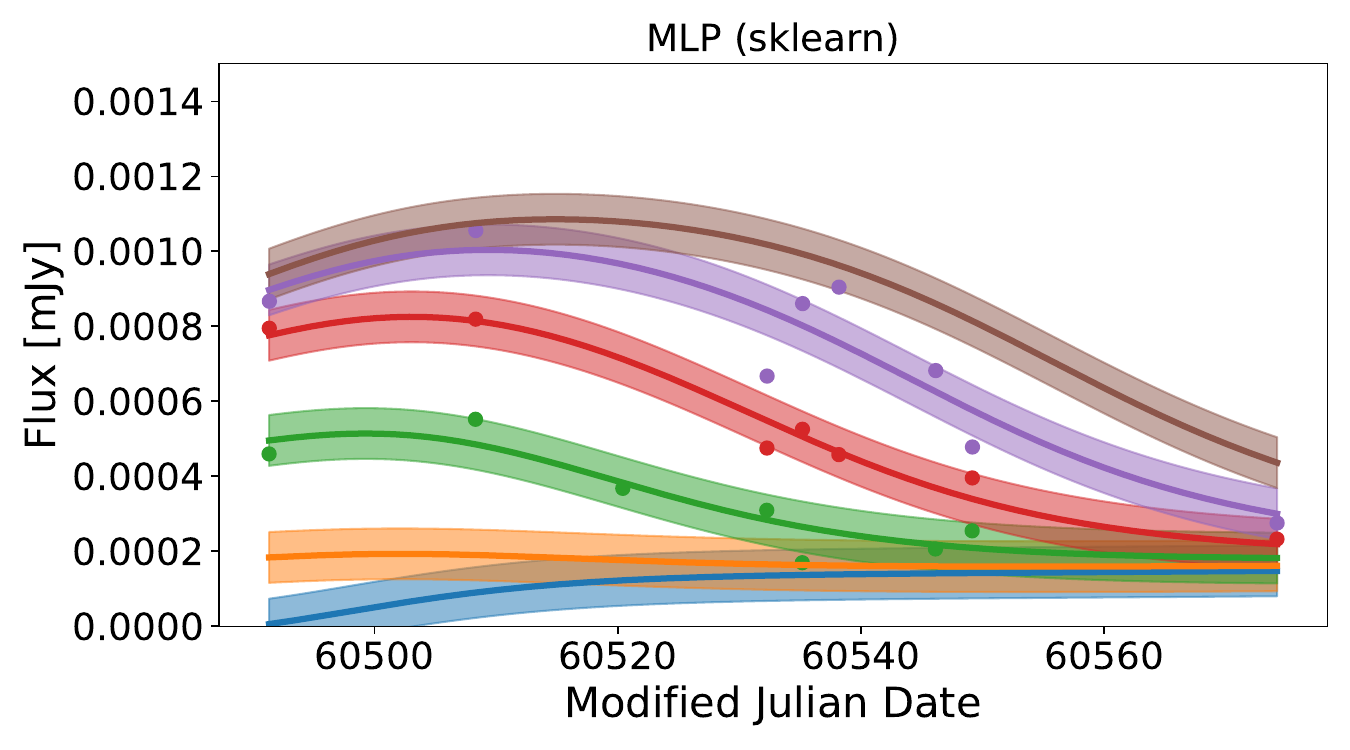}{(c)}
\includegraphics[width=0.47\linewidth]{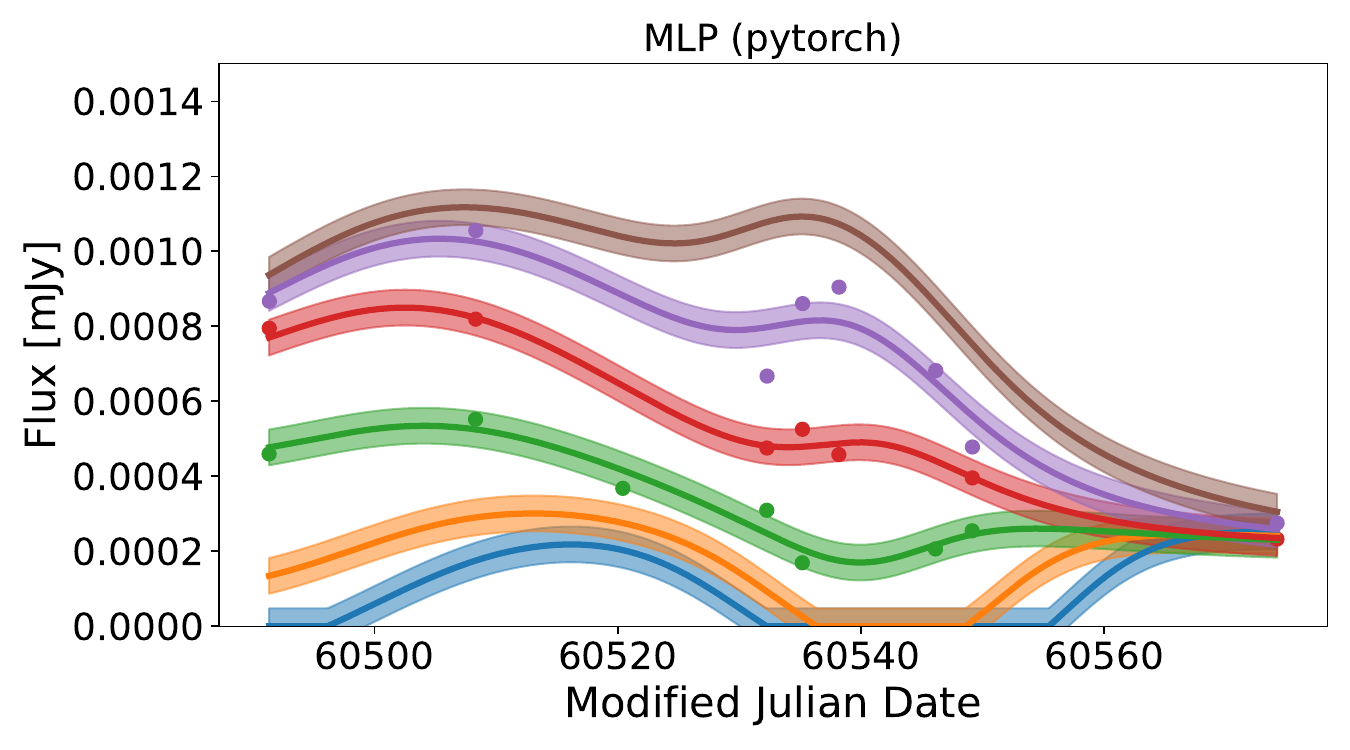}{(d)}
\includegraphics[width=0.47\linewidth]{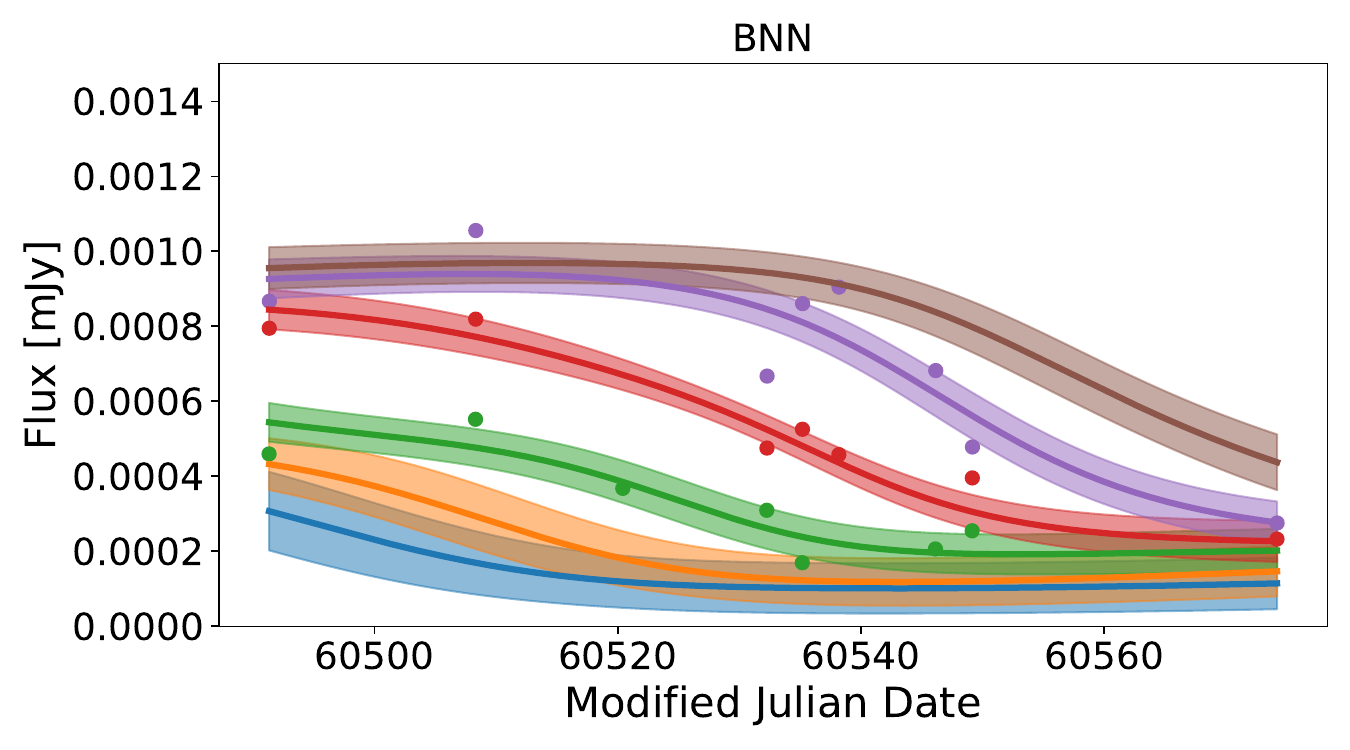}{(e)}
\includegraphics[width=0.47\linewidth]{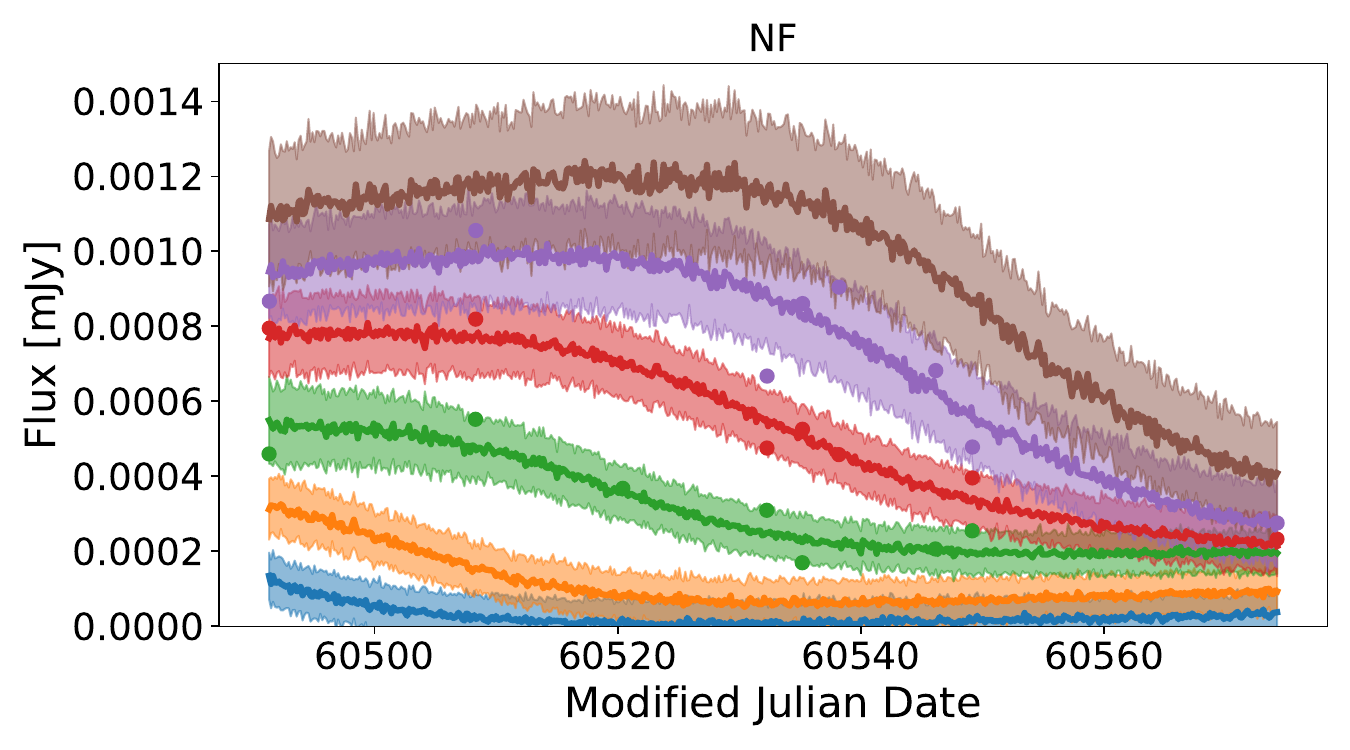}{(f)}

\caption{Examples of PLAsTiCC light curve (ID 4551) approximations using different methods: (a) the light curve before approximation; (b) GP; (c) MLP (sklearn); (d) MLP (pytorch); (e) BNN; and (f) NF. The points represent measurements in the corresponding passbands. Solid lines are the estimated mean $\mu(x)$ values. The shaded areas represent the $\pm 3\sigma(x)$ uncertainty band.}
\label{fig:plastic_4551_appendix}
\end{figure*}

\begin{figure*}
\includegraphics[width=0.47\linewidth]{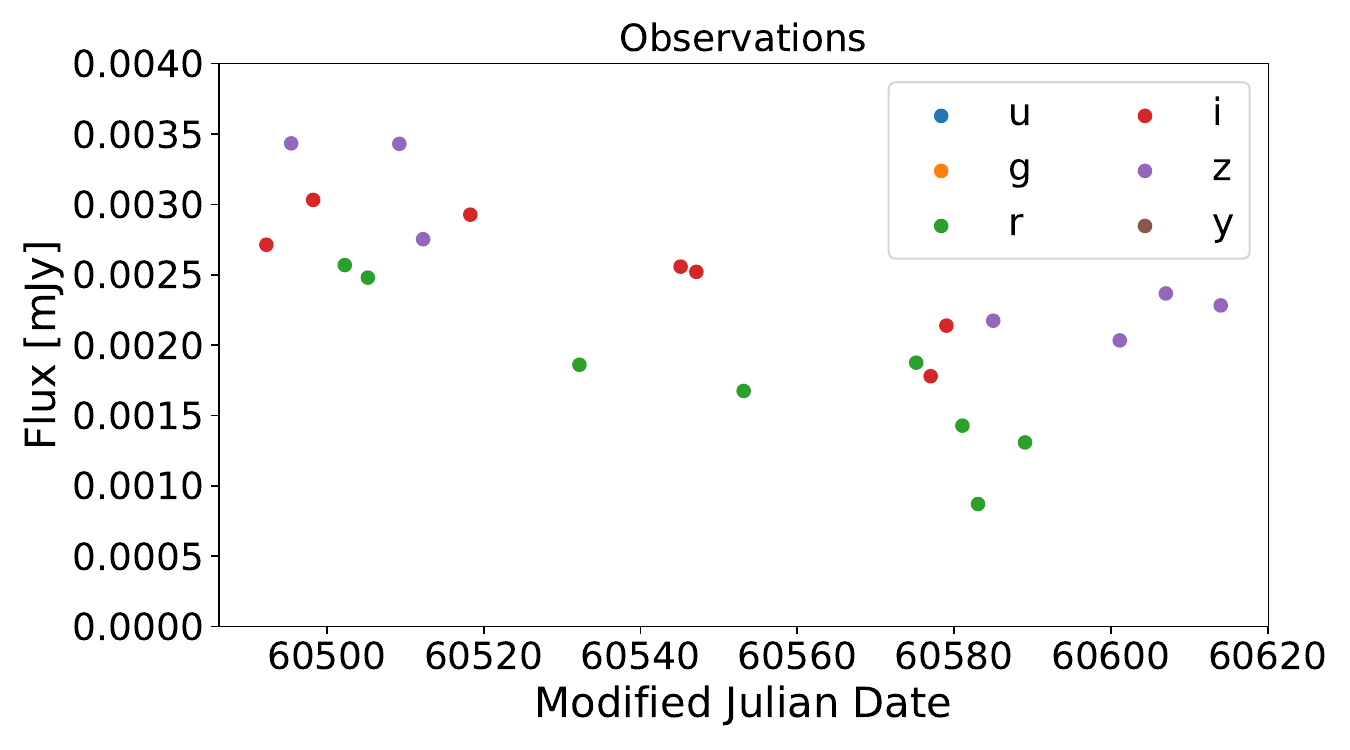}{(a)}
\includegraphics[width=0.47\linewidth]{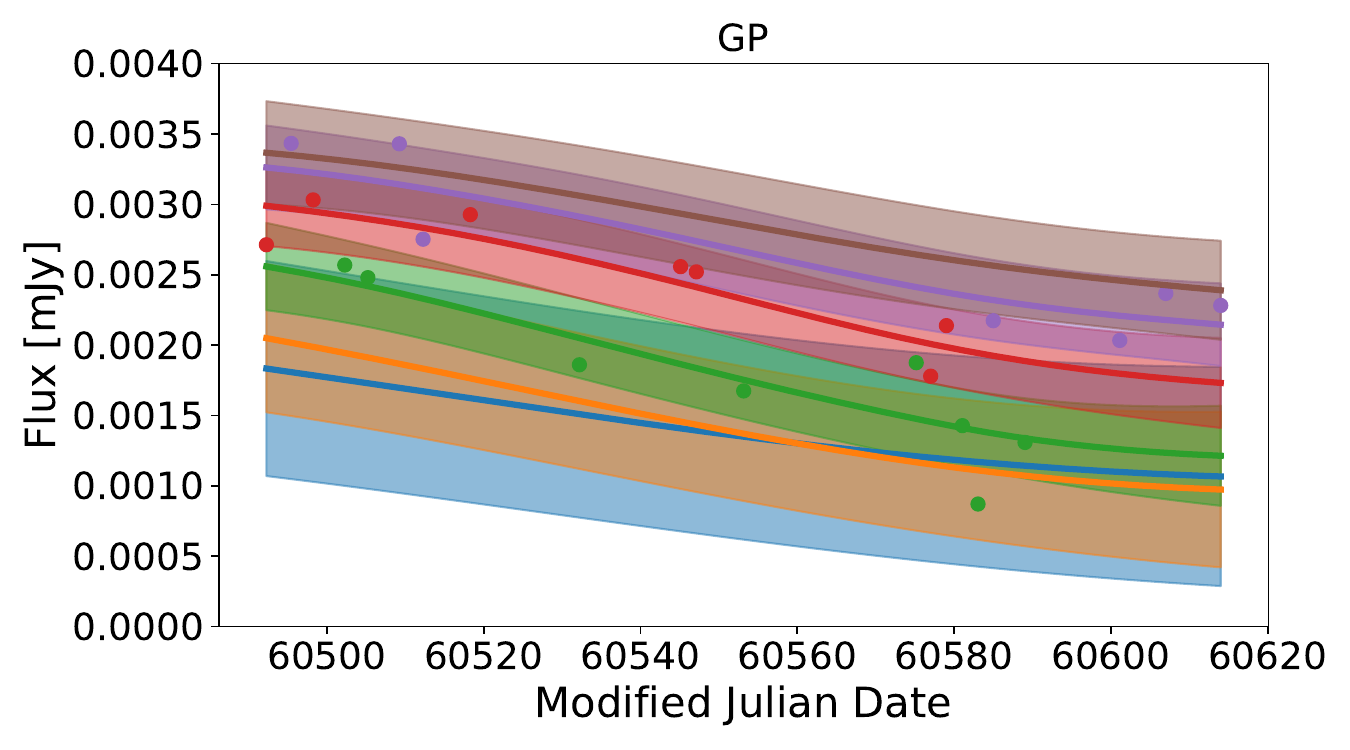}{(b)}
\includegraphics[width=0.47\linewidth]{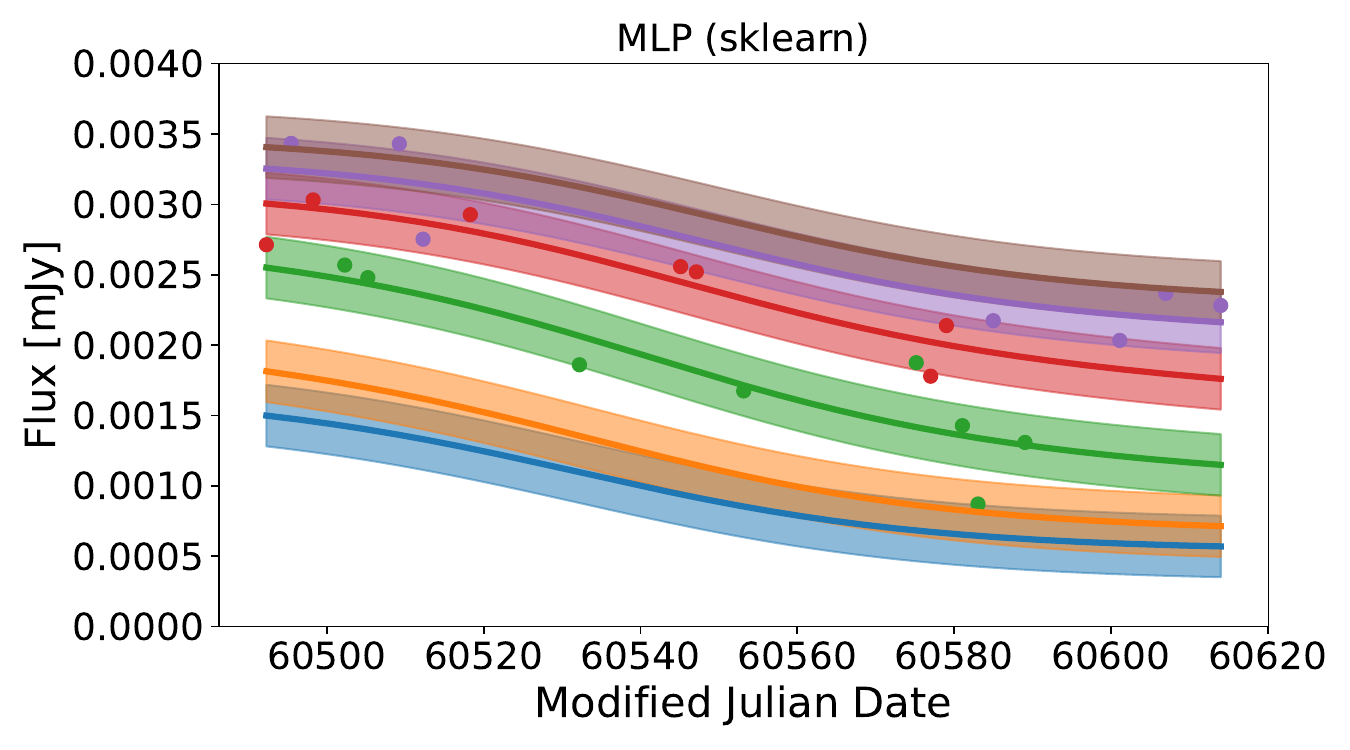}{(c)}
\includegraphics[width=0.47\linewidth]{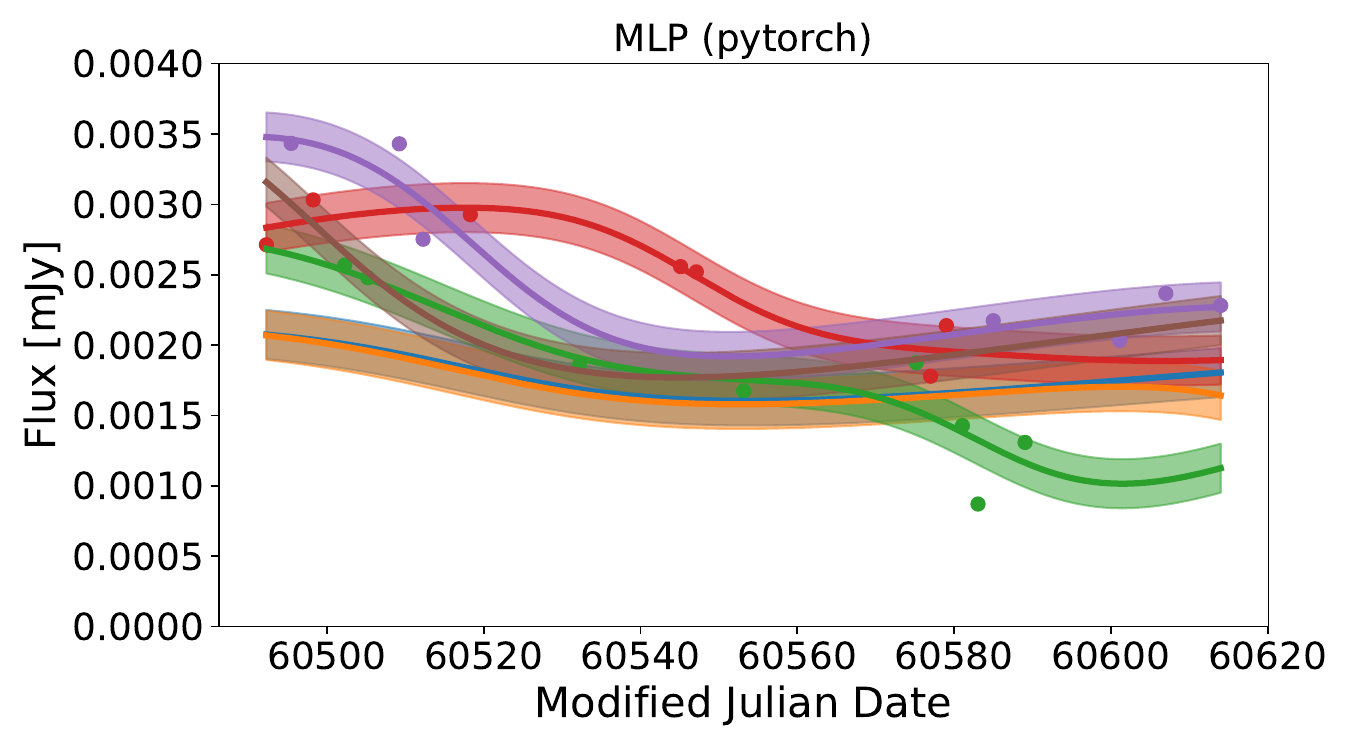}{(d)}
\includegraphics[width=0.47\linewidth]{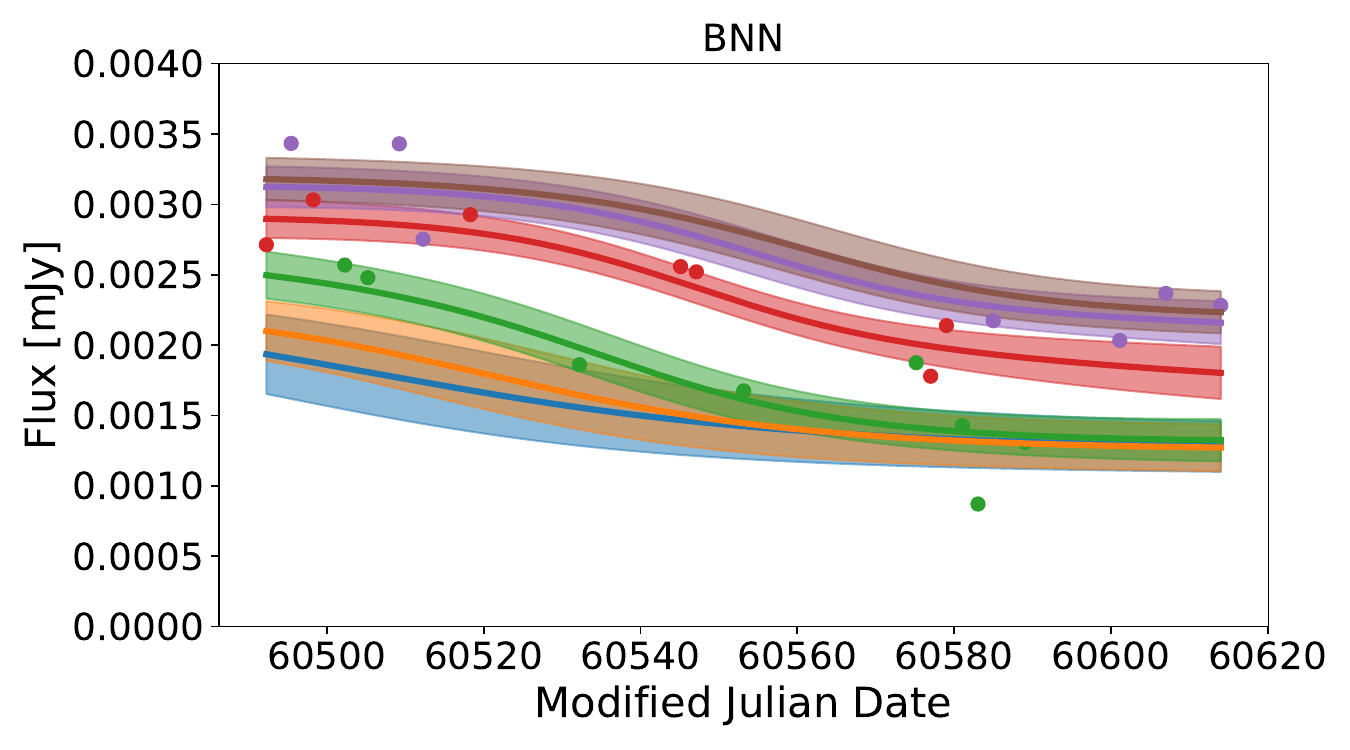}{(e)}
\includegraphics[width=0.47\linewidth]{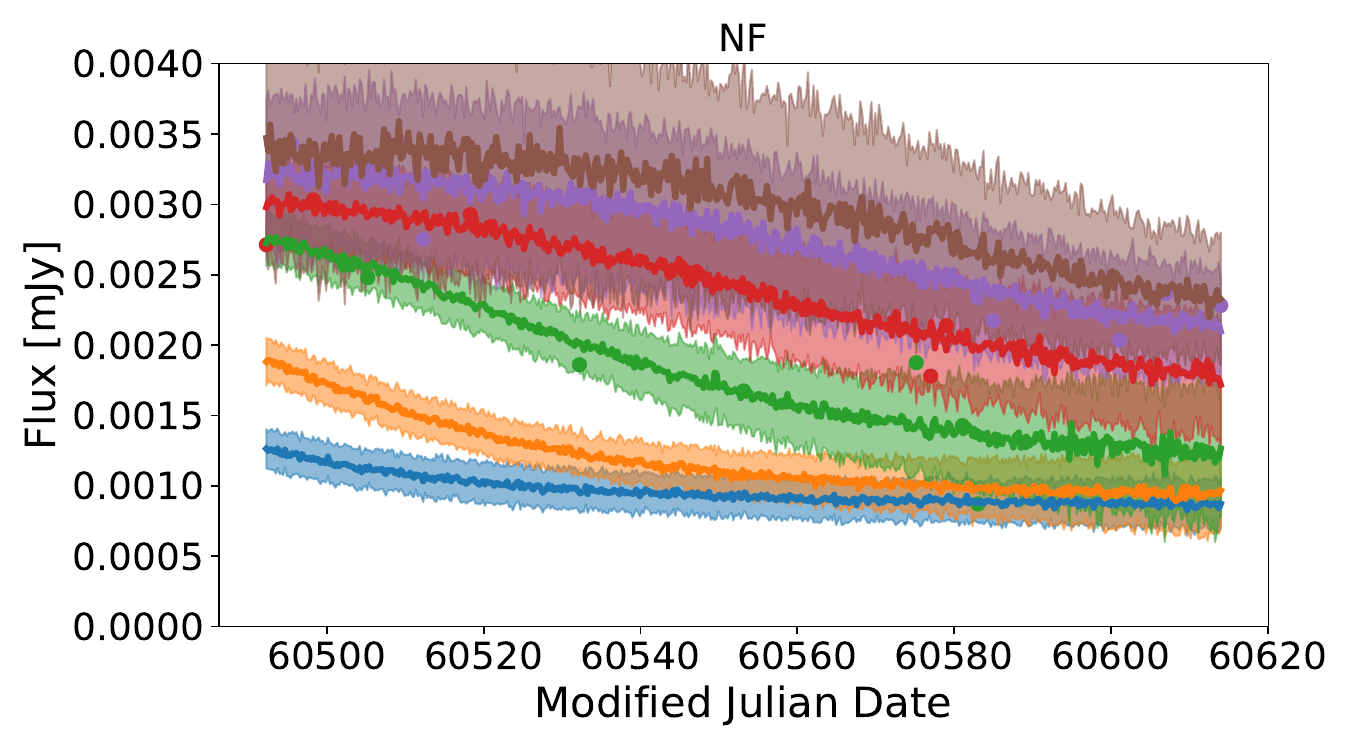}{(f)}

\caption{Examples of PLAsTiCC light curve (ID 31087708) approximations using different methods: (a) the light curve before approximation; (b) GP; (c) MLP (sklearn); (d) MLP (pytorch); (e) BNN; and (f) NF. The points represent measurements in the corresponding passbands. Solid lines are the estimated mean $\mu(x)$ values. The shaded areas represent the $\pm 3\sigma(x)$ uncertainty band.}
\label{fig:plastic_31087708_appendix}
\end{figure*}

\begin{figure*}
\includegraphics[width=0.47\linewidth]{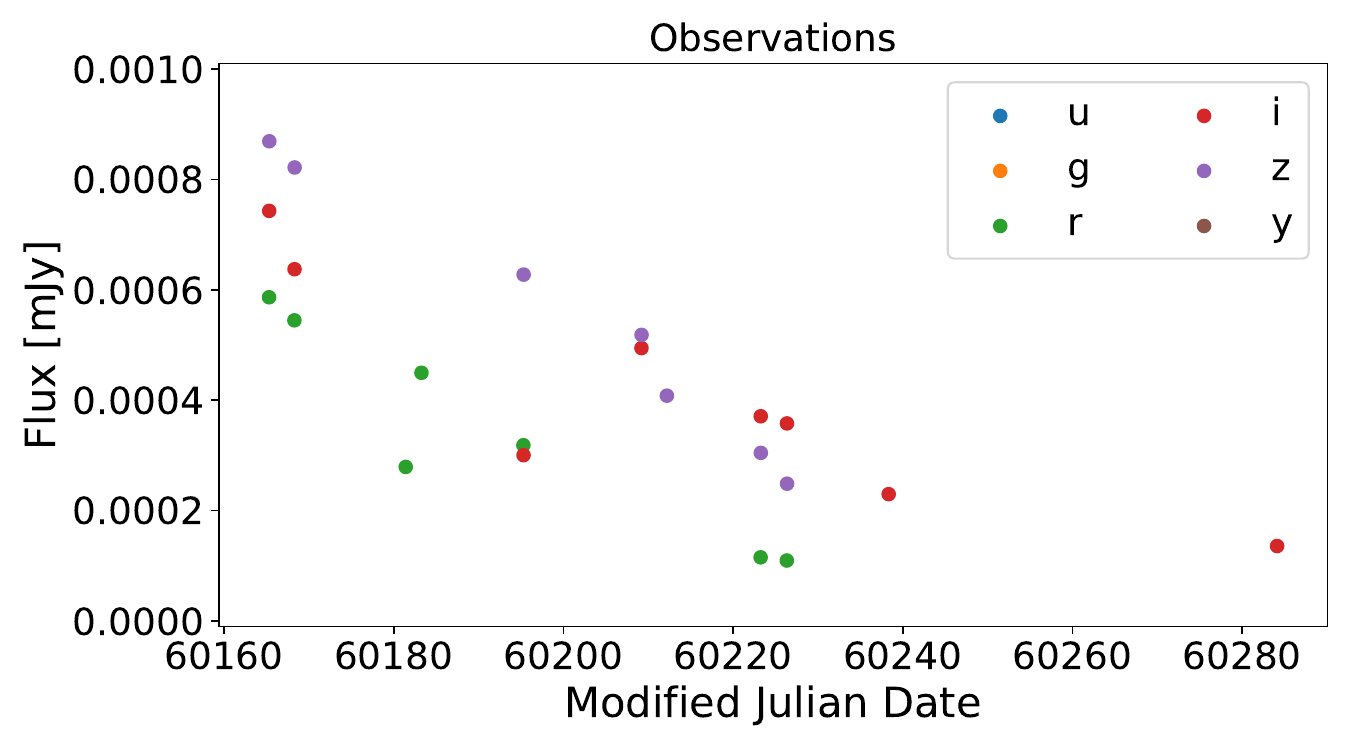}{(a)}
\includegraphics[width=0.47\linewidth]{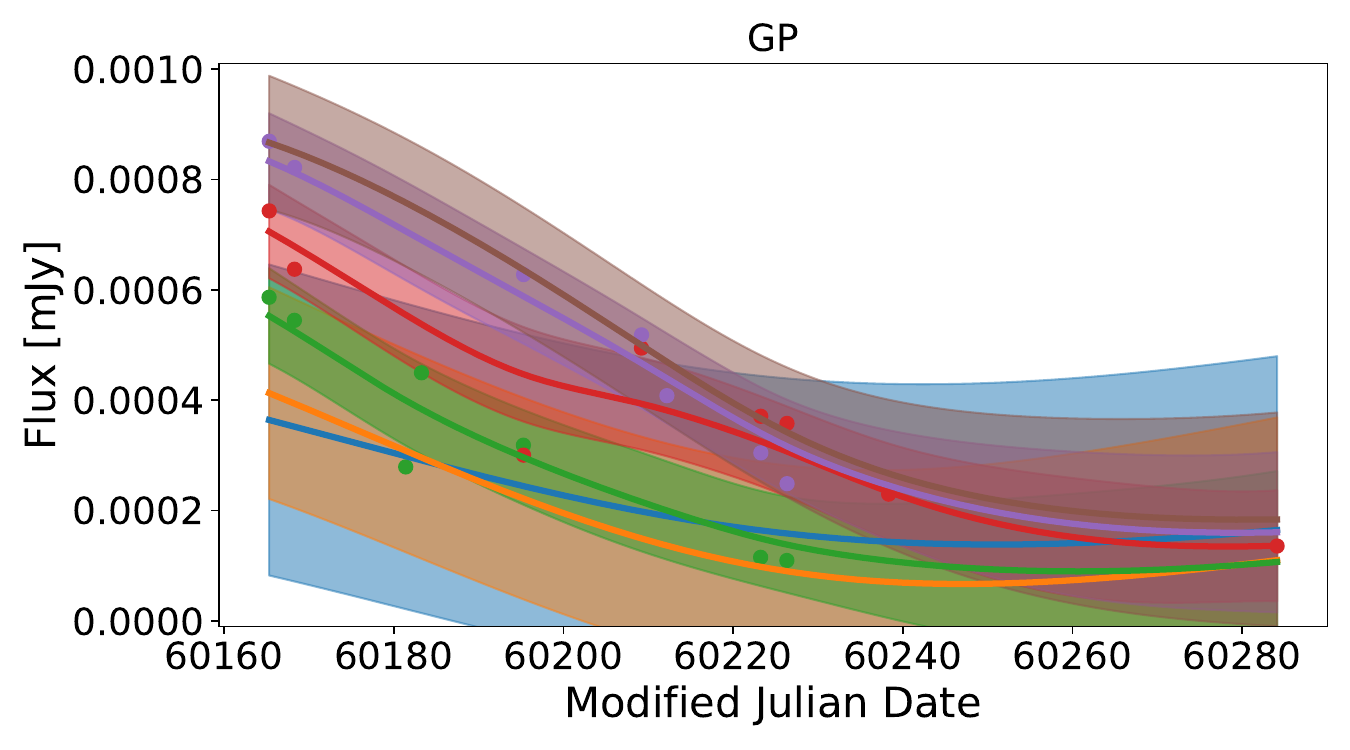}{(b)}
\includegraphics[width=0.47\linewidth]{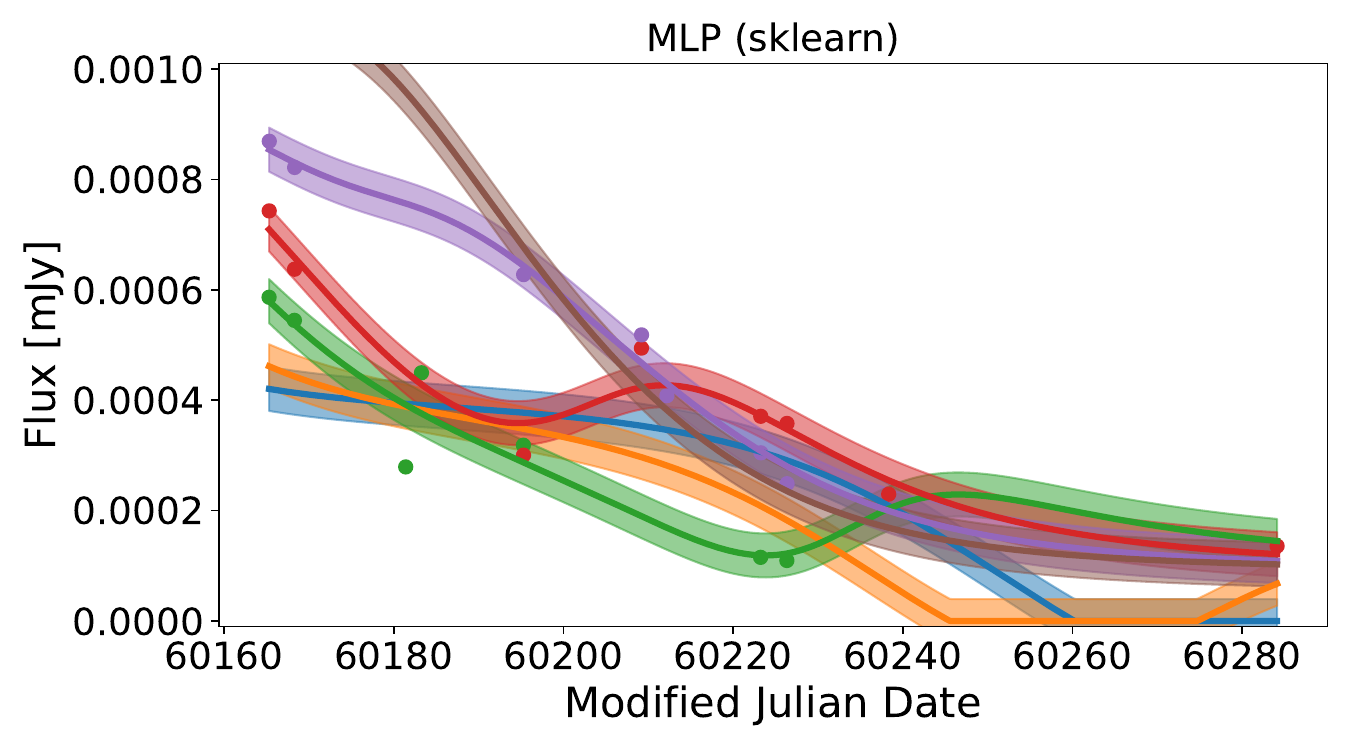}{(c)}
\includegraphics[width=0.47\linewidth]{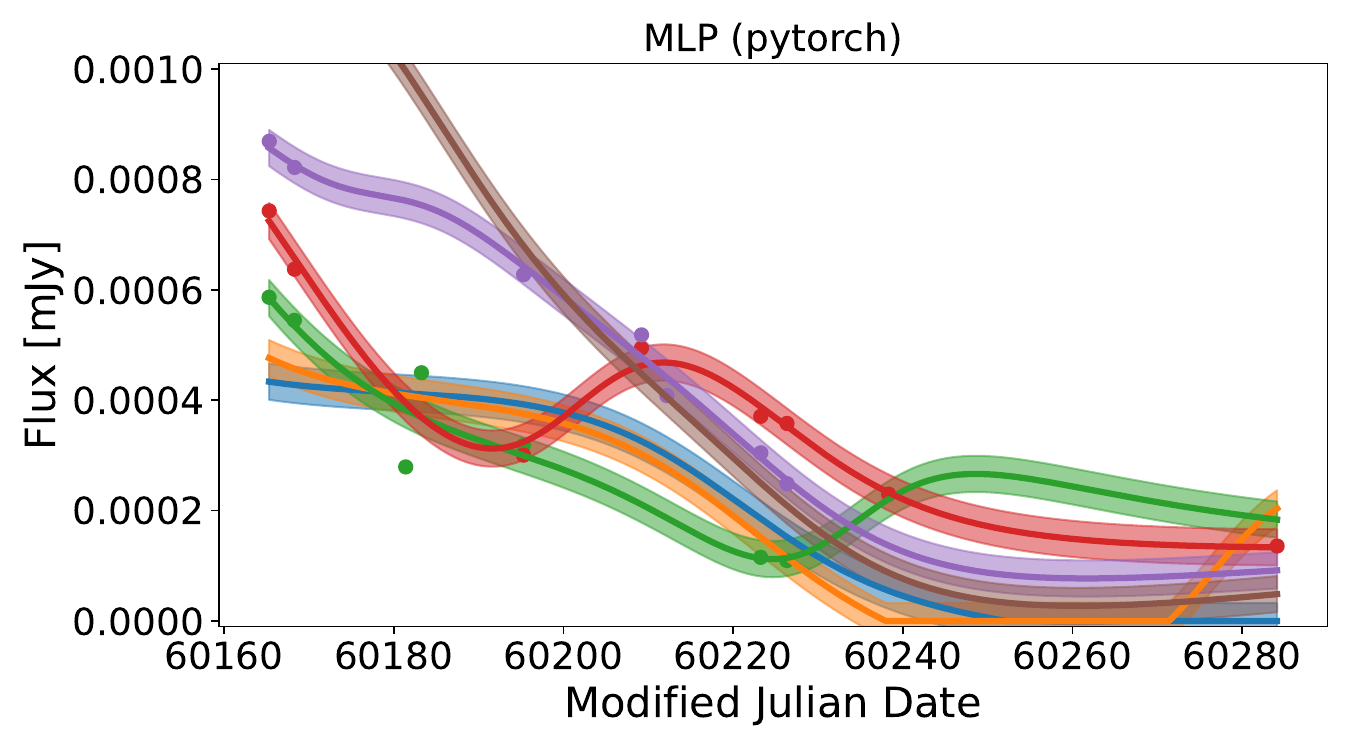}{(d)}
\includegraphics[width=0.47\linewidth]{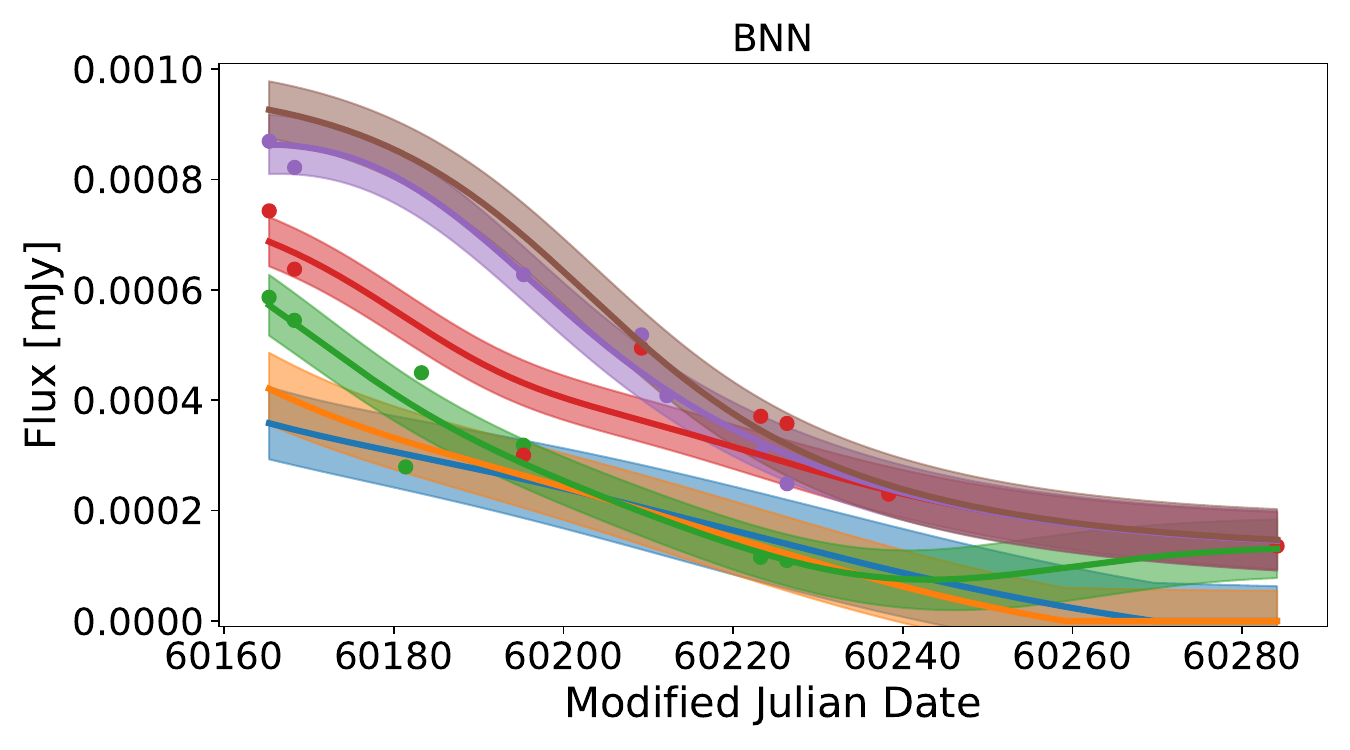}{(e)}
\includegraphics[width=0.47\linewidth]{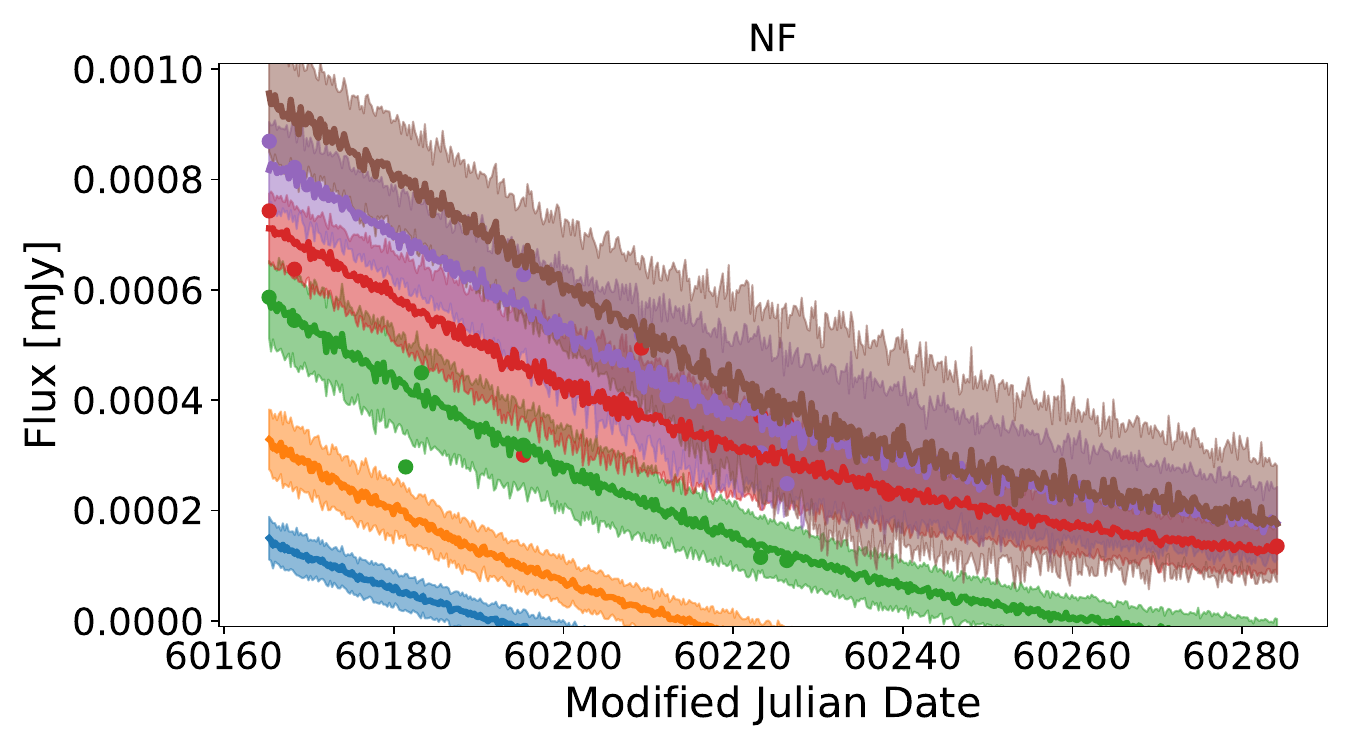}{(f)}

\caption{Examples of PLAsTiCC light curve (ID 68731) approximations using different methods: (a) the light curve before approximation; (b) GP; (c) MLP (sklearn); (d) MLP (pytorch); (e) BNN; and (f) NF. The points represent measurements in the corresponding passbands. Solid lines are the estimated mean $\mu(x)$ values. The shaded areas represent the $\pm 3\sigma(x)$ uncertainty band.}
\label{fig:plastic_68731_appendix}
\end{figure*}

\begin{figure*}
\includegraphics[width=0.47\linewidth]{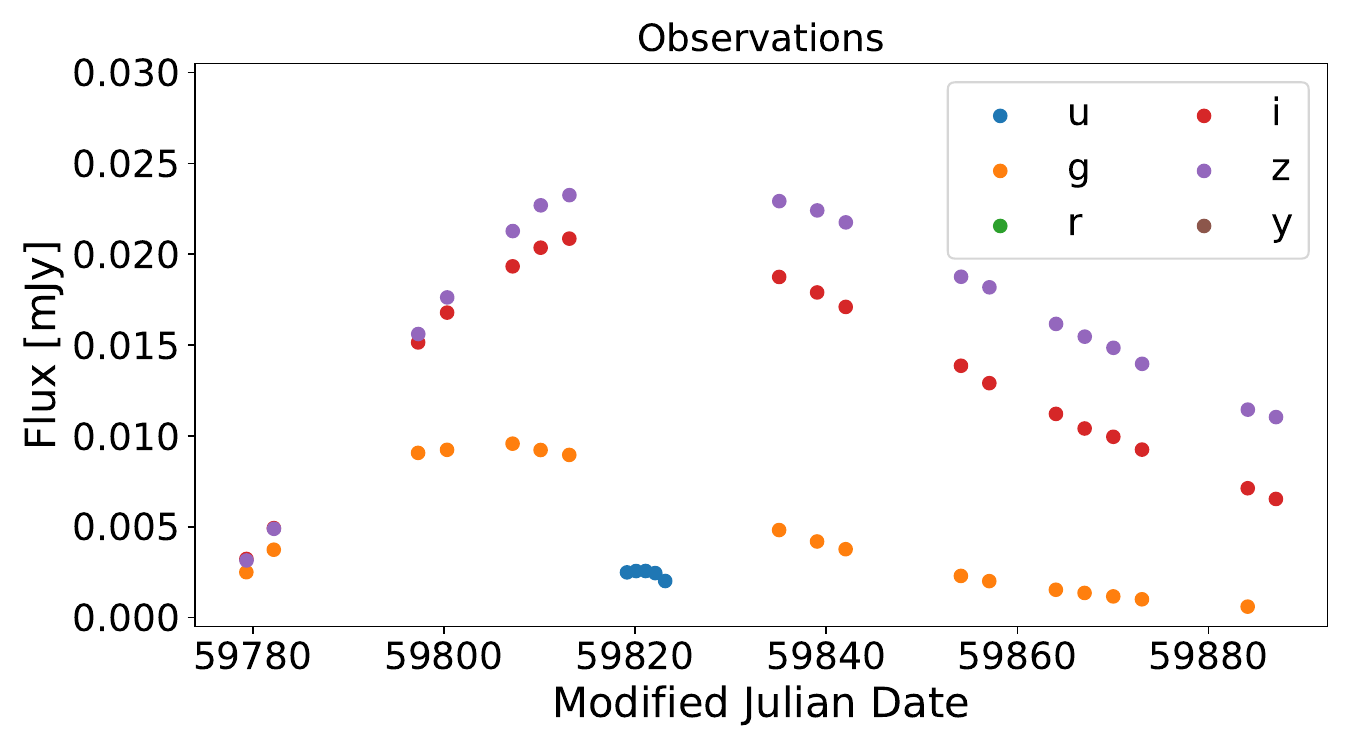}{(a)}
\includegraphics[width=0.47\linewidth]{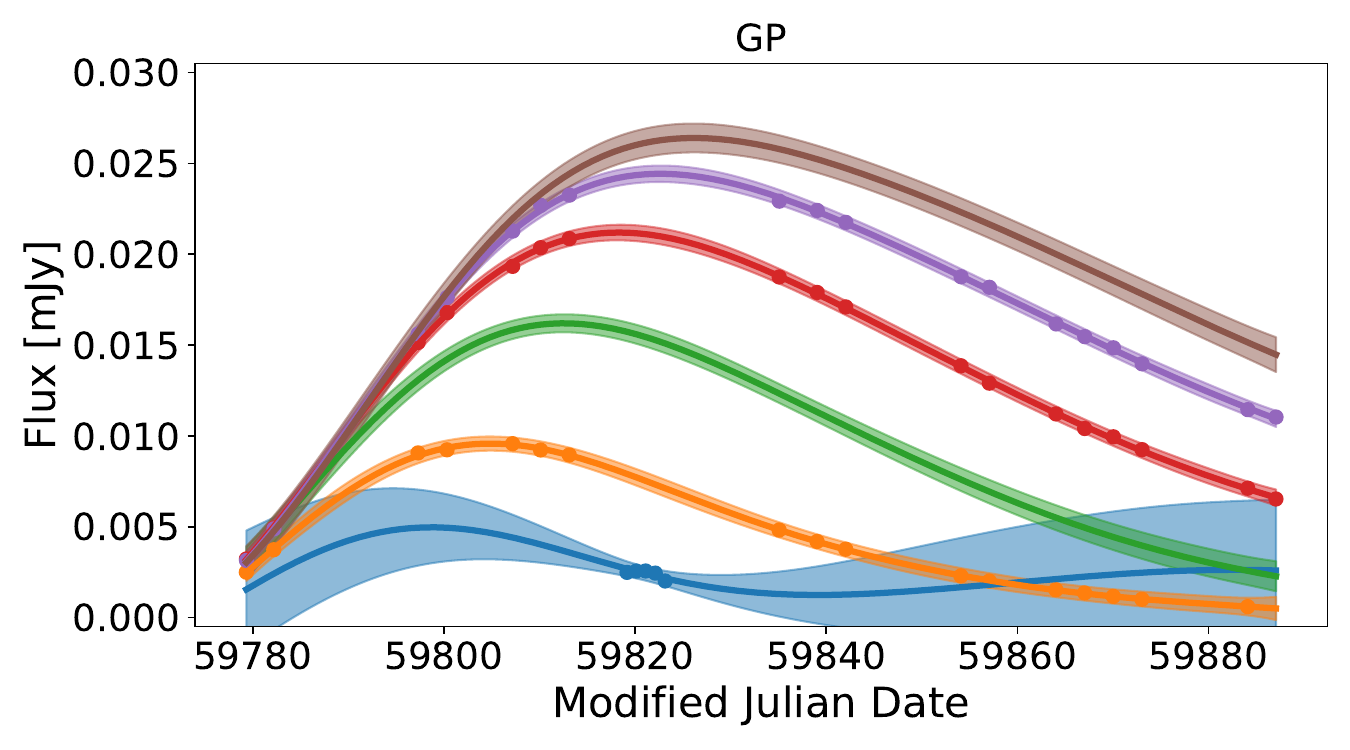}{(b)}
\includegraphics[width=0.47\linewidth]{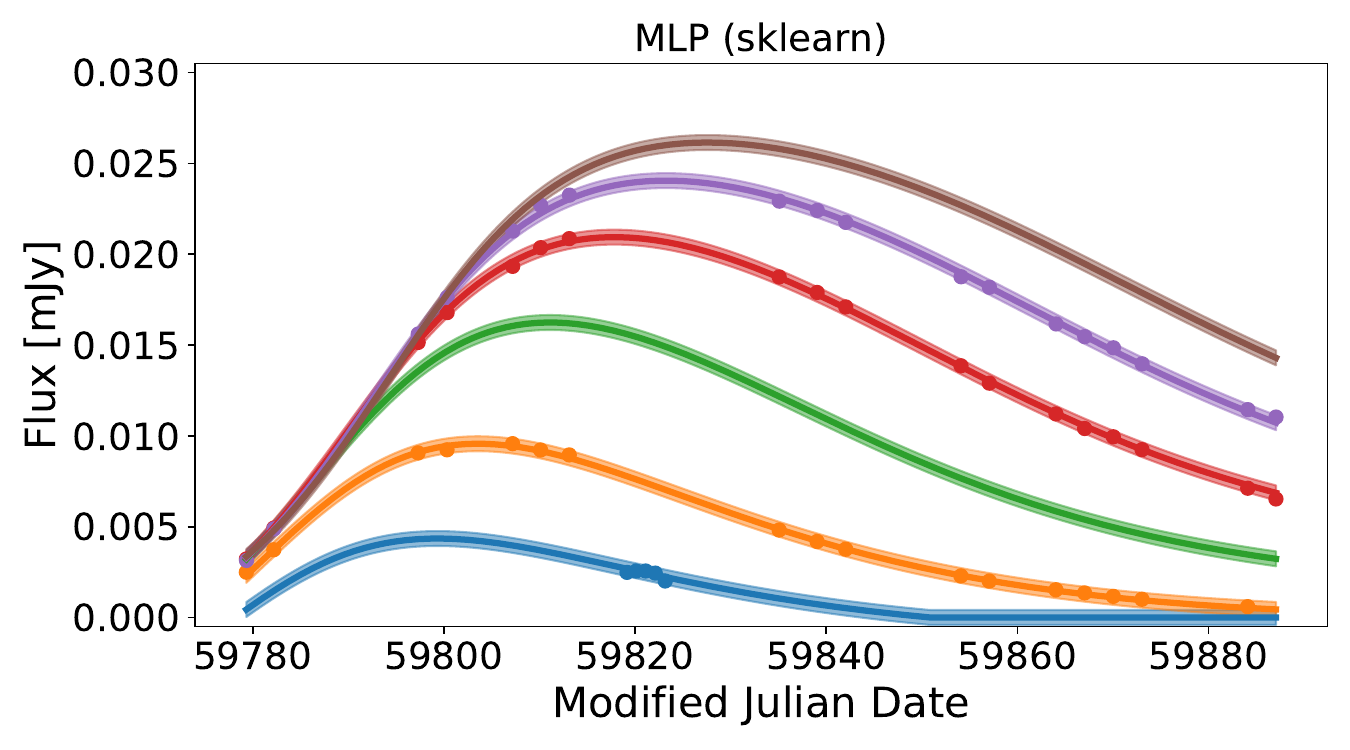}{(c)}
\includegraphics[width=0.47\linewidth]{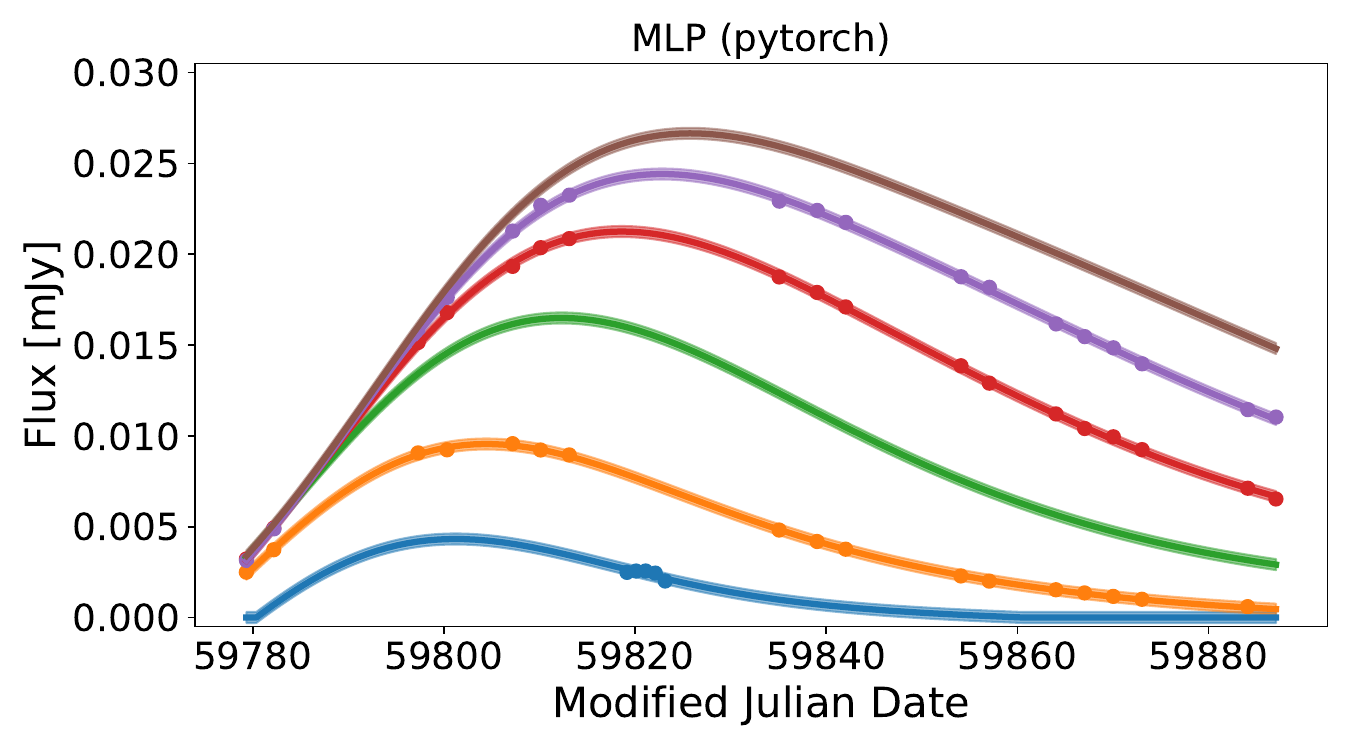}{(d)}
\includegraphics[width=0.47\linewidth]{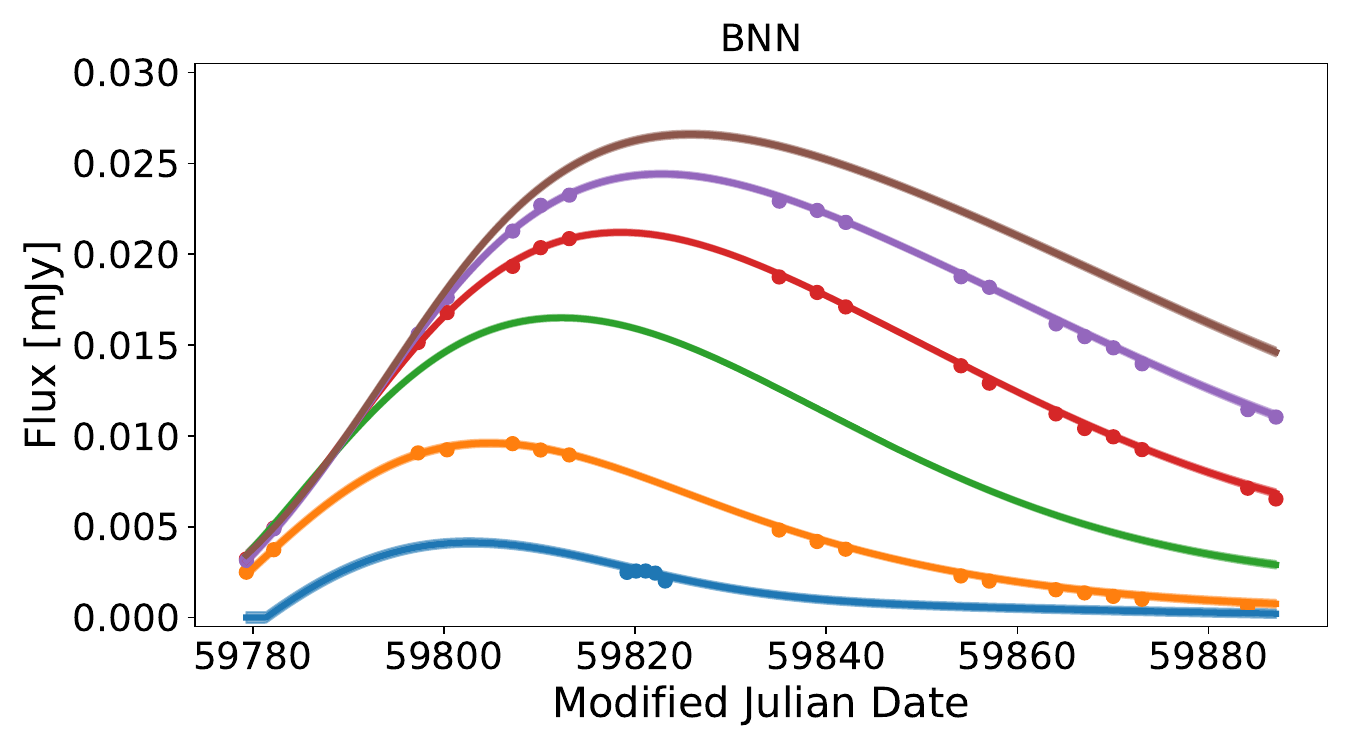}{(e)}
\includegraphics[width=0.47\linewidth]{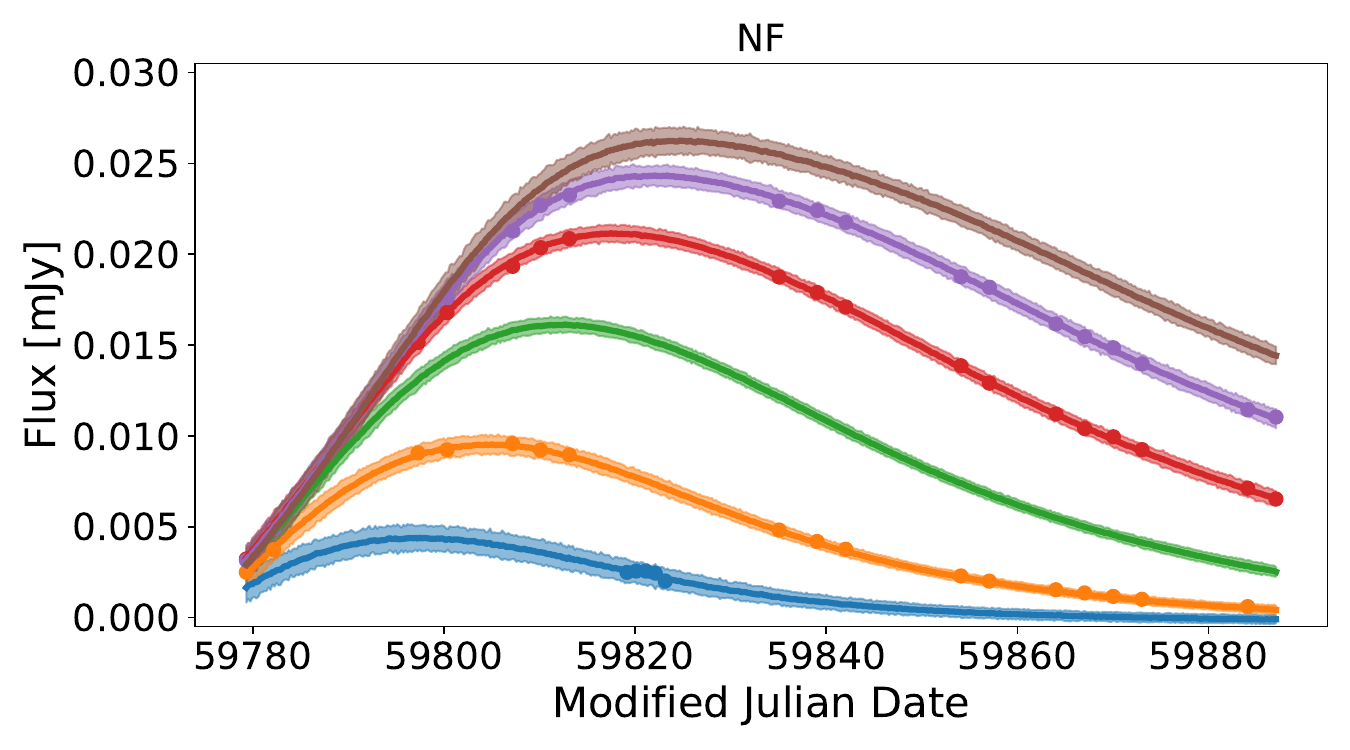}{(f)}

\caption{Examples of PLAsTiCC SN Ibc (ID 34299) light curve approximations using different methods: (a) the light curve before approximation; (b) GP; (c) MLP (sklearn); (d) MLP (pytorch); (e) BNN; and (f) NF. Measurements in the $r$ and $y$ passbands are removed. The methods interpolate and extrapolate the predictions in these passbands, respectively. The points represent measurements in the corresponding passbands. Solid lines are the estimated mean $\mu(x)$ values. The shaded areas represent the $\pm 3\sigma(x)$ uncertainty band.}
\label{fig:empty_passpands_plastic}
\end{figure*}

\begin{figure*}
\includegraphics[width=0.47\linewidth]{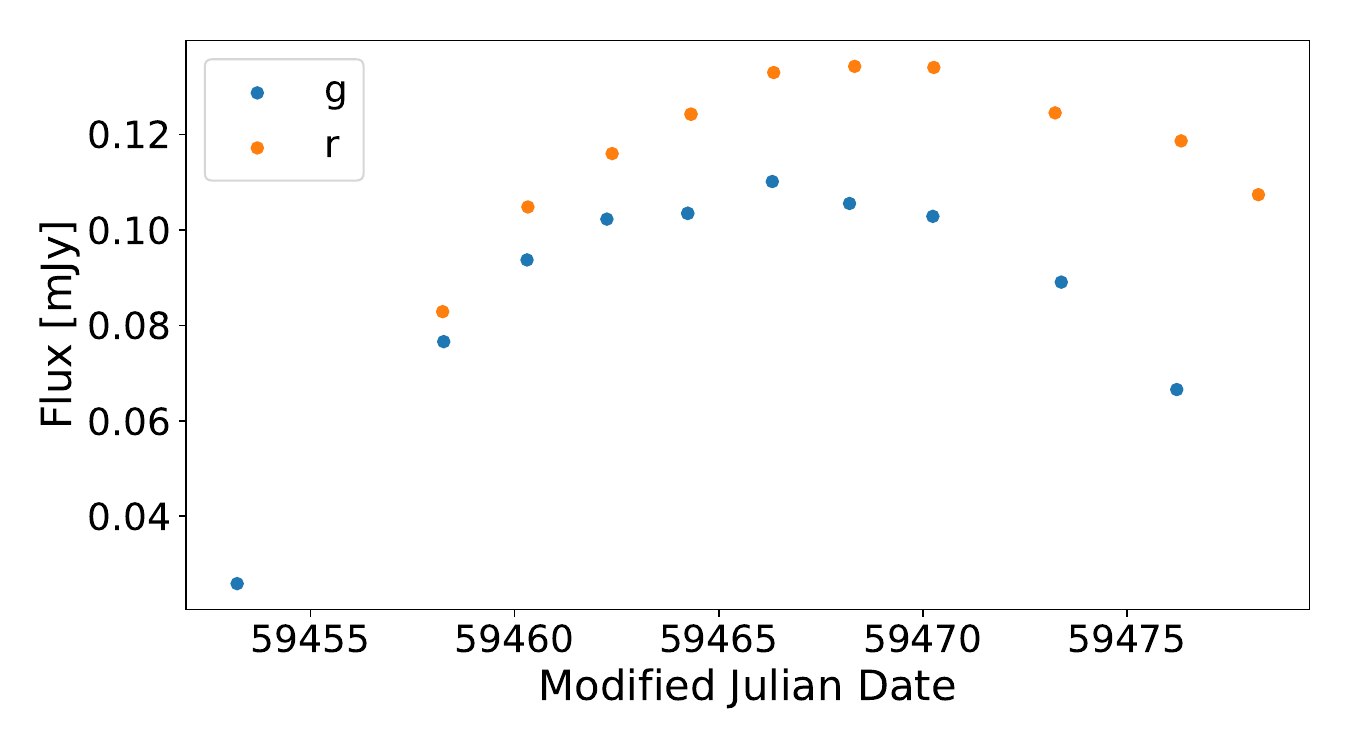}{(a)}
\includegraphics[width=0.47\linewidth]{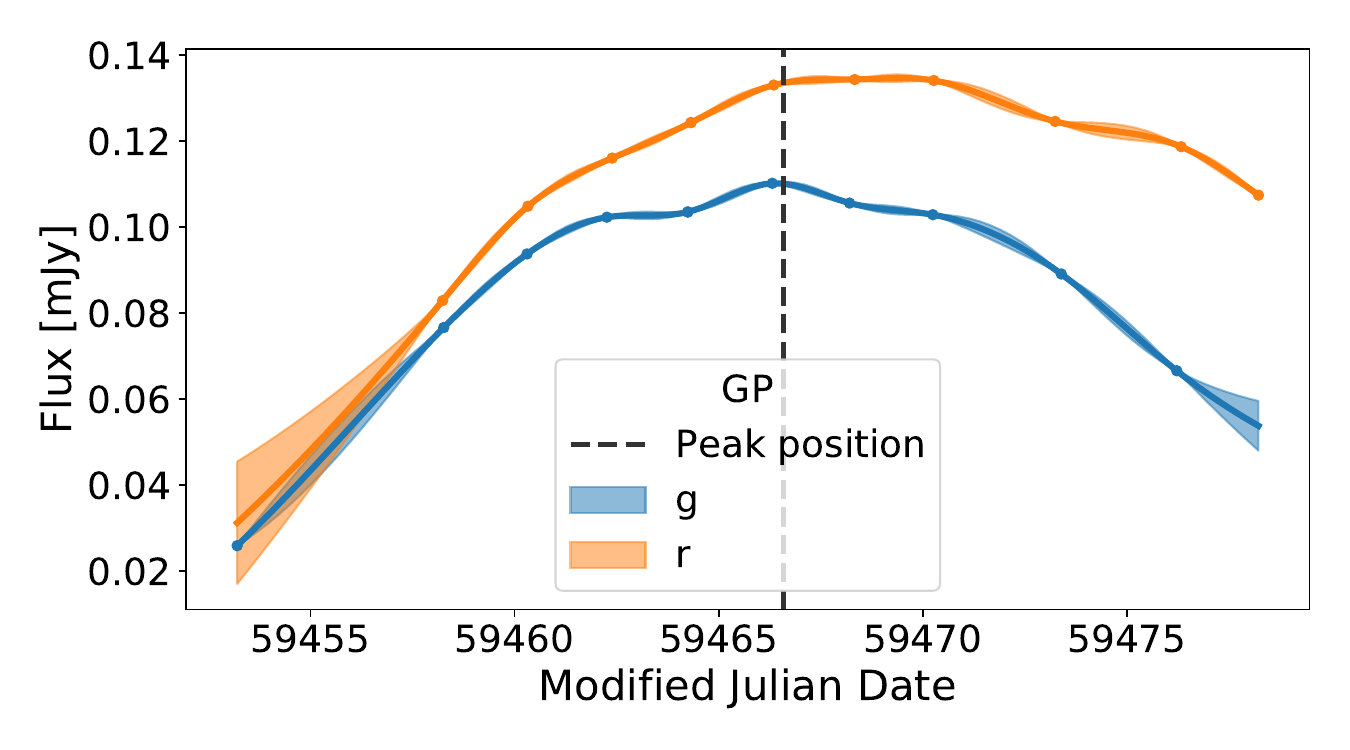}{(b)}
\includegraphics[width=0.47\linewidth]{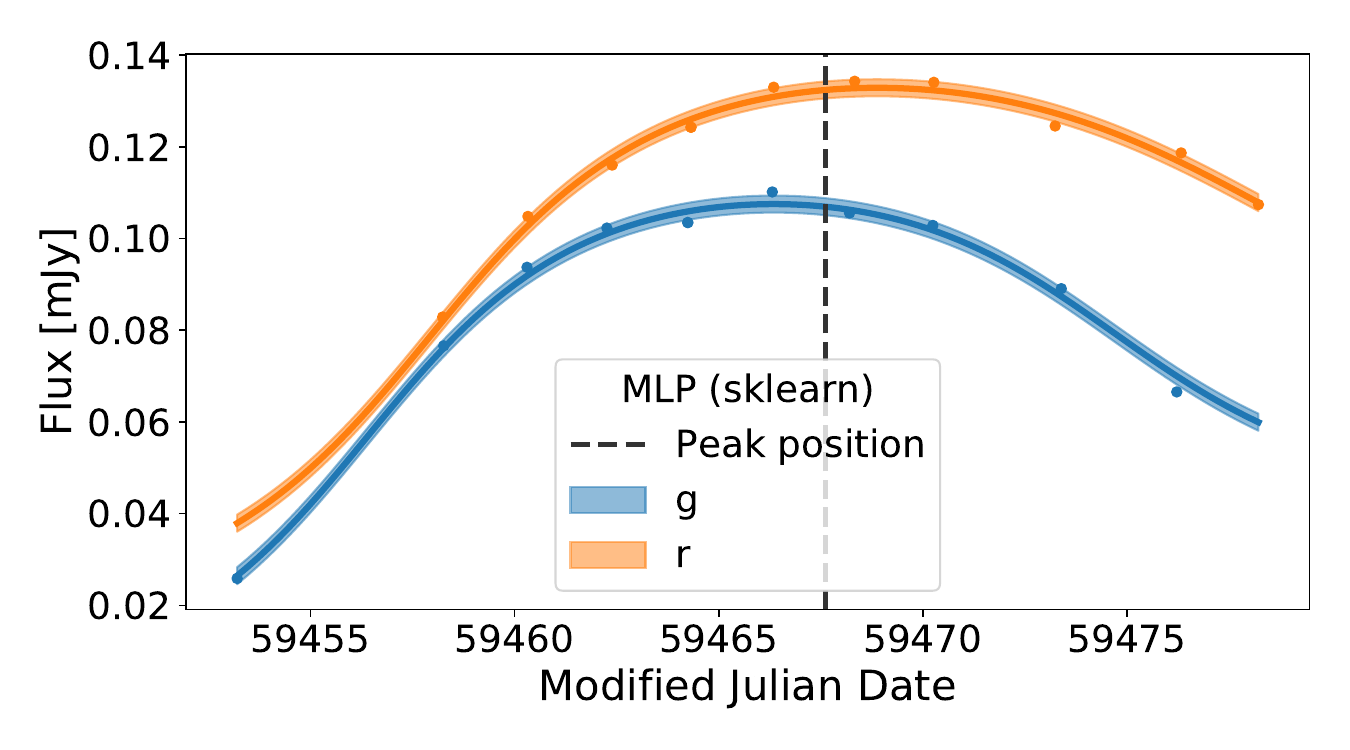}{(c)}
\includegraphics[width=0.47\linewidth]{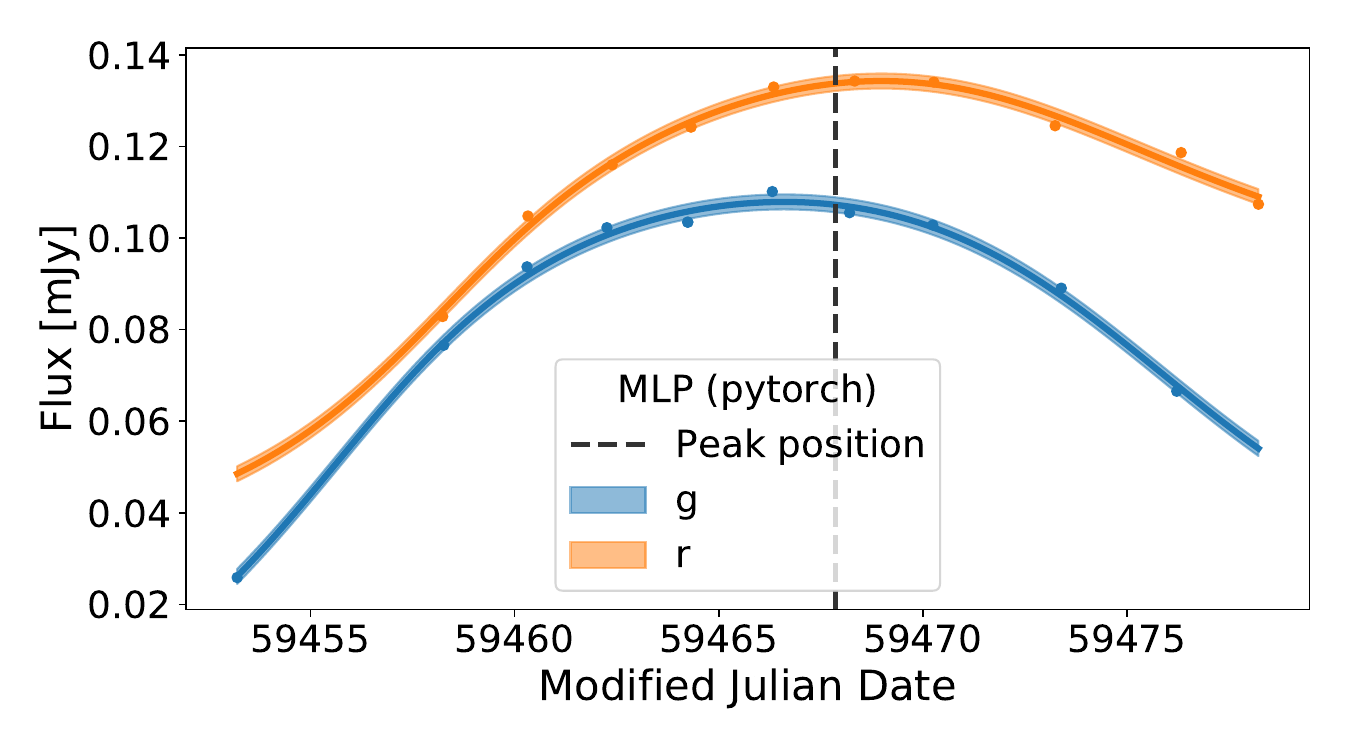}{(d)}
\includegraphics[width=0.47\linewidth]{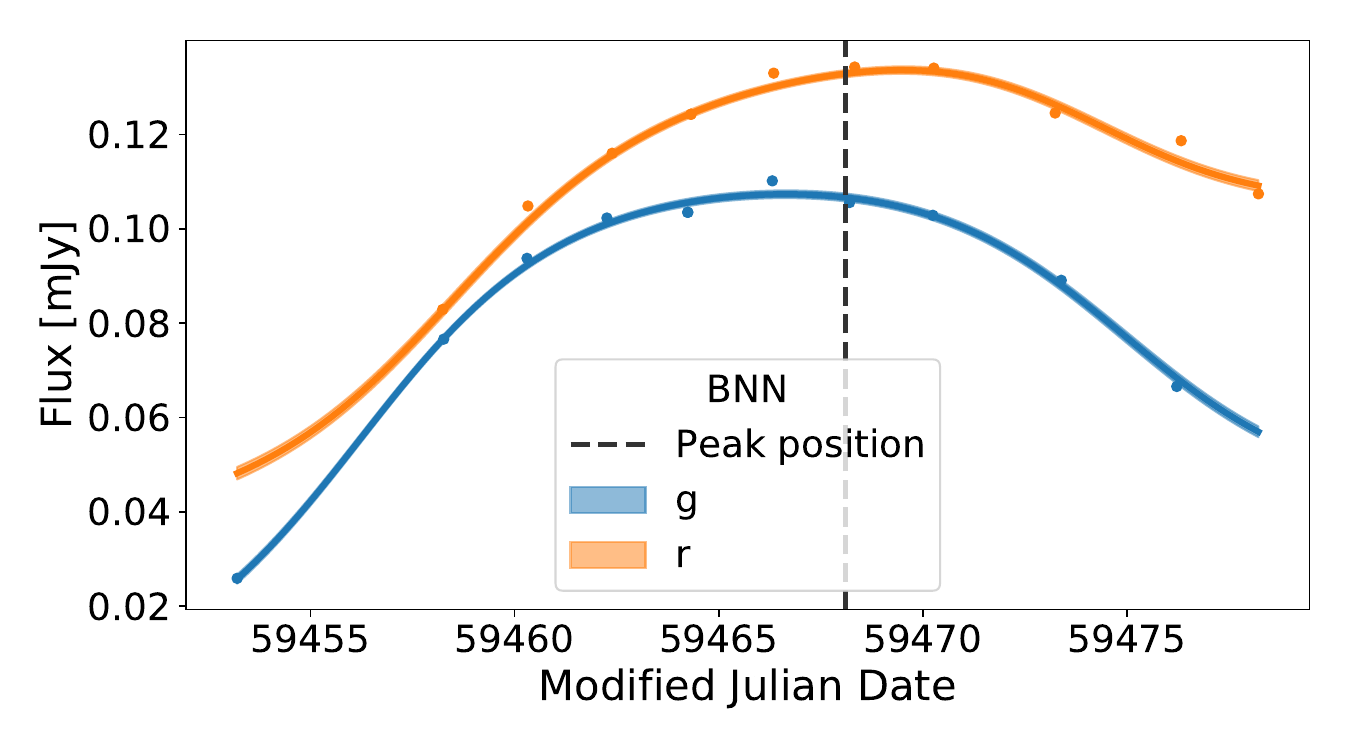}{(e)}
\includegraphics[width=0.47\linewidth]{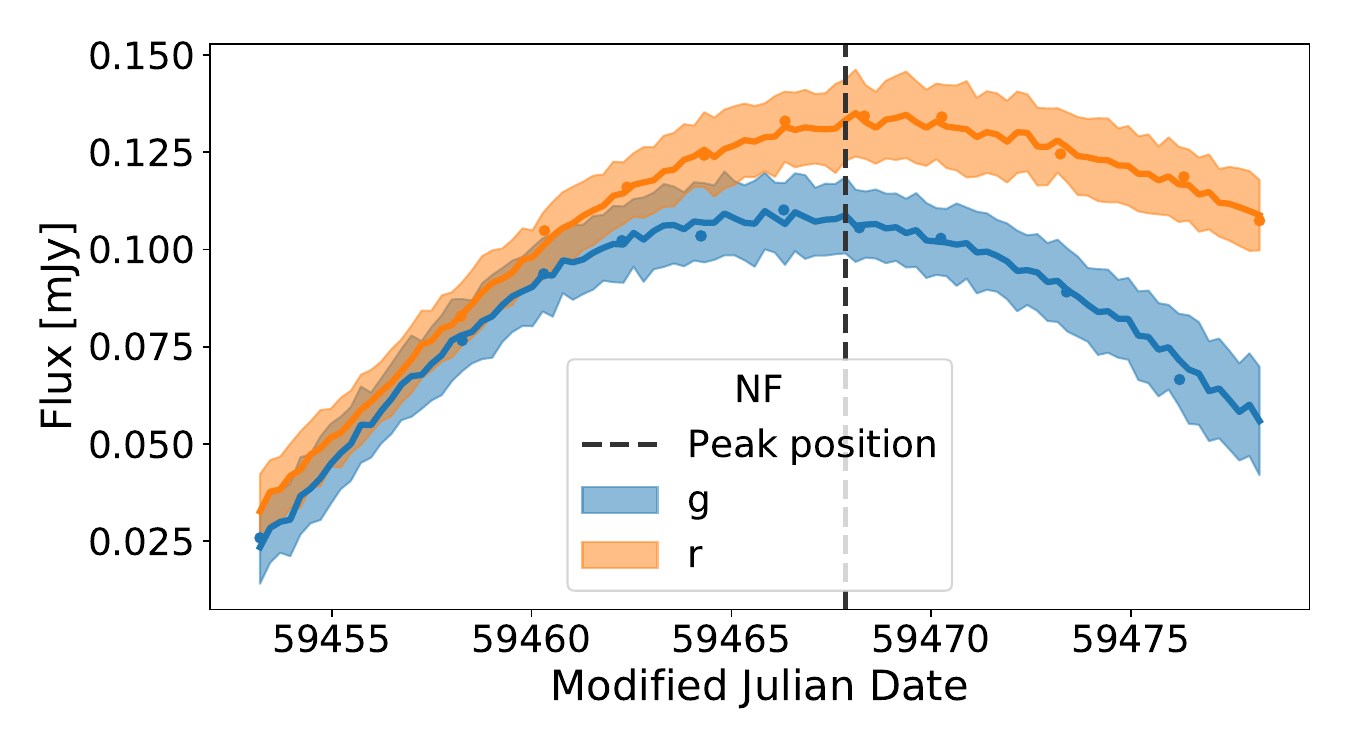}{(f)}

\caption{Examples of SN~Ia ZTF21abvrfsr light curve approximations using different methods: (a) light curve before approximation; (b) GP; (c) MLP (sklearn); (d) MLP (pytorch); (e) BNN; and (f) NF. The points represent measurements in the corresponding passbands. The solid lines are the estimated mean $\mu(x)$ values. The shaded areas represent the $\pm 1\sigma(x)$ uncertainty band.}
\label{fig:ZTF21abvrfsr_22_points}
\end{figure*}

\end{appendix}
\end{document}